\documentclass[%
 reprint,
 amsmath,amssymb,
pra,
]{revtex4-2}
\pdfoutput=1
\usepackage{graphicx}%
\usepackage{dcolumn}
\usepackage{bm}

\usepackage[english]{babel}
\usepackage{ulem}

\usepackage{xcolor}
\newcommand{\rf}[1]{(\ref{#1})}

\newcommand{\beq}{\begin{equation}}
\newcommand{\eeq}{\end{equation}}
\newcommand{\beqr}{\begin{eqnarray}}
\newcommand{\eeqr}{\end{eqnarray}}

\newcommand{\lb}[1]{\label{#1}}

\newcommand{\bc}{\begin{center}}
\newcommand{\ec}{\end{center}}
\newcommand{\ct}[1]{\cite{#1}}

\usepackage{natbib}
\usepackage[]{nth} 
\usepackage[colorinlistoftodos,prependcaption,textsize=scriptsize]{todonotes} %

\begin{document}

\preprint{APS/123-QED}

\title{Quantum Langevin approach for superradiant nanolasers}

\author{Igor Protsenko$^{1,2}$, Alexander Uskov$^{1}$, Emil C.  Andr\'{e}$^{2}$, Jesper M{\o}rk$^{2,3}$, and  Martijn Wubs$^{2,3}$}
\affiliation{%
$^1$Lebedev Physical Institute of RAS, Leninsky prospect, 53, Moscow, 119991, Russia}
\affiliation{$^2$Department of Photonics Engineering, Technical University of Denmark, DK-2800 Kgs. Lyngby, Denmark}%
\affiliation{$^3$ NanoPhoton - Center for Nanophotonics, Technical University of Denmark, Ørsteds Plads 345A, DK-2800 Kgs. Lyngby, Denmark}

\date{\today}

\begin{abstract}
A new approach for analytically solving quantum nonlinear Langevin equations is proposed and applied to calculations of spectra of superradiant  lasers where collective effects play an important role.  
We calculate lasing spectra for arbitrary pump rates and recover well-known results such as the pump dependence of the laser linewidth across the threshold region. We predict new sideband peaks in the spectrum of superradiant lasers with large  relaxation oscillations as well as new nonlinear  structures in the lasing spectra for weak pump rates. 
Our approach sheds new light on the importance of 
 population fluctuations in the narrowing of the laser linewidth,   in the 
 structure of the lasing spectrum, and in the  transition to coherent operation. 
\begin{description}
\item[Keywords]
laser spectra, superradiance, nanolasers 
\item[OCIS codes] 140.0140, 140.3430, 270.0270
\end{description}
\end{abstract}

\maketitle


\section{\label{sec:level1}Introduction}

Progress in various technologies has enabled considerable size reductions of lasers. Nowadays quantum dot photonic crystal~\ct{7907235,Noda260,Prieto:15,Takiguchi:16,Ota:17,Nozaki:07,Yu2017}, 
micropillar~\ct{Li2019,doi:10.1063/1.4791563,Kreinberg2017}, plasmonic~\ct{Suh2012} and other kinds of nanolasers \ct{Khajavikhan2012} are intensively investigated. This research is motivated by fundamental questions, such as the minimum size of lasers and the role of quantum effects.   The miniaturization of nanolasers is also driven by applications, in nano-electronics for example, where energy-efficient nanolasers are directly incorporated  into nano-chips~\ct{Kurosaka2010,8658143,Crosnier2017}. 
The high density of photon states in nanocavities  leads to Purcell enhancement~\ct{PhysRev.69.674} of spontaneous emission into the nanolaser mode, large gain and to the rapid increase of laser power even at small pump rates. 

Nowadays there is great interest in {\it superradiant} lasers,  
which are lasers that combine a large gain with a small  cavity  operating in the so-called bad-cavity regime~\ct{Khanin, Belyanin_1998,PhysRevA.71.053818}. In this regime, the polarization cannot necessarily be adiabatically eliminated and collective spontaneous emission into the lasing mode is significant.  
Superradiant lasers  have been experimentally realized, for example  with cold alkaline earth atoms~\ct{PhysRevX.6.011025, PhysRevA.96.013847,PhysRevA.81.033847,PhysRevA.98.063837},  rubidium atoms~\ct{Bohnet}, and with quantum dots~\ct{Jahnke} as the active medium. Superradiant lasers are less sensitive  to cavity-length fluctuations, which is important for atomic clocks~\ct{PhysRevX.6.011025,PhysRevA.96.013847,Bohnet}. Superradiance leads to  interesting collective effects, such as  excitation trapping~\ct{Bohnet, PhysRevA.81.033847} and superthermal photon statistics~\ct{corr, Jahnke,Kreinberg2017}, with possible applications in high-visibility optical imaging~\ct{PhysRevA.95.053809}. 

An analytical description of  superradiant nanolasers and their spectra is complicated by the facts that their  quantum noise is not a perturbation, that the equations are nonlinear,  and  that the polarization of the active medium cannot be adiabatically eliminated. We will address these issues in this paper, where we present an 
analytical approach to understanding 
superradiant lasers. 

The quantum theory of lasers began  with applications of methods of classical statistical radio-physics first for lasers comprising a cavity with high quality (Q) factor~\ct{PhysRev.112.1940,PhysRevLett.13.329,PhysRev.159.208} and later also for low-Q cavity lasers~\ct{Haken_book1984,Lax_book1966}. In many papers the fluctuations of amplitude and phase of laser radiation are considered separately, in the frame of rate equations, where the active-medium polarization is adiabatically eliminated~\ct{Henry1986,1072058,1071986,1071726, McKinstrie:20}. This approach is satisfactory  for usual semiconductor lasers and  leads to various analytical results, but the approximations leading to the usual rate equations are not always justified for superradiant nanolasers.  

Presently nanolasers are theoretically modeled either by rate equations as in~\ct{McKinstrie:20,doi:10.1063/1.5022958}, by numerical solution of the density matrix  equations as in~\ct{PhysRevA.81.033847,Auff_ves_2011,PhysRevA.88.063825,6603264}, or by systems of equations for  correlations as in the cluster expansion~\ct{Jahnke,PhysRevA.75.013803} or cumulant expansion~\ct{PhysRevLett.102.163601, Kirton_2018} methods. 
Numerical analysis of  superradiant emission and lasing has recently led to new and interesting results, such as  mechanical effects in photon-atom interactions~\ct{PhysRevA.101.023616}, lasing with a millihertz linewidth and rapid emitter number fluctuations~\ct{Zhang_2018},  Wigner functions for semiconductor heterostructures~\ct{Vasil_ev_2020}, transition from superradiance to regular lasing by varying the coherent and incoherent driving~\ct{Kirton_2018}, sub- and superradiance in multimode optical
waveguides~\ct{Ostermann_2019}, and photon-antibunching in the fluorescence from an optical nanofiber-tip~\ct{Suarez_2019}. However, complementary analytical methods to model nanolasers without adiabatic elimination of polarization, that would apply to superradiant nanolasers, are not well developed.  

Here we use Heisenberg-Langevin equations, which are very convenient for the description of lasers~\ct{Scully}, to describe systems ranging from LEDs to superradiant nanolasers.
The method also describes the limit of non-superradiant lasers, where it will reproduce some well-known results. We follow the approach by Lax, see for example Ref.~\ct{Lax_book1966}: operators are treated as stochastic variables, while quantum properties such as non-vanishing commutation relations are taken into account by diffusion coefficients.  

Our first new application of the method will be the description of  both the lasing field and the active-medium polarization by symmetric (S) and anti-symmetric (A) combinations of quadrature operators.  Quadratures of the polarization have been used in Lamb's semiclassical laser theory~\ct{PhysRev.159.208}, while quadratures of the electromagnetic field were used, for example, in the analysis of the driven Van der Pol oscillator 
applied to lasers in~\ct{1071726} and in quantum optics to squeezed states of the radiation field~\ct{Scully}.  Symmetric and anti-symmetric combinations of the quadratures of a quantum field  are also used in entanglement  criteria~\ct{PhysRevLett.84.2722, Feyisa2020}. Our work is an extension since, to our knowledge, quadratures {as well as their S/A combinations for both the lasing field and polarization combined have not been used in laser theory before. The approach has several advantages: it does not require a quantum phase operator~\ct{PhysRevA.39.1665},  and furthermore we find the equations for the  S/A quadrature combinations less cumbersome than for the density matrix~\ct{Scully}. 

Our second new application is the linearization of the quantum Langevin equations, where fluctuations are not necessarily small compared to mean values. 
This goes beyond a small-signal analysis as, for example, in Refs.~\ct{RevModPhys.68.127, PhysRevA.47.1431, 1071726} and requires other approximations. We describe our method in detail, and it may be useful also for other physical systems with resonant nonlinear interactions of light with matter, see examples in Ref.~\ct{Butylkin}.

Section~\ref{S1} introduces  
the quantum Maxwell-Bloch equations  with dissipation for a two-level laser and we rederive some key  results of semiclassical laser theory~\ct{Yariv, Scully} to be used for comparison later in the paper. 

In Sec.~\ref{Below_th} we demonstrate a method, used before in~\ct{PhysRevA.59.1667, Andre:19}, for approximately solving the nonlinear quantum Maxwell-Bloch equations by use of a Fourier transform. With this method, we extend the semiclassical approach by taking into account spontaneous emission into the lasing mode below the semiclassical threshold, where population fluctuations can be neglected. In subsequent sections we do take  population fluctuations into account. 
We obtain the expression for the laser linewidth  as a function of the population inversion, as in~\ct{Andre:19}, reproduce the well-known laser linewidth at small excitation, as in~\ct{Yariv} and derive the beta-factor for bad-cavity lasers, as in~\ct{doi:10.1063/1.5022958}, three results to illustrate the  efficiency of our method. 

Secs.~\ref{LLE_Lf}--~\ref{above threshold} are the main parts of the paper. In Sec.~\ref{LLE_Lf} we  represent the lasing field and polarisation by the S- and A-combinations of quadrature-operators, and  derive the central linear equations of our approach.  
We will show that in linear approximation only the S-combinations interact with population fluctuations,
while the A-combinations do not.  We  justify the approximations made to  linearize the initial nonlinear Maxwell-Bloch equations. 

In Sec.~\ref{P_quad_sp} we solve the equations for the A-combinations, show that they describe the laser output power and the linewidth in the high-pump limit, and reproduce the formula for the laser linewidth in that limit. 

In Sec.~\ref{above threshold} we derive expressions for lasing spectra and 
show examples of the analysis of
spectra for superradiant and non-superradiant lasers.
The final two sections contain  
discussions and our conclusions.

The novelty of our method is that we analyticaly describe the laser below, near and above the threshold by the same set of stochastic equations, taking into account the field, polarisation and population quantum fluctuations, spontaneous emission into the lasing mode and full laser dynamics  without adiabatic elimination of polarization. Such accurate treatment is in particular important for superradiant lasers, where collective effects among the emitters need to be taken into account. With our method we reproduce well-known results  and identify features, in particular in the laser spectra, that largely went unnoticed. In particular, we calculate the full spectrum of the lasing field below as well as above threshold and identify and explain the appearance of a broad spectral background above threshold and a multi-peak structure above and below threshold. In particular, we focus on the role of population fluctuations and nonlinear polarisation dynamics in superradiant lasers.
\section{Quantum Maxwell-Bloch equations.  Semiclassical laser model}\label{S1}
In order to keep the analysis simple, we consider a stationary single-mode laser, shown schematically in Fig.~\ref{FIG1N}, with a large number  $N_0 \gg 1$  of homogeneously broadened  identical two-level emitters,  with their transition frequency $\omega_0$  equal to the cavity mode frequency.  
%
%
\begin{figure}[thb]\bc
\includegraphics[width=5cm]{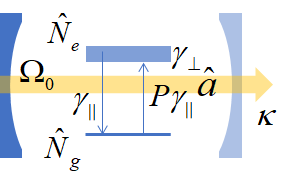}
\caption{Scheme of the two-level laser, with parameters and operators as defined in the main text. It is common to assume that $\gamma_{\perp} \gg 2\kappa,\gamma_{\parallel}$ so that the polarization  can be adiabatically eliminated and the laser is well described by rate equations. This does not work for the  superradiant nanolasers considered here, where $\gamma_{\perp} < 2\kappa$. }
\label{FIG1N}\ec
\end{figure}
%
%
We write the quantum Maxwell-Bloch equations (MBE) for such a laser~\ct{Scully} in the rotating-wave approximation with carrier frequency $\omega_0$, 
\begin{subequations}\lb{1}
\beqr
   \dot{\hat{a}}& = & -\kappa \hat{a}+{{\Omega }_{0}}\hat{v}+{{{\hat{F}}}_{a}} \lb{1_1}\\ 
 \dot{\hat{v}}& =& -(\gamma_{\perp}/2)\hat{v}+{{\Omega }_{0}}f\hat{a}\hat{N}+{{{\hat{F}}}_{v}} \lb{1_2}\\
  \dot{\hat{N}}_e&=&-{{\Omega }_{0}}({{{\hat{a}}}^{+}}\hat{v}+{{{\hat{v}}}^{+}}\hat{a})+\gamma_{\parallel}(P\hat{N}_g-\hat{N}_e)  +\hat{F}_{N_e}. \lb{1_3}
\eeqr\end{subequations}
Here $\hat{a}$ is the annihilation operator of the laser cavity mode, the operator $\hat{v} = i\sum_{i=1}^{N_0}f_i\hat{\sigma}_i$ describes the polarization of the emitters, $\hat{\sigma}_i$ describes transitions from the excited to the ground state of the $i^{\rm th}$ emitter; 
$f_i$ characterizes the coupling of the  $i^{\rm th}$ emitter to the lasing mode. We also define the average coupling $f = N_0^{-1}\sum_i f_i^2$. Furthermore,   $\Omega_0$ is the vacuum Rabi frequency and the total excited- and ground-state population operators    $\hat{N}_{e,g}$ are defined as the sums $\sum_{i=1}^{N_0}\hat{n}_i^{(e,g)}$, where $\hat{n}_i^e$ ($\hat{n}_i^g$) are operators of populations of excited (ground) states of the $i^{\rm th}$  emitter. The operator  $\hat{N} = \hat{N}_e-\hat{N}_g$ is then the population inversion.  In general we use 'hats' to denote operators, while mean values of operators are indicated by the absence of a hat, for example $\langle\hat{N}\rangle=N$. We will consider the stationary case, so mean values do not depend on time. The laser field leaves the cavity through the mirror at the cavity decay rate $2\kappa$;   $\gamma_{\parallel}$ is the population relaxation rate of the upper lasing level, $\gamma_{\parallel}P$ is the pump rate from the lower to the upper level; $\gamma_{\perp}/2$ is the polarization relaxation rate (so that $\gamma_{\perp}$ is the width of the lasing transition).  The total number of emitters is assumed preserved, $\hat{N}_e + \hat{N}_g = {N}_0$, so we can rewrite $\hat{N}_g = {N}_0-\hat{N}_e$ and  $\hat{N} = 2\hat{N}_e-{N}_0$. 

The quantum operators in Eq.~\rf{1} are interpreted as stochastic variables, where Langevin forces and their associated correlation strengths ensure correct quantum properties. In more detail, we introduced the Langevin forces $\hat{F}_{\alpha}$, with $\alpha$ taken from the set $\{a,v,a^+,v^+,N_e\}$. These  describe white noise, have zero mean (i.e.  $\langle\hat{F}_{\alpha}\rangle=0$), and are
delta-correlated in time: $\langle\hat{F}_{\alpha}(t)\hat{F}_{\beta}(t')\rangle =D_{\alpha\beta}\delta(t-t')$, where 
$D_{\alpha \beta }$ are the diffusion constants. 
In the Fourier-domain the cross-correlation of these Langevin forces is given by 
\beq 
    \left\langle {{{\hat{F}}}_{\alpha }}(\omega ){{{\hat{F}}}_{\beta }}(\omega ') \right\rangle = 2{{D}_{\alpha \beta }}\delta (\omega +\omega '). \lb{nn6}
\eeq
From Eq.~\rf{1_1} and its Hermitian conjugate we then obtain 
\beq
    0=-2\kappa n +\Omega_0(\langle\hat{a}^+\hat{v}\rangle+\langle\hat{v}^+\hat{a}\rangle), \lb{dop}
\eeq
in terms of the stationary mean photon number $n=\langle\hat{a}^+\hat{a}\rangle$. In combination with  Eq.~\rf{1_3}   we can eliminate the atom-field correlations and  obtain the energy conservation law 
\begin{equation}
2\kappa n+\gamma_{\parallel}{N}_e=\gamma_{\parallel}P{N}_g. \lb{nn2}
\end{equation}
Stationary solutions of Eq.~\rf{1} are readily obtained, 
if we neglect the Langevin forces and replace  operators by c-numbers: $\hat{a}\rightarrow a$, $\hat{v}\rightarrow v$,  $\hat N \rightarrow N$. This gives the conventional stationary equations for the stationary solutions   of the semiclassical laser model~\ct{Scully} 
\begin{subequations}\lb{nn21}\beqr
   0&=&-\kappa a+{{\Omega }_{0}}v\lb{nn21_1} \\ 
  0&=&-(\gamma_{\perp}/2)v+{{\Omega }_{0}}faN \lb{nn21_2}\\ 
 0&=&-{{\Omega }_{0}}(a^*v+v^*a)+\gamma_{\parallel}(P{N}_g-{N}_e). \lb{nn21_3}
\eeqr\end{subequations}
 Eqs.~\rf{nn21_1}, \rf{nn21_2} have  non-vanishing solutions if 
\beq
	N  > N_{\rm th} \equiv \kappa\gamma_{\perp}/(2\Omega _{0}^{2}f). \lb{nn22} 
\eeq 
From the energy conservation law~\rf{nn2}, we find the stationary number of photons in the semiclassical model to be
\beq 
	n = \frac{\gamma_{\parallel}}{2\kappa}(N_0+N_{\rm th})(P/P_{\rm th} - 1). \lb{tls_1}
\eeq 
So the semiclassical model predicts lasing, $n>0$,  when $N_0>N_{\rm th}$ and when the dimensionless pump rate $P$ exceeds the semiclassical lasing threshold  $P_{\rm th}$ 
\beq 
		P>P_{\rm th} = {(N_0+N_{\rm th})}/{(N_0-N_{\rm th})}. \lb{tls_2}
\eeq 
For a dimensionless pump rate $P$ smaller than $P_{\rm th}$, lasing is absent and $n=0$ in the semiclassical laser model. %

\section{Analysis neglecting population fluctuations}\label{Below_th}
%
Nanolasers have large beta factors, so that spontaneous emission 
into the lasing mode is non-negligible~\ct{Noda260, Prieto:15,Takiguchi:16,Ota:17}. Below and close to the semiclassical threshold, nanolasers are not well described by the standard semiclassical model, which neglects  spontaneous emission and predicts zero photons. One can improve upon this when $\gamma_{\perp} \gg 2\kappa,\gamma_{\parallel}$,  in which case   polarization  can be adiabatically eliminated. This is typically the case for semiconductor lasers. In this case, the laser can be described, for all pump rates, by quantum rate equations (QRE)~\ct{Siegman,PhysRevA.50.4318,doi:10.1063/1.5022958}, that do take into account spontaneous emission into the lasing mode. 
The intensity noise spectra can then also be found by small-signal analysis of the QRE~\ct{doi:10.1063/1.5022958}. However, the QRE cannot be  applied to superradiant lasers, where $\gamma_{\perp} < 2\kappa$ and the active-medium polarization cannot be adiabatically eliminated. 

In this Section we go beyond the rate-equation approach in the low-pump limit by including the dynamics of the polarization, rather than eliminating it adiabatically.
Instead,  following~\ct{PhysRevA.59.1667,Andre:19}, our main approximation will be that we neglect population fluctuations. This is a good approximation for small pump rates, when the material gain is much smaller than the cavity loss, such that fluctuations of the populations, and thereby of the material gain, do not significantly change the net cavity gain~\ct{1072058}. 
We  also take into account spontaneous emission into the lasing mode and introduce
a Fourier-expansion method that is used throughout the paper. We will reproduce the well-known formula for the lasing linewidth in the low-pump limit and introduce the $\beta$-factor for lasers with low-quality cavities. These results will be used as a reference for comparison  in the following sections, where we do take population fluctuations into account. 

Neglecting population fluctuations, we replace population operators in Eq.~\rf{1_2} by their mean values
\beq 
\hat{N}_{e,g} \approx N_{e,g}, \hspace{1cm} {\hat{N}} \approx N. \lb{app_pop}
\eeq
Thereby  Eqs.~\rf{1_1} and~\rf{1_2} 
turn into a set of linear equations   
\begin{subequations}\lb{1n}
\beqr
   \dot{\hat{a}}& = & -\kappa \hat{a}+{{\Omega }_{0}}\hat{v}+{{{\hat{F}}}_{a}} \lb{1_1n}\\ 
 \dot{\hat{v}}& =& -(\gamma_{\perp}/2)\hat{v}+{{\Omega }_{0}}f\hat{a}N+{{{\hat{F}}}_{v}}. \lb{1_2n}
\eeqr\end{subequations}
We express $\hat{a}(t)$ and $\hat{v}(t)$ and their corresponding Langevin forces through Fourier-component operators
\beq
        \hat{\alpha}(t) = \frac{1}{\sqrt{2\pi}}\int_{-\infty}^{\infty}\hat{\alpha}(\omega)e^{-i\omega t}d\omega \lb{FCOp}
\eeq
for $\hat{\alpha} = \{\hat{a},\hat{v},\hat{F}_a,\hat{F}_v\}$, and obtain from Eqs.~\rf{1n} linear algebraic equations for all $\alpha(\omega)$ 
and find from them
\beq
\hat{a}(\omega )=\frac{\left(  \gamma_{\perp}/2 - i\omega \right){{{\hat{F}}}_{{{a}}}}(\omega )+{{\Omega }_{0}}{{{\hat{F}}}_{{{v}}}}(\omega )}{\left( i\omega -\kappa  \right)\left( i\omega -\gamma_{\perp}/2 \right)-\Omega _{0}^{2}fN }.\lb{nn13_1}
\eeq
Coming back from $\hat{a}(\omega )$ to $\hat{a}(t)$ by an inverse Fourier-transformation, we calculate the mean number of photons in the cavity as 
\beq 
n = \langle\hat{a}^+(t)\hat{a}(t)\rangle = \frac{1}{2\pi}\int_{-\infty}^{\infty}n(\omega)d\omega,\lb{nn34_n}
\eeq
where $n(\omega)$ is the  spectral power density of  the field in the lasing mode, or  optical spectrum, which is related to $\hat{a}(\omega)$ as  
\beq
    \langle\hat{a}^+(\omega)\hat{a}(\omega')\rangle = n(\omega)\delta(\omega+\omega'). \lb{PSp}
\eeq
We will determine $n(\omega)$ and then find $n$ by Eq.~\rf{nn34_n}. 
In order to find $n(\omega)$ we must know the relevant diffusion coefficients~\ct{RevModPhys.68.127,trove.nla.gov.au/work/21304573}. After neglecting, as is usual, any thermal radiation in the lasing mode, since $k_BT\ll\hbar\omega$, we take the diffusion coefficient $2D_{a^+a}=0$. When population fluctuations are neglected, the diffusion coefficient $D_{v^+v}$ becomes
\beq
	2D_{v^+v}=f\gamma_{\perp}{N}_{e}, \lb{nn320}
\eeq
as shown in  Appendix~\ref{App_A}. With these diffusion coefficients, we find the optical spectrum
\beq
	n(\omega )=\frac{ (\kappa\gamma_{\perp}^2/2){{N}_{e}/N_{\rm th}}}{[(1-N/N_{\rm th})(\kappa\gamma_{\perp}/2)-\omega^2]^2+\omega^2(\kappa +\gamma_{\perp}/2)^2}. 	\lb{nn34}
\eeq
This spectrum may either have one or two peaks. Two peaks occur when all emitters are collectively and strongly coupled to the lasing mode, under the condition  
\beq
	N_c \equiv \frac{1}{2}\left(\frac{2\kappa}{\gamma_{\perp}}+\frac{\gamma_{\perp}}{2\kappa}\right)N_{\rm th}<N_0,  \lb{splt_b_th}
\eeq
and when $P<P_c$, where $P_c$ is such that $N(P_c) = -N_c$. The two peaks in  $n(\omega)$ are then caused by and a signature of collective Rabi splitting~\ct{Andre:19}. Otherwise,  $n(\omega)$ has a single peak, with full width at half maximum $\gamma_{\rm las}$,  defined by  $n(\gamma_{\rm las}/2) = n(\omega =0)/2$, with the value
\beq 
	\gamma_{\rm las} = \frac{2\kappa+\gamma_{\perp}}{{\sqrt{2}}}\left\{r-1+\sqrt{(r-1)^2+r^2}\right\}^{1/2}, 	\lb{nn36}
\eeq
where the parameter $r$ is given by 
\[
	r=\frac{4\kappa\gamma_{\perp}}{(2\kappa+\gamma_{\perp})^2}(1-N/N_{\rm th}).
\]
For $r\ll1$,  as obtained for pumping levels where $N$ is close to $N_{\rm th}$, we expand Eq.~\rf{nn36} as a series in $r$ and to first order in $r$ obtain 
\beq 
\gamma_{\rm las} \approx \frac{2\kappa+\gamma_{\perp}}{2}r =\gamma_c(1-N/N_{\rm th}), \lb{linewidth}
\eeq
where $\gamma_c=2\kappa\gamma_{\perp}/(2\kappa +\gamma_{\perp})$. Examples of optical spectra calculated according to Eq.~\rf{nn34} are given in Fig.~\ref{Fig2_1ab}, showing both cases of single- and double-peaked spectra.
%
%
\begin{figure}[t]\bc
\includegraphics[width=8cm]{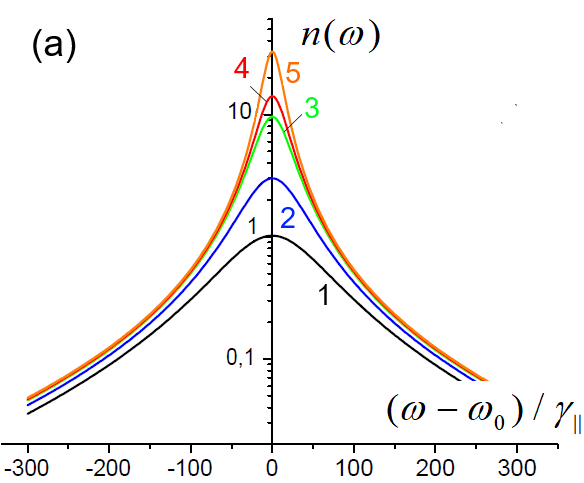}\\\includegraphics[width=8cm]{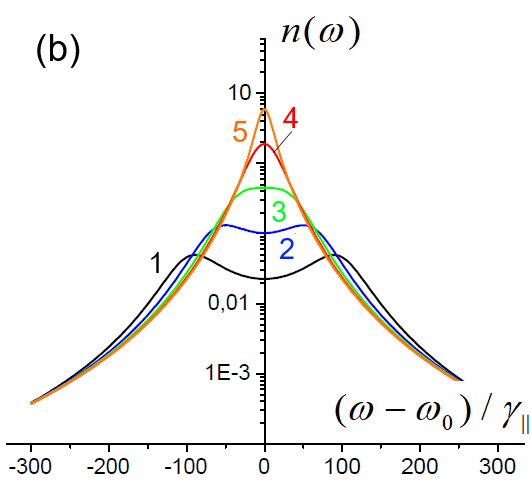}
\caption{Optical spectrum $n(\omega)$,  obtained by neglecting  population fluctuations, as given by Eq.~\rf{nn34}. Parameter values: $2\kappa = 100\gamma_{\parallel}$, $\Omega_0 = 34\gamma_{\parallel}$, $N_0 = 100$. The scaled pump rates are  $P=2$, $4$, $8$, $10$, and $16$ for the curves labelled 1 through 5, respectively.
Panel (a): $\gamma_{\perp} = 700\gamma_{\parallel}$, $N_c=108>N_0$. The optical spectrum has only one peak. Panel (b): $\gamma_{\perp} = 50\gamma_{\parallel}$, $N_c=2.7<N_0$. The optical spectrum has two peaks only for the curves~1 and~2, since only for these two are the 
pump rates $P$  below $P_c=7.4$.  } 
\label{Fig2_1ab}\ec
\end{figure}
%
%

Our goal in the remainder of this section is to express the linewidth $\gamma_{\rm las}$ in terms of familiar  laser parameters. In order to do so we first determine the population inversion $N$. (Incidentally, the same procedure to calculate $N$ will be used later again, when we also take population fluctuations into account.)  
From Eqs.~\rf{nn34_n} and~\rf{nn34} we find for the mean number of photons in the cavity 
\beq
	n=\frac{\gamma_{\perp}{{N}_{e}}}{(2\kappa +\gamma_{\perp})(N_{\rm th}-N)}. 	\lb{nn35}
\eeq
By inserting this 
into the energy conservation law~\rf{nn2}, we obtain a quadratic equation for the population inversion, 
\beq
	 \tilde{\beta}_c(N_0+N)=[P(N_0-N)-N_0-N](1-N/N_{\rm th}), \lb{nn36a}
\eeq
where we introduced the parameters
\begin{subequations}
\beqr
\tilde{\beta}_c & = & \tilde{\beta}/(1+2\kappa/\gamma_{\perp}), \quad\mbox{with} \\
 \tilde{\beta} & = & 4\Omega_0^2f/(\gamma_{\perp}\gamma_{\parallel}),
\lb{betac}
\eeqr
\end{subequations}
following  Refs.~\ct{doi:10.1063/1.5022958} and~\ct{PhysRevA.50.4318}, respectively.  
In the special case $\tilde{\beta}_c = 0$ the two solutions of Eq.~\rf{nn36a} for $N$  coincide with the stationary population inversion found in semiclassical laser theory. The general solution of Eq.~\rf{nn36a} with  $\tilde{\beta}_c \neq 0$ is different, 
because  the approach that led to Eq.~\rf{nn36a} takes into account spontaneous emission into the lasing mode. 
Indeed, in the limit  $2\kappa/\gamma_{\perp} \rightarrow 0$, the coefficient $\tilde{\beta}_c$ tends to $\tilde{\beta}$, which was introduced in Ref.~\ct{PhysRevA.50.4318} as the ratio of the rate of spontaneous emission into the lasing mode to
the rate of all other emission processes (i.e. background emission). 

Solving Eq.~\rf{nn36a},
we find the pump-dependent population inversion $N(P)$, given by Eq.~\rf{N_bel_th_1} of Appendix~\ref{App_0}. By inserting $N(P)$ into Eq.~\rf{nn36} we obtain an explicit expression for the pump-dependent linewidth $\gamma_{\rm las}(P)$. Similarly, by inserting $N(P)$ into the energy conservation law~\rf{nn2} we obtain the pump-dependent photon number $n(P)$, given in  Eq.~\rf{N_bel_th_2} of Appendix~\ref{App_0}.

We can now express the linewidth~\rf{linewidth} in terms of the laser output  power $W_{\rm out} = 2\kappa\hbar\omega_0n$ by using Eq.~\rf{nn35} to express the factor $(1-N/N_{\rm th})$ in terms of $W_{\rm out}$, giving
\beq
\gamma_{\rm las} = \left(\frac{2\kappa\gamma_{\perp}}{2\kappa+\gamma_{\perp}}\right)^2N_{\rm sp}\frac{\hbar\omega_0}{W_{\rm out}}, \lb{lw_bel_th}
\eeq
where  $N_{\rm sp} = {N_e}/{N_{\rm th}}$ 
is the so-called spontaneous-emission factor~\ct{PhysRevLett.72.3815}. Eq.~\rf{lw_bel_th} is the well-known result for the laser linewidth {\it below threshold}, which (apart from notations) coincides  with, for example, results in Refs.~\ct{PhysRevLett.72.3815, Yariv}. 

It is generally accepted that the laser linewidth {\it far above threshold}  
is suppressed by a factor of two compared to Eq.~\rf{lw_bel_th} ~\ct{Yariv}. We notice, though, that recent work \ct{POLLNAU2020100255}  challenges this result, based on a semiclassical analysis. In the next sections we will show that a fully quantum mechanical theory for the lineshape far above threshold agrees with adding the extra factor of $1/2$ to Eq.~\rf{lw_bel_th},  and that it can be ascribed to the effect of population fluctuations,  in particular to relaxation oscillations induced by population fluctuations.
\section{Linearization of equations}\label{LLE_Lf}
Our aim is now to develop a theory for the optical spectrum of a laser, valid at any pump rate, without making the assumption that fluctuations in the lasing field and polarization  are always small. Our linearization procedure is therefore different from the usual small-signal analysis as presented, for example, in Refs.~\ct{RevModPhys.68.127, PhysRevA.47.1431, 1071726,McKinstrie:20}. The theory is still approximative and we shall seek to clearly identify the approximations made.

We  begin with the linearization of Eq.~\rf{1_2} by writing the population operators as the sum of their mean values $N_{e,g}$ and population fluctuations $\delta\hat{N}_{e,g}$,  
\beq 
\hat{N}_{e,g}=N_{e,g} +\delta\hat{N}_{e,g}, \hspace{0.35cm}\delta\hat{N}_g=-\delta\hat{N}_e,  \lb{nn3}
\eeq 
whereby the population fluctuations are defined. We consider a large number $N_0\gg 1$ of emitters and suppose small fluctuations of populations  $\langle\delta\hat{N}_{e,g}^2\rangle^{1/2} \ll N_{e,g}$. We also suppose weak coupling,  $2\Omega_0^2f/\gamma_{\perp}\gamma_{\parallel} \ll 1$, and, for superradiant lasers, low-Q cavities with $2\kappa \leq \gamma_{\perp}$. The mean photon number for such a laser below and near the semiclassical threshold is of the order of unity or less, which we see in Fig.~\ref{Fig2_rev}(a). So we do {\it not} assume that  fluctuations of the lasing field and polarisation are small compared to their mean values. 

We next insert Eq.~\rf{nn3} into Eqs.~\rf{1} and obtain
\begin{subequations}\lb{N_e pert}\beqr
  \dot{\hat{a}}&=&-\kappa \hat{a}+{{\Omega }_{0}}\hat{v}+{{{\hat{F}}}_{a}}, \lb{N_e pert_1}\\ 
  \dot{\hat{v}}&=&-({{\gamma }_{\bot }}/2)\hat{v}+{{\Omega }_{0}}f\left( \hat{a}N+2\hat{a}\delta {{{\hat{N}}}_{e}} \right)+{{{\hat{F}}}_{v}}, \lb{N_e pert_2}\\ 
  \delta {{{\dot{\hat{N}}}}_{e}}&=&-{{\Omega }_{0}}\left( {{{\hat{v}}}^{+}}\hat{a}+{{{\hat{a}}}^{+}}\hat{v}-\left\langle {{{\hat{v}}}^{+}}\hat{a}+{{{\hat{a}}}^{+}}\hat{v} \right\rangle  \right)\nonumber\\& &-{{\gamma }_{\parallel }}(P+1)\delta {{{\hat{N}}}_{e}}+{{{\hat{F}}}_{{{N}_{e}}}}. \lb{N_e pert_3}
\eeqr\end{subequations}
We shall now show that instead of the conventional representation of the laser field in terms of amplitude and phase, it is convenient to represent the field and polarization by their quadratures   
\beq
\hat{\alpha}_x  =  (\hat{\alpha}+\hat{\alpha}^+)/\sqrt{2}, \hspace{0.5cm}\hat{\alpha}_p  =  i(\hat{\alpha}^+-\hat{\alpha})/\sqrt{2}, \lb{quadr}
\eeq
where $\hat{\alpha}$ stands for $\hat{a}$ or $\hat{v}$. In our stochastic approach, $\hat{\alpha}_{x,p}$ are represented by real-valued stochastic variables. The equations of motion for the quadratures follow from Eqs.~\rf{N_e pert},
\begin{subequations}\lb{quadratures_eq}\beqr
   \dot{\hat{a}}_{x,p}&=&-\kappa {{{\hat{a}}}_{x,p}}+{{\Omega }_{0}}{{{\hat{v}}}_{x,p}}+{{{\hat{F}}}_{{{a}_{x,p}}}}, \lb{quadratures_eq_1}\\ 
 {{{\dot{\hat{v}}}}_{x,p}}&=&-({{\gamma }_{\bot }}/2){{{\hat{v}}}_{x,p}}+{{\Omega }_{0}}f\left( {{{\hat{a}}}_{x,p}}N+2{{{\hat{a}}}_{x,p}}\delta {{{\hat{N}}}_{e}} \right)\nonumber\\ & & +{{{\hat{F}}}_{{{v}_{x,p}}}}, \lb{quadratures_eq_2}\\ 
 \delta {{{\dot{\hat{N}}}}_{e}}&=&-{{\Omega }_{0}}\left( {{{\hat{a}}}_{p}}{{{\hat{v}}}_{x}}+{{{\hat{v}}}_{p}}{{{\hat{a}}}_{x}}-\left\langle {{{\hat{a}}}_{p}}{{{\hat{v}}}_{x}}+{{{\hat{v}}}_{p}}{{{\hat{a}}}_{x}} \right\rangle  \right)\nonumber\\& &-{{\gamma }_{\parallel }}(P+1)\delta {{{\hat{N}}}_{e}}+{{{\hat{F}}}_{{{N}_{e}}}}. \lb{quadratures_eq_3}
 \eeqr\end{subequations}
The Langevin forces in Eqs.~\rf{quadratures_eq_1} and~\rf{quadratures_eq_2} are
\[
\hat{F}_{{\alpha}_x}  =  (\hat{F}_{\alpha}+\hat{F}_{{\alpha}^+})/\sqrt{2}, \hspace{0.5cm}\hat{F}_{{\alpha}_p}  =  i(\hat{F}_{{\alpha}^+}-\hat{F}_{\alpha})/\sqrt{2},
\]
where $\alpha$ stands for $a$ or $v$. 

We linearize Eqs.~\rf{quadratures_eq} in several steps. In the first step we consider {\it only the nonlinear terms} in Eqs.~\rf{quadratures_eq_2} and~\rf{quadratures_eq_3}. In these terms, 
we neglect  low-frequency fluctuations of the field and polarization,  in a frequency range  $\Delta\omega$ around $\omega=0$. In more detail, we approximate 
\begin{subequations}\lb{int_app}
\beqr
  {{{\hat{a}}}_{\mu}}(t)&\approx& \sqrt{n}+\frac{1}{\sqrt{2\pi }}\left( \int\limits_{-\infty }^{-\Delta \omega /2}+\int\limits_{\Delta \omega /2}^{\infty } \right){{{{\hat{a}}}_{\mu}}(\omega ){{e}^{-i\omega t}}d\omega }\nonumber\\ & &\equiv \sqrt{n}+{{{\hat{a}}}_{\mu}}' \lb{int_app_1}\\ 
 {{{\hat{v}}}_{\mu}}(t)&\approx& V+\frac{1}{\sqrt{2\pi }}\left( \int\limits_{-\infty }^{-\Delta \omega /2}+\int\limits_{\Delta \omega /2}^{\infty } \right){{{{\hat{v}}}_{\mu}}(\omega ){{e}^{-i\omega t}}d\omega }\nonumber\\ & &\equiv V+{{{\hat{v}}}_{\mu}}'  
\eeqr\end{subequations}
where the index $\mu$ stands for $x$ or $p$, 
and the c-number $V$ will be determined below.  The “cut-off” frequency $\Delta \omega$ is chosen such that $\left\langle \hat{a}_{\mu}'^2 \right\rangle \ll n$ and $\left\langle \hat{v}_{\mu}'^2 \right\rangle \ll V^2$, i.e. $\hat{a}'_{\mu}$ and ${{\hat{v}'}_{\mu }}$ are small perturbations relative to $\sqrt{n}$ and $V$. 
In Sec.~\ref{above threshold} we will calculate the optical spectra [see Eq.~\rf{fullspectrum}] and find that  the approximation~\rf{int_app} can  be made in the high-excitation limit, where almost all energy of the lasing field resides in a narrow spectral peak $n_A(\omega)$ of  width $\gamma_{\rm las}$} around the optical frequency $\omega_0$, with only a small part of the energy in the wide spectral background $n_S(\omega)$ of width $\gamma_{\rm bg}\gg\gamma_{\rm las}$.   The peak $n_A(\omega)$ and the background $n_S(\omega)$ of the full optical spectrum  are depicted in Fig.~\ref{Fig5lfc_ab}. 
%
%
\begin{figure}[thb]\bc
\includegraphics[width=7cm]{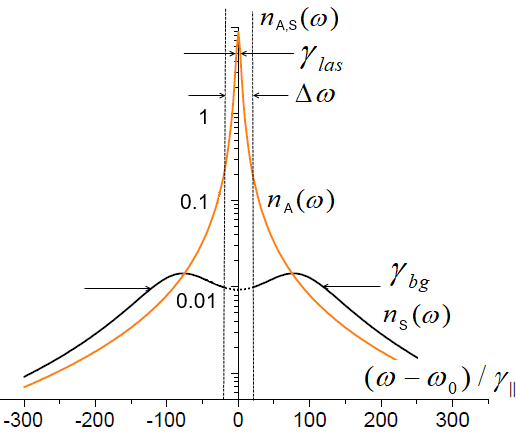}
\caption{Optical spectra of the field components 
residing in the asymmetric (A) combination of quadratures, $n_A(\omega)$ (orange curve) and in the symmetric combination $n_S(\omega)$ (black curve). 
Spectra are calculated for a high excitation rate $P=40$,   $\gamma_{\perp} = 0.7$~THz and with other parameters given in Sec.~\ref{above threshold}. In the approximation~\rf{int_app},  fluctuations inside the region restricted by the vertical dashed lines are neglected. The cut-off frequency $\Delta\omega$, introduced in Eq.~\rf{int_app}, equals $40\gamma_{\parallel}$. The S-combinations contain only 9\%  of the energy of the field. In the approximation~\rf{app_prime},  $\hat{a}'_S(\omega)$ is extended to the frequency interval $|\omega|< \Delta\omega/2$, which is the small dotted segment in the $n_S(\omega)$ curve.}
\label{Fig5lfc_ab}\ec
\end{figure}
%
%
The $n_A(\omega)$ and $n_S(\omega)$ correspond to (but do not exactly coincide with), respectively, the field  $W_{\varepsilon}$ and the amplitude $W_{\delta}$ spectra calculated, for example, in Ref.~\ct{1071986}.

Figure~\ref{Fig5lfc_ab} shows that $\Delta\omega$ can be chosen
from the interval $\gamma_{\rm las}<\Delta\omega \ll \gamma_{\rm bg}$, and we shall see that the final results 
are independent of $\Delta\omega$. Later we will explain that the approximation~\rf{int_app} 
can safely be made in the 
nonlinear terms in Eq.~\rf{quadratures_eq_2} also at low or moderate laser excitation. 

We insert the approximation~\rf{int_app} into  Eq.~\rf{quadratures_eq_1} and find %
\beq
	V=(\kappa /{{\Omega }_{0}})\sqrt{n}.	\lb{7_wd}
\eeq
One may expect
that in Eqs.~\rf{int_app}, different c-numbers should be defined for the different quadratures, instead of only $\sqrt{n}$ and $V$. However, the different c-numbers 
reduce to $\sqrt{n}$ and $V$
by the replacements   
	$\hat{a}\to \hat{a}{{e}^{i\varphi }}$ and   $\hat{v}\to \hat{v}{{e}^{i\varphi }}$ 	
with real-valued phase $\varphi$, which is the constant phase of the lasing field. The solution of the initial MBE~\rf{1} does not depend on such a replacement, which adds only a constant phase multiplier to the Langevin forces $\hat{F}_a$ and $\hat{F}_v$. 
Therefore  $\sqrt{n}$ and $V$ in Eqs.~\rf{int_app} can be chosen to be the same for both quadratures in the general case.  

In the second linearization step, we insert the approximation~\rf{int_app} into {\it only the nonlinear terms} in Eqs.~\rf{quadratures_eq}. Then, by neglecting the products of fluctuations in these terms, we obtain linear equations for $\hat{a}_{x,p}$ and $\hat{v}_{x,p}$. This step is  different from the usual small-signal analysis, where linear equations should be written  for the perturbations $\hat{a}'_{x,p}$ and $\hat{v}'_{x,p}$.

\begin{figure}[t]
\includegraphics[width=6cm]{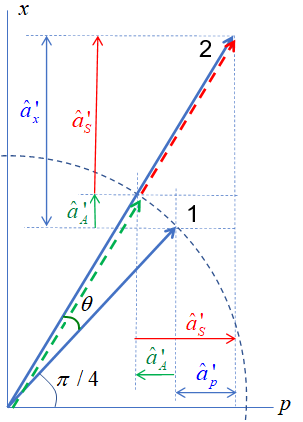}
\caption{Sketch of the lasing field and its fluctuations. The initial lasing field, represented by  the blue vector~1 at an angle of $45^{\circ}$, changes into vector~2 due to a fluctuation of its direction and length during a short time $\Delta t\sim \gamma_{\rm bg}^{-1}$
much smaller than the characteristic time scale for direction (phase) fluctuations $\gamma_{\rm las}^{-1}$. 
Direction fluctuations are much slower than length fluctuations, so after $\Delta t$ the difference angle $\theta\ll 1$. Also shown are the quadratures $\hat{a}_{x,p}'$ as well as the symmetric $\hat{a}_S'$ and anti-symmetric $\hat{a}_A'$ combinations that are defined by Eqs.~\rf{sym_ass} and used in the approximation~\rf{int_app}. For $\theta\ll 1$, the $\hat{a}_S'$ and $\hat{a}_A'$ combinations determine, correspondingly, the amplitude and phase fluctuations of the field. The green (the red) dashed vectors denote the change of the field vector~1 due to only phase (amplitude) fluctuations. 
}  
\label{Fig2_3}
\end{figure}

In the third and final step of the linearization, the set of Eqs.~\rf{quadratures_eq} can be further simplified, if we extract the symmetric (S)  ${{\hat{\alpha }}_S}$ and anti-symmetric (A) ${{\hat{\alpha }}_A}$ combinations   of the quadratures and their high-frequency fluctuations  (denoted by primes) 
\beq
\hat{\alpha}_S=(\hat{\alpha }_x+\hat{\alpha}_p)/2,  \hspace{0.5cm}\hat{\alpha }_A=(\hat{\alpha}_x-\hat{\alpha}_p)/2,\lb{sym_ass}\eeq  
\[\hat{\alpha}_S'=(\hat{\alpha }_x'+\hat{\alpha}_p')/2,  \hspace{0.5cm}\hat{\alpha }_A'=(\hat{\alpha}_x'-\hat{\alpha}_p')/2, 
\]
where $\alpha $ means $a$  or $v$. 
A-combinations change their signs upon exchange of the indices  $x\rightleftarrows p$ in Eqs.~\rf{sym_ass}, while S-combinations do not. From the inverse relations $\hat{\alpha}_x' = \hat{\alpha}_S'+\hat{\alpha}_A'$ and $\hat{\alpha}_p' = \hat{\alpha}_S'-\hat{\alpha}_A'$, we see that S-combinations contribute in the same way to the x- and p-quadratures, whereas A-combinations contribute with the same absolute values to both quadratures, but with opposite signs. These properties lead to the important physical interpretation that {\it S-combinations correspond to amplitude fluctuations while A-combinations correspond to phase fluctuations}, as illustrated in  Figure~\ref{Fig2_3}. 
%
%
%
%

After the three linearization steps, we 
obtain, finally, from Eqs.~\rf{quadratures_eq} as one of our main results the approximate linear equations for the S- and A-combinations of quadratures 
\begin{subequations}\lb{sym_antisym_eq}
\beqr
  \dot{\hat{a}}_{S,A} &=&-\kappa {{{\hat{a}}}_{S,A }}+ {{\Omega }_{0}}{{{\hat{v}}}_{S,A}}+{{{\hat{F}}}_{{{a}_{S,A}}}}, \lb{sym_antisym_eq_1}\\
 {{{\dot{\hat{v}}}}_S}&=&-({{\gamma }_{\bot }}/2){{{\hat{v}}}_S}+\lb{sym_antisym_eq_2}\\
 & & {{\Omega }_{0}}f\left[ {{{\hat{a}}}_S}\left( 2{{N}_{e}}-{{N}_{0}} \right)+2\sqrt{n}\delta {{{\hat{N}}}_{e}} \right]+ {{{\hat{F}}}_{{{v}_S}}},\nonumber \\
 \dot{\hat{v}}_A &=& -(\gamma_{\bot }/2)\hat{v}_A-\Omega_0f\hat{a}_A\left( 2N_e-N_0 \right)+\hat{F}_{v_A}, \lb{sym_antisym_eq_3}\\
\delta\dot{\hat{N}}_e & = & -2\sqrt{n}\left(\Omega_0\hat{v}_S+\kappa\hat{a}_S\right)\lb{sym_antisym_eq_4}\\ & & \hspace{2cm}-\gamma_{\parallel}(P+1)\delta \hat{N}_e+\hat{F}_{N_e},\nonumber 
\eeqr\end{subequations}
with
\beq
    \hat{F}_{\alpha_{S,A}}=(\hat{F}_{\alpha}e^{\mp{i\pi/4}} + \hat{F}_{\alpha^+}e^{\pm{i\pi/4}})/2, \lb{LF_s_as}
\eeq
and where $\alpha$ stands for $a$ or $v$. The derivation of Eqs.~\rf{sym_antisym_eq} is given in Appendix~\ref{App_C0}. We see that Eqs.~\rf{sym_antisym_eq} are split into two sets of equations: the two Eqs.~\rf{sym_antisym_eq_1} and \rf{sym_antisym_eq_2} for the A-combinations of the quadratures and the three  equations~\rf{sym_antisym_eq_1}, \rf{sym_antisym_eq_3} and~\rf{sym_antisym_eq_4} for the S-combinations and the population fluctuations. Both sets can be solved independently,
but their solutions are related through their common c-number variables $N_e$ and $n$. 

For clarity, let us summarize
the approximations made in the derivation of
the system of equations  Eq.~\rf{sym_antisym_eq}: 
we neglected low-frequency fluctuations in the approximation~\rf{int_app}; 
in  Eq.~\rf{sym_antisym_eq_2} we neglected the small term ${{\hat{a}}_S}'$ with respect to the large term $\sim\sqrt{n}$; similarly, in  Eq.~\rf{sym_antisym_eq_3} we neglected a term ${{\hat{a}}_A}'\delta {{\hat{N}}_{e}}$ since the term  ${{\hat{a}}_A}N$ is much larger because $\langle\delta N_e^2\rangle^{1/2}\ll N$; 
in Eq.~\rf{sym_antisym_eq_4} we neglected the second-order correlation 
\beq
\Omega_0\left[ \left( \hat{a}_S'\hat{v}_S'-\left\langle \hat{a}_S'\hat{v}_S' \right\rangle  \right)+\left( \hat{a}_A'\hat{v}_A'-\left\langle \hat{a}_A'\hat{v}_A' \right\rangle  \right) \right]  \lb{soc_1}
\eeq
with respect to 
$2\sqrt{n}\left( {{\Omega }_{0}}{{{\hat{v}}}_S}'+\kappa {{{\hat{a}}}_S}' \right)$ 
and we approximated
\beq
	{{\Omega }_{0}}{{\hat{v}}_S}'+\kappa {{\hat{a}}_S}'\approx {{\Omega }_{0}}{{\hat{v}}_S}+\kappa {{\hat{a}}_S}, 	\lb{app_prime}
\eeq
as described in detail in Appendix~\ref{App_C0}. 

Now let us find and discuss the solutions of the equations Eqs.~\rf{sym_antisym_eq}. We solve them by Fourier transformation, and we find $\left\langle \hat{a}_{S,A }^{2} \right\rangle \equiv {{n}_{S,A}}({{N}_{e}})$. In order to find the mean photon number $n$, we express 
\beq
\hat{a} = \hat{a}_Se^{i\pi/4}+\hat{a}_Ae^{-i\pi/4}, \lb{a_AS}
\eeq
insert this Eq.~\rf{a_AS}  into  $n=\left<\hat{a}^+\hat{a}\right>$ and obtain
\beq
n= {{n}_S}({{N}_{e}})+{{n}_A}({{N}_{e}}) +i\left<[\hat{a}_A,\hat{a}_S]\right>. \lb{comm_AS}
\eeq
Inserting Eq.~\rf{a_AS} into $[\hat{a},\hat{a}^+] = 1$ we find $[\hat{a}_A,\hat{a}_S] = i/2$, while for the mean photon number we obtain
\beq
	n= {{n}_S}({{N}_{e}})+{{n}_A}({{N}_{e}}) - {1}/{2}. 	\lb{Pn_nm_sym_ass}
\eeq
By inserting this Eq.~\rf{Pn_nm_sym_ass} into the energy conservation law~\rf{nn2}, we obtain an equation for ${{N}_{e}}$. 

Though Eqs.~\rf{sym_antisym_eq_1},  \rf{sym_antisym_eq_3} for the A-combinations do not depend on population fluctuations  explicitly, the A-combinations do depend on $n_S$ and, therefore, on $\delta\hat{N_e}$ implicitly, through  the energy conservation law~\rf{nn2}, where $n(N_e)$ is given by Eq.~\rf{Pn_nm_sym_ass}. Because of this, we will obtain different relations between $n_A$, $n_S$ and $n$ at low and at high excitation.

In order to justify the  approximation~\rf{app_prime}, we note that the Fourier expansions of ${{\hat{a}}_S}'$ and ${{\hat{v}}_S}'$ by their definition in Eq.~\rf{int_app} have the same Fourier components as  ${{\hat{a}}_S}$ and ${{\hat{v}}_S}$ for  frequencies $\left| \omega  \right|>\Delta \omega $, but instead have 
zero Fourier components for $\left| \omega  \right|\leq\Delta \omega $. The Fourier components of ${{\hat{a}}_S}'$ and ${{\hat{v}}_S}'$ are smooth functions of $\omega $, as confirmed by Fig.~\ref{Fig5lfc_ab}, so as an approximation we extend them to the small interval $\left| \omega  \right|\leq\Delta \omega $, which is exactly the approximation~\rf{app_prime}. 

The approximation~\rf{int_app} was made during the linearization of equations~\rf{sym_antisym_eq} 
in the high-excitation limit, when the spectrum of the lasing field has a narrow peak and a broad background, as in Fig.~\ref{Fig5lfc_ab}. Most of the energy is concentrated in the peak and the cut-off frequency $\Delta \omega $ is about as small as the width $\gamma_{\rm las}$ of the peak, so  $\hat{a}' \ll \sqrt{n}$ and $\hat{v}' \ll V$. 
These inequalities do not hold for  low and moderate excitations. 
For such excitations, however, population fluctuations are so small that, as a zero-order   approximation one may neglect them altogether, as in Sec.~\ref{Below_th}. The linearized equations~\rf{sym_antisym_eq} are still approximate, but they are the next-order approximation  at low excitation, where they describe population fluctuations incompletely, but do not neglect them. This is an argument for using the  equations~\rf{sym_antisym_eq} for
all excitation levels as the first approximation  to incorporate small population fluctuations. 
This approach is analogous to the one in Ref.~\ct{doi:10.1063/1.5022958}, where it was shown that in the rate-equation limit, the standard linear small-signal analysis, when carried out for large number of emitters and weak coupling, agrees very well with results of exact numerical nonlinear analysis  also below the lasing threshold. 

For these reasons, below we will use the linearized equations~\rf{sym_antisym_eq} at all levels of excitation. 
%
\section{Solving the  combinations of the  field quadratures}\label{P_quad_sp} 
%
In this section we  first solve the equations~\rf{sym_antisym_eq_1} and~\rf{sym_antisym_eq_3}  for the A-combinations of the field and polarization quadratures,  these equations being the ones that do not explicitly depend on population fluctuations. 
In doing so,  well-known results for the laser linewidth in the high-pump limit  will be reproduced.

We replace $\hat{a}_A$ and $\hat{v}_A$ in Eqs.~\rf{sym_antisym_eq_1} and~\rf{sym_antisym_eq_3} by their Fourier expansions~\rf{FCOp}, then solve the linear equations for the Fourier components $\hat{a}_A(\omega)$ and $\hat{v}_A(\omega)$, and find
\beq
\hat{a}_A(\omega )=\frac{( {{\gamma }_{\bot}/2-i\omega})\hat{F}_{a_A}(\omega ) +\Omega_0\hat{F}_{v_A}(\omega )}{\left( i\omega -\kappa  \right)\left( i\omega -{{\gamma }_{\bot }}/2 \right)-\Omega _{0}^{2}fN}.\lb{nn13}\eeq
From $\hat{a}_A(\omega )$ we can calculate the corresponding optical spectrum  $n_A(\omega)$  by 
\beq
    \langle\hat{a}_A(\omega)\hat{a}_A(\omega')\rangle = n_A(\omega)\delta(\omega+\omega'), \lb{PSp_qu}
\eeq
analogous to Eq.~\rf{PSp}. 
The relevant diffusion coefficients are given by
\beq
2D_{a_A a_A}=\kappa/2, \hspace{0.25cm} 2D_{v_A v_A}=f{{\gamma }_{\bot }}{{N}_0}/4,  \lb{nn14}
\eeq
as calculated in Appendix~\ref{App_A}. 
With the help of Eq.~\rf{nn13}, the diffusion coefficients~\rf{nn14},  and the correlations of Langevin forces~\rf{nn6}, we find the spectrum of the A-combinations of the photon field to be  
\beq
n_A(\omega ) = \frac{(\kappa/2)[(1+N_0/N_{\rm th}){\gamma }_{\bot }^2/4+\omega^2]}{{{[(1-N /{N_{\rm th}})(\kappa {{\gamma }_{\bot }}/2)-{{\omega }^{2}}]}^{2}}+{{\omega }^{2}}{{(\kappa +{{\gamma }_{\bot }}/2)}^{2}}}. \lb{sp_above_th}
\eeq
The  photon numbers $n_{A,S}$ residing in the A- or S-combinations are in general given by the integrated spectra 
\beq
n_{A,S} = \frac{1}{2\pi}\int\limits_{-\infty }^{\infty }n_{A,S}(\omega )\mbox{d}\omega. \lb{n_x_int}
\eeq
By carrying out this integration 
with $n_A(\omega )$ given by Eq.~\rf{sp_above_th}, we find the number of photons in the A-combinations of the field quadratures as
\beq 
n_A   = 
\frac{\gamma_{\bot}}{4(2\kappa+ \gamma_{\bot})}\left(\frac{N_0+N_{\rm th}}{N_{\rm th}-N }+\frac{2\kappa}{\gamma_{\bot }}\right).\lb{nn18}
\eeq
So here we find, perhaps not surprisingly, that $n_A$ can grow very large when the laser is pumped strongly and the population inversion  $N$ approaches its semiclassical stationary value $N_{\rm th}$. 
However,  in Sec.~\ref{above threshold} we will see that in the same high-pump limit, the number of photons $n_S$ in the S-combinations of the quadratures is much smaller than $n_A$ in Eq.~\rf{nn18}. Mathematically, this is the case because  $n_S$ does not have a corresponding term $\sim 1/(N_{\rm th}-N)$.  Physically, because $n_S$ is suppressed by relaxation oscillations introduced by  population fluctuations. So  almost all lasing photons reside in the A-combinations and  $n_A\approx n$. 
Anticipating these results for the spectra and for the number of photons in  the A-combinations, we now replace $n_A$ by $n$ in Eq.~\rf{nn18}, and use this to derive an expression for the linewidth as a function of the laser output power, as we did before in Eq.~\rf{lw_bel_th}: in Eq.~\rf{nn18} we neglect the term $2\kappa/\gamma_{\perp}$, which above threshold is small compared to the large first term $\propto (N_{\rm th} - N)^{-1}$. The linewidth $\gamma_{\rm las}$ is again  defined as the full width at half maximum of the optical spectrum, i.e. $n_A(\gamma_{\rm las}/2) = n_A(\omega = 0)/2$. This leads to exactly the same expressions~\rf{nn36}  and~\rf{linewidth} for the linewidth  $\gamma_{\rm las}$ as found previously for the spectrum~\rf{nn34}, when expressed in terms of the average population inversion, $N$. However, the crucial difference is that the variation of $N$ with the pump level $P$ changes quantitatively as the laser threshold is passed, leading to different dependencies of the laser linewidth on power above and below the laser threshold.

To see this, we express $(N_{\rm th} - N)$ in terms of $n$ and the laser output power $W_{\rm out}$. We insert the result into Eq.~\rf{linewidth}, 
and obtain for the {\it laser linewidth in the high-excitation limit}
\beq
\gamma_{\rm las}=\frac{1}{2}\left( \frac{2\kappa \gamma_{\bot}}{2\kappa +\gamma_{\bot }} \right)^{2}N_{\rm sp}\frac{\hbar\omega_{0}}{W_{\rm out}}.\lb{nn19}
\eeq
This looks a lot like Eq.~\rf{lw_bel_th} for the laser linewidth in the {\it low-pump limit}. It differs only  by the prefactor of  $1/2$ in Eq.~\rf{nn19}, and by a different expression for the spontaneous-emission factor, which for  Eq.~\rf{nn19} reads  
\beq
    N_{\rm sp} = (N_0+N_{\rm th})/2N_{\rm th}. \lb{Sp_em_f_ab_th}
\eeq
The result~\rf{nn19} is the same as, for example,  in Ref.~\ct{van_Exter}. Our approach gives a new interpretation to this remarkable result: the linewidths~\rf{lw_bel_th} and~\rf{nn19} at low and at high pump rates are different due the different roles of population fluctuations below and above threshold. As shown  in the next section, above threshold the population fluctuations reduce the number of photons in S- and increase it in A-combinations, which govern only slow frequency fluctuations and thereby narrow the linewidth. We will also see that the total number of photons $n(P)$ in the field practically does not depend on population fluctuations. 
\section{Calculation and analysis of optical spectra}\label{above threshold}
Equations~\rf{sym_antisym_eq_1}, \rf{sym_antisym_eq_2}  for the symmetric combinations $\hat{a}_S$ and $\hat{v}_S$ of the field and polarization  and Eq.~\rf{sym_antisym_eq_4} for population fluctuations  lead to algebraic equations for the Fourier-component operators $\hat{a}_S(\omega )$, $\hat{v}_S(\omega )$, and $\delta\hat{N}_e(\omega )$. 
By solving them, 
we obtain the Fourier-component $\hat{a}_S(\omega )$ of the S-combinations of the field quadratures. Then, using relation~\rf{PSp_qu} with indices "S" instead of "A", we find the spectrum of the S-combinations as
\begin{widetext}\beq
n_S(\omega)  =  \frac{\kappa[(\omega_{\rm ro}^2-\omega^2+\gamma_P\gamma_{\perp}/2)^2+\omega^2(\gamma_{\perp}/2+\gamma_P)^2] + \kappa\gamma_{\perp}^2(\omega^2+\gamma_P^2)N_0/4N_{\rm th} + \omega_{\rm ro}^2\kappa\gamma_{\parallel}\gamma_{\perp}(PN_g+N_e)/N_{\rm th}}{2|(i\omega - \gamma_P)[(1-N/N_{\rm th})\kappa\gamma_{\perp}/2 - \omega^2 -i\omega(\kappa+\gamma_{\perp}/2)]   +\omega_{\rm ro}^2(i\omega - 2\kappa )|^2}. \lb{sp_x00}
\eeq\end{widetext}
where $\gamma_P \equiv \gamma_{\parallel}(P+1)$  and
\beq
{\omega}_{\rm ro}^2 \equiv  4\Omega_0^2 f n \lb{-omega_ro-}
\eeq
is the squared relaxation oscillation frequency at high pump in the rate-equation limit $2\kappa \ll \gamma_{\perp}$~\ct{McKinstrie:20}. For the derivation of Eq.~\rf{sp_x00} we used the diffusion coefficients 
$2D_{a_Sa_S}=2D_{a_Aa_A}$ and $2D_{v_Sv_S}=2D_{v_Av_A}$ 
given by Eq.~\rf{nn14}, and the diffusion coefficient
\beq
	2D_{N_eN_e} = \gamma_{\parallel}(PN_g + N_e). \lb{dif1_x}
\eeq
We 
have found that correlations between the Langevin forces representing fluctuations of the carrier population and the polarization 
give only a very small contribution  to the spectrum 
for a large number of emitters $N_0\gg 1$ and at weak coupling $\Omega_0/(2\kappa + \gamma_{\perp}) \ll 1$, which is the case considered here. For that reason
we shall in the following neglect these correlations and put the corresponding diffusion coefficients to zero 
(i.e. $2D_{v_{A,S}N_e}  =0$).
The full spectrum of the lasing mode is
\beq 
	n(\omega) = n_A(\omega)+n_S(\omega)+n_{AS}(\omega). \lb{fullspectrum}
\eeq 
where $n_A(\omega)$ and $n_S(\omega)$ are given by Eqs.~\rf{sp_above_th} and~\rf{sp_x00}. So we are left with calculating $n_{AS}(\omega)$.   By inserting the Fourier expansions \rf{FCOp} for $\hat{a}_S$ and $\hat{a}_A$ into the commutator $\left<[\hat{a}_A,\hat{a}_S]\right>$ in Eq.~\rf{comm_AS},  we find for  $n_{AS}(\omega)$ the relation
\[
i\langle[\hat{a}_A(\omega),\hat{a}_S(\omega')]\rangle = n_{AS}(\omega)\delta(\omega+\omega'),
\]
We 
calculate $n_{AS}(\omega )$  approximately by noting that the contribution of $n_{AS}(\omega)$ is only important when the number of photons is small, in which case population fluctuations and their contribution  to S-combinations are also small. So we calculate $n_{AS}(\omega)$ while neglecting  population fluctuations. By setting $\delta\hat{N}_e =0$ (that is the case considered in Section~\ref{Below_th}) in Eqs.~\rf{sym_antisym_eq},  we see that the equations for $\hat{a}_S$ and $\hat{v}_S$ are  identical to the equations for $\hat{a}_A$ and $\hat{v}_A$, so that   $n_A(\omega)= n_S(\omega)$ and therefore 
\beq
n_{AS}(\omega) \approx 2n_A(\omega) - \left.n(\omega)\right|_{\delta\hat{N}_e=0},  \lb{n_xp}
\eeq
where $\left.n(\omega)\right|_{\delta\hat{N}_e=0}$ is obtained in Sec.~\ref{Below_th} and given by Eq.~\rf{nn34}. By inserting Eq.~\rf{n_xp} into Eq.~\rf{fullspectrum} and applying there Eqs.~\rf{nn34} and~\rf{sp_above_th}, we obtain the lasing spectrum 
\beqr
& &n(\omega)  =  {n_S(\omega)} + \lb{full_las_sp}\\ 
& & \frac{(\kappa\gamma_{\perp}^2/4N_{\rm th})(N+N_0/2) - 0.5\kappa(\gamma_{\perp}^2/4+\omega^2)}{{{[(1-N /{N_{\rm th}})(\kappa {{\gamma }_{\bot }}/2)-{{\omega }^{2}}]}^{2}}+{{\omega }^{2}}{{(\kappa +{{\gamma }_{\bot }}/2)}^{2}}}, \nonumber
\eeqr
with $n_S(\omega)$ still given by Eq.~\rf{sp_x00}.

The spectrum $n(\omega)$   depends on the population inversion $N$, which can be found from the energy conservation law~\rf{nn2} in the same way as in Sec.~\ref{Below_th}. We express $n$ [entering $n_S(\omega)$ through Eqs.~\rf{-omega_ro-}] through $N$ by the same energy conservation law~\rf{nn2}. Then, the population inversion $N$ is the only unknown variable in Eq.~\rf{nn2}.  Written in terms of the numbers of photons in the S- and A-combinations, the conservation law  becomes
\beq
        n_S(N)+n_A(N)-1/2 = \frac{\gamma_{\parallel}}{4\kappa}[P(N_0-N) - N_0-N]. \lb{eq_N_fin}
\eeq
Here, $n_A(N)$ is given by Eq.~\rf{nn18}, and $n_S(N)$ is found by integrating  Eqs.~\rf{n_x_int} with $n_S(\omega)$ given by Eq.~\rf{sp_x00}. We find the mean population inversion $N$ by solving the integral equation~\rf{eq_N_fin} numerically.

In our calculation examples 
we choose parameters close to typical ones for photonic crystal nanolasers with  quantum-dot active media~\ct{doi:10.1063/1.5022958}: for the  wavelength of the lasing transition we pick $\lambda_0 = 1.55$~$\mu$m, for the background refractive index $n_r = 3.3$, the cavity mode volume $V_c = 10(\lambda_0/n_r)^3$ with $N_0 = 100$ emitters; a population relaxation rate $\gamma_{\parallel} = 10^9~s^{-1}$; a vacuum Rabi frequency $\Omega_0 = (d/n_r)[\omega_0/(\varepsilon_0\hbar V_c)]^{1/2}$ with a dipole moment of the lasing transition $d = 10^{-28}$~Cm so that $\Omega_0 = 34\gamma_{\parallel}$; the average atom-lasing mode-coupling factor $f=1/2$; finally, we choose the cavity quality factor $Q = 1.2\cdot 10^4$ so that $2\kappa = 100\gamma_{\parallel}$. 

In the examples 
below we vary the dephasing rate $\gamma_{\perp}$ and the pump $P$ while keeping all other parameters fixed. The value for  $\gamma_{\perp}$ is varied between $\gamma_{\perp}^{\rm min} = 50$~GHz (so that $2\kappa/\gamma_{\perp}^{\rm min} = 2$) to $\gamma_{\perp}^{\rm max} = 1.5$~THz (with $2\kappa/\gamma_{\perp}^{\rm max} = 0.07$). This is a realistic  region of  $\gamma_{\perp}$ for quantum dots~\ct{PhysRevB.46.15574}. Within this range for $\gamma_{\perp}$, the conventional beta-factor ${\beta}$ varies from 0.98 to 0.6, while the beta-factor $\tilde{\beta}_c$ varies from $15$  to $1.4$, so lasers with the chosen  parameters have significant amounts of spontaneous emission into the lasing mode. 

Lasers with high $\beta$-factors and low dephasing rates, $2\kappa/\gamma_{\perp}^{\rm min} = 2$, are superradiant, while lasers with $2\kappa/\gamma_{\perp}^{\rm max} = 0.07\ll 1$ are not superradiant even if $\tilde{\beta}_c>1$.  Upon variation of $\gamma_{\perp}$ between $\gamma_{\perp}^{\rm min}$ and $\gamma_{\perp}^{\rm max}$, we will thus be able to compare results for superradiant and for non-superradiant  lasers.
%
\subsection{Photon numbers and population inversions}\label{MV}
%
Calculations of mean values of photon numbers and population inversions are helpful for the identification of different lasing regimes and for understanding the role of population fluctuations. Our procedure to find the mean photon numbers and population inversions is as follows: first we calculate $n_S(N)$ by inserting the spectrum $n_S(\omega)$ from Eq.~\rf{sp_x00} into Eq.~\rf{n_x_int}. Then we  insert $n_S(N)$ found from Eq.~\rf{n_x_int} and $n_A(N)$ from Eq.~\rf{nn18} into the energy conservation law~\rf{eq_N_fin}. By solving the latter equation, we can determine the population inversion $N$. By inserting this $N$ back into $n_{S,A}(N)$, we find the mean photon number $n$ from Eq.~\rf{Pn_nm_sym_ass}. 

\begin{figure}[thb]\bc
\includegraphics[width=8cm]{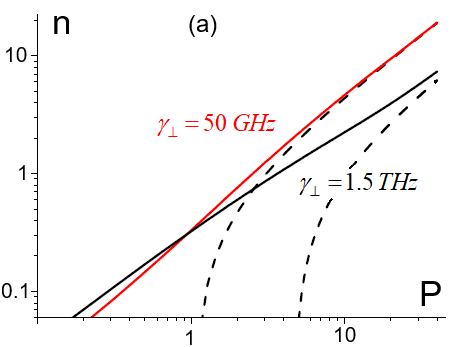}\\
\includegraphics[width=8cm]{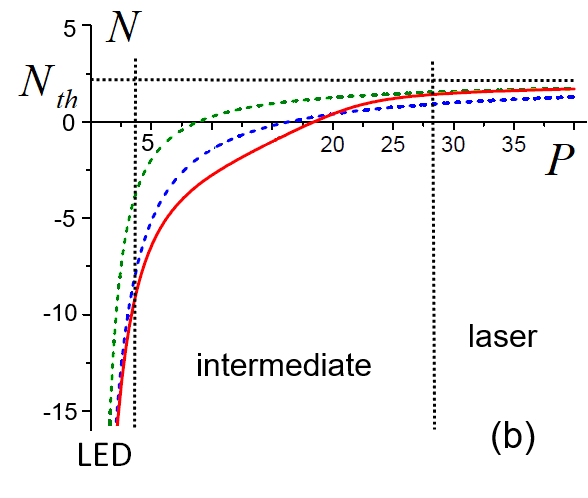}
\caption{(a) Mean photon number for a  superradiant laser  with $\gamma_{\perp}^{\rm min} = 50$~GHz (red curve) and for a non-superradiant laser with $\gamma_{\perp}^{\rm max}=1.5$~THz (black curve).  Black dashed curves are $n(P)$ as calculated by the  semiclassical model~of Eq.~\rf{nn21}.  
(b) Population inversion $N(P)$ for superradiant laser (red curve). Green (blue) dashed curves show the asymptotics of $N(P)$ at high (low) pump. The green curve is $N(P)$ at $n=n_A$ for large pump; the blue curve is $N(P)$ without population fluctuations. 
Vertical dotted lines  separate the LED and laser regions from an intermediate cross-over region.}
\label{Fig2_rev}\ec
\end{figure}

The red curve in Fig.~\ref{Fig2_rev}(a)
%
%
shows the mean photon number $n(P)$ for a superradiant laser,  
while the corresponding black curve is for a non-superradiant laser.
For low pump rates the superradiant laser  exhibits ``subradiance'' and  excitation trapping~\ct{Bohnet,PhysRevA.81.033847}, i.e. the photon number is smaller than for a non-superradiant laser. At high pump rates, by contrast, the superradiant laser is seen to generate more photons than the non-superradiant laser. There are two reasons for this: first, excitation trapping becomes weaker as the average emitter population grows and is suppressed when population inversion is achieved; second, in our case the superradiant laser has a smaller polarization relaxation rate $\gamma_{\perp}$ than the non-superradiant laser. 

Panel~\ref{Fig2_rev}(b) shows the population inversion $N(P)$  for a superradiant laser   (red curve). The blue dashed curve depicts  $N(P)$ as found by neglecting population fluctuations, which should be a good approximation for low pump rates. The dashed green curve shows  $N(P)$ in the approximation  $n\approx n_A$, which should be valid for high pump rates. The exact red curve indeed approaches the blue (green) curves at low (high) pump rates. 

By following the red curve in Fig.~\ref{Fig2_rev}(b) and observing where it approaches its asymptotics at low and at high pump rates, we can roughy identify three regions:
the LED region at small pump rates, where fluctuations of populations are negligible; the lasing region at high pump rates, when  almost all photons reside in the A-combinations of quadratures; and the remaining  intermediate region between the lasing and the LED regions.  These regions are  separated by vertical dotted lines in Fig.~\ref{Fig2_rev}(b).

\begin{figure}[thb]
\includegraphics[width=8cm]{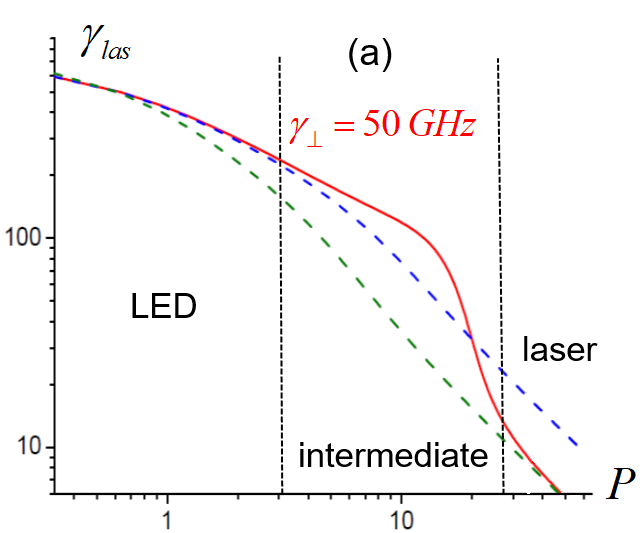}\\
\includegraphics[width=8cm]{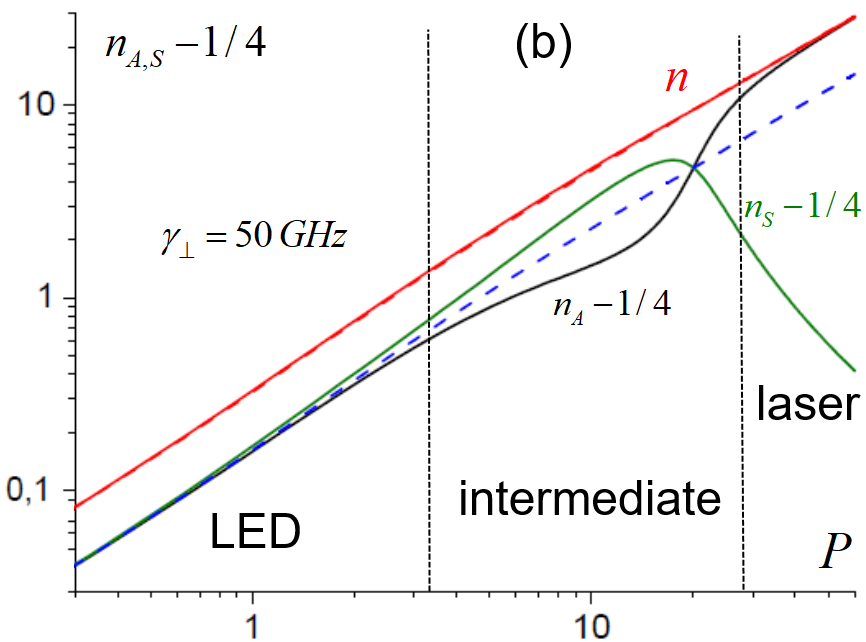}
\caption{(a) Linewidth $\gamma_{\rm las}(P)$ of Eq.~\rf{nn36}  for  a superradiant laser with $\gamma_{\perp} = 50$~GHz, with $N$ found from Eq.~ \rf{eq_N_fin}  (red curve). An approximation for $\gamma_{\rm las}(P)$ in the low-pump limit is obtained by neglecting  population fluctuations   (dashed blue curve) and  in the high-pump limit (dashed green curve), with  $N$ given by Eq.~\rf{N_bel_th_1} for both these limits;  (b)  Photon numbers in the S- (green curve) and A-combinations (black curve)   of the field of a superradiant laser. The same three radiation regions as in Fig.~\ref{Fig2_rev}(b) are shown,  separated by vertical dotted lines. The red line is the mean photon number $n$ \rf{Pn_nm_sym_ass}, which practically overlaps the red dotted line depicting $n$ without population fluctuations.} 
\label{Fig3_rev}
\end{figure}

Fig.~\ref{Fig3_rev}(a) 
%
%
shows the linewidth $\gamma_{\rm las}(P)$, given by  Eq.~\rf{nn36} with $N$ determined from Eq.~\rf{eq_N_fin} for a superradiant laser (red curve). The LED, intermediate and lasing regions shown in Fig.~\ref{Fig3_rev} are the same as found in Fig.~\ref{Fig2_rev}(b). The effects of the same approximations as in Fig.~\ref{Fig2_rev} are now shown for the linewidth:   the blue dashed curve in Fig.~\ref{Fig3_rev}(a), given by Eq.~\rf{lw_bel_th} for a laser below threshold \ct{PhysRevLett.72.3815, Yariv}, represents the approximation of neglecting population fluctuations, which is again shown to be valid for small pump rates, i.e. in the LED region.
The green dashed curve in Fig.~\ref{Fig3_rev}(a) is given by the laser linewidth  Eq.~\rf{nn19} for a laser far above threshold~\ct{van_Exter} that was found by taking  $n\approx n_A$. As before in Fig.~\rf{Fig2_rev}, this approximation is shown to be accurate at large pump rates, i.e. in the lasing region.

Fig.~\ref{Fig3_rev}(b) shows the photon numbers $(n_{S,A}-1/4)$  of the S- and  A-combinations of the field quadratures for a superradiant laser. The reason to display $(n_{S,A}-1/4)$ on the vertical logarithmic axis is that  in the low-pump limit  $P\rightarrow 0$, when $n_{S}\rightarrow n_A$, we have $n_{S,A}\rightarrow 1/4$,   in accordance with Eq.~\rf{eq_N_fin}. 
We see that population fluctuations (from now on abbreviated as PF) 
have different influences on $n_S$ and $n_A$. Removing $\delta \hat N_e$ from Eqs.~\rf{sym_antisym_eq_1}--\rf{sym_antisym_eq_3}, we find that these equations are the same for A- and for S-combinations, so that without PF, the $n_A$ and $n_S$ would be identical and follow the blue dashed curve in Fig.~\ref{Fig3_rev}~(b). Preserving $\delta \hat N_e$ in Eqs.~\rf{sym_antisym_eq_1}--\rf{sym_antisym_eq_3} and plotting $n_{S,A}(P)$ [the black and red curves in Fig.~\ref{Fig3_rev}(a)], we see that photon numbers $n_{S,A}$ are hardly affected by PF in the LED region, but  PF lead to $n_S > n_A$ in the beginning of the intermediate region. From the end of the intermediate region and onwards, the PF suppress $n_S$ making $n_S \ll n_A$ in the lasing region, where we also see that indeed $n_A \simeq n$, as we anticipated in Sec.~\ref{P_quad_sp} to derive the central result Eq.~\rf{nn19} for the linewidth. Thus, due to PF above threshold, $n_S$ is strongly suppressed and $n_A$ increased by a factor of two.

\begin{figure}[thb]\bc
\includegraphics[width=8cm]{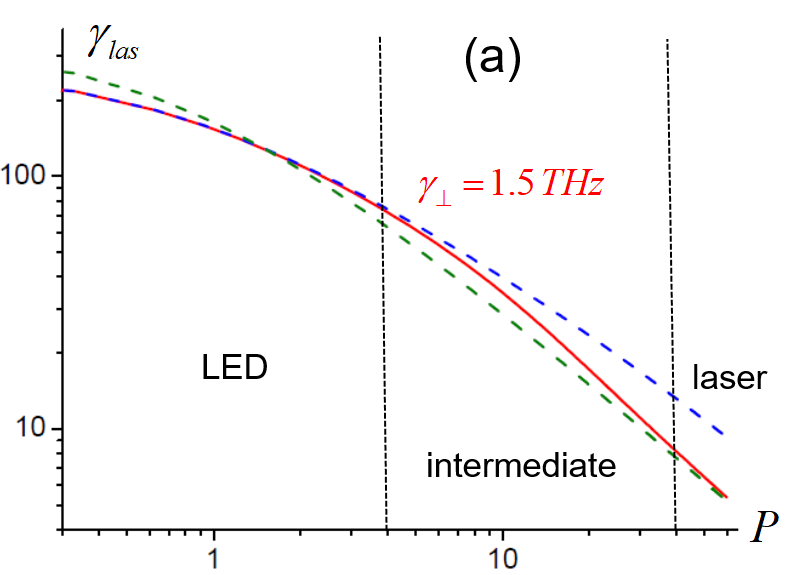}\\
\includegraphics[width=8cm]{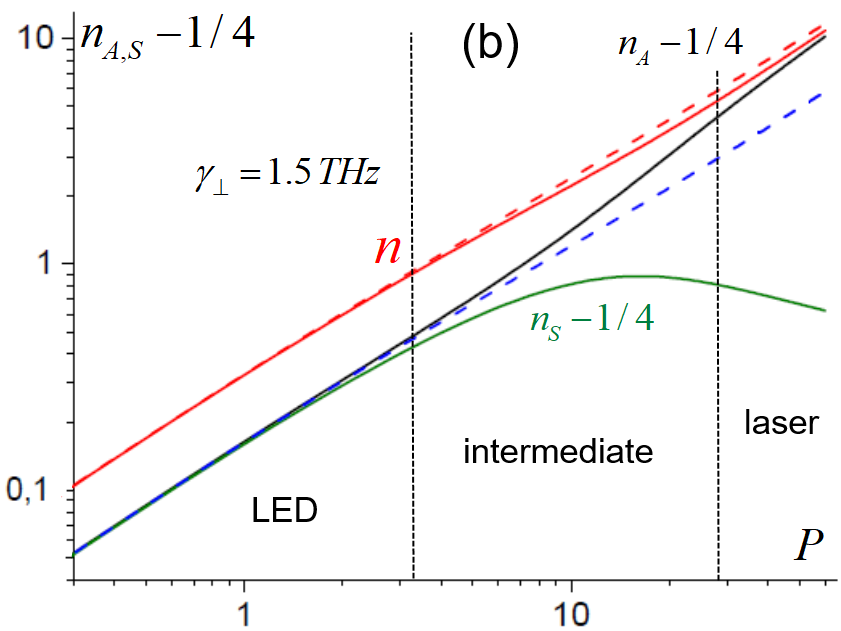}
\caption{(a) Linewidth $\gamma_{\rm las}(P)$ for  non-superradiant laser with $\gamma_{\perp}^{\rm max} = 1.5$~THz  (red curve) and approximations for $\gamma_{\rm las}(P)$ without population fluctuations (dashed blue curve) and for high pump (dashed olive curve). (b)  Energies $n_S$ of S-  and $n_A$ of  A-fluctuations 
 of  the field of a non-superradiant laser. The blue dashed curve shows $n_A=n_S$ as found without population fluctuations. In contrast with the superradiant laser in Fig.~\ref{Fig3_rev}(b), here  $n_S<n_A$ everywhere. The same radiation regions as in Fig.~\ref{Fig2_rev}(b) are  separated by vertical dotted lines. The red solid line is the mean photon number $n$ of Eq.~\rf{Pn_nm_sym_ass}, while the red dotted line is $n$ without population fluctuations.}
\label{Fig4_rev}\ec
\end{figure}

Fig.~\ref{Fig4_rev}(a,b) shows the analogous results for a conventional (non-superradiant) laser, to be contrasted with the case of Fig.~\ref{Fig3_rev}(a,b).
%
%
%
The linewidth $\gamma_{\rm las}(P)$ of the conventional laser  departs more gradually from its asymptotics at low pump to the asymptotics for high pump. An even more conspicuous difference with superradiant lasers is that for the non-superradiant  laser $n_S$ never exceeds $n_A$. This is an indication that PF have a stronger effect on the superradiant lasers. But $n_{S}$ and $n_{A}$ are both clearly affected by PF for both types of lasers, and for both lasers $n_S \simeq 0$ and $n_A \simeq n$ at high pump rates.  By contrast, both for the superradiant laser in  Fig.~\ref{Fig3_rev}(b) and for the conventional laser in Fig.~\ref{Fig4_rev}(b), we see that the total number of photons practically does not depend on PF.  
 
In the next section we will see that the larger PF of superradiant lasers will make their spectra qualitatively different from those of non-superradiant lasers.
%
\subsection{Optical spectra}\label{SPC}
%
The different shapes of optical spectra for low and for high pump rates reflect different physical effects. It is therefore convenient to consider the spectra for high and for low pump rates separately.

\begin{figure}[thb]\bc
\includegraphics[width=8cm]{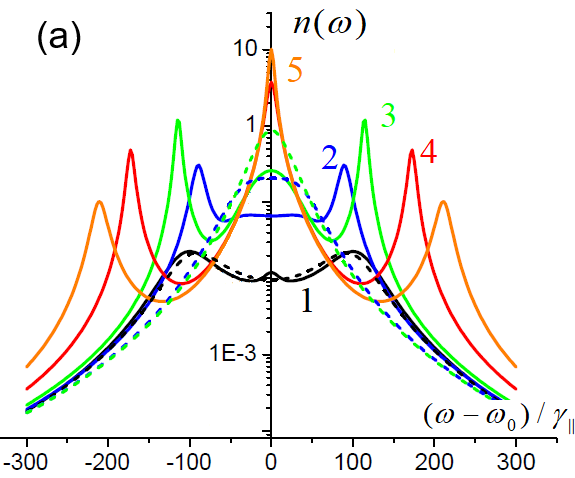}\\
\includegraphics[width=8cm]{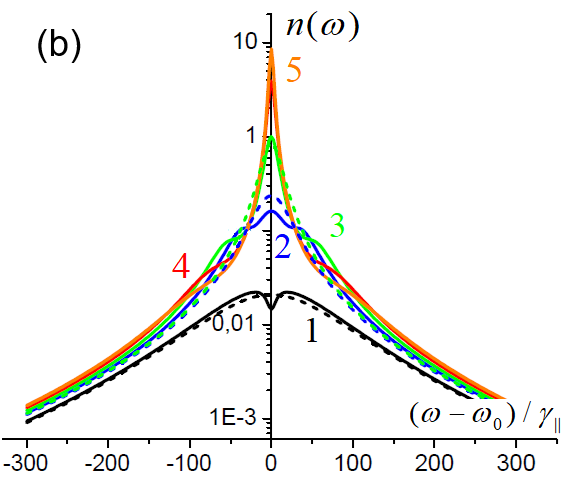}
\caption{(a) Optical spectra of a superradiant laser with $\gamma_{\perp}^{\rm min}=5\cdot 10^{10}~s^{-1}$ and $2\kappa/\gamma_{\perp} = 2$  and (b) of a non-superradiant laser with $\gamma_{\perp}=5\cdot 10^{11}~s^{-1}$ and $2\kappa/\gamma_{\perp} = 0.2$,  for pump rates $P=2$ (curves 1); 8 (curves 2); 16 (curves 3); 28 (curves 4) and 40 (curves 5). The dashed curves are spectra  found without population  fluctuations, with the same parameters as for the solid curves 1, 2 and 3 of the same color. }
\label{Fig5_rev}\ec
\end{figure}

Figure~\ref{Fig5_rev}(a) shows optical spectra $n(\omega)$, given by Eq.~\rf{full_las_sp}, for a superradiant laser, with  $2\kappa/\gamma_{\perp}^{\rm min} = 2$, for high pump rates $P\geq 2$. 
Fig.~\ref{Fig5_rev}(b) shows the same for a non-superradiant laser, with  $2\kappa/\gamma_{\perp} = 0.2$. Only in Fig.~\ref{Fig5_rev}(a) for the  superradiant laser do we see sideband peaks in  the spectra. 
%
%
%

These sideband peaks (or spikes) in $n(\omega)$ for Curves 2 to 5 in Fig.~\ref{Fig5_rev}(a) have the same nature as  relaxation oscillations  in lasers with $2\kappa \ll \gamma_{\perp}$~\ct{1071726} that are described by rate equations.  Sideband peaks  appear because  the carrier population  reacts with some delay to changes in the field and polarization. The delay causes oscillatory energy exchange between the field, polarization and population with a resonance at the  relaxation oscillation  frequency. For $2\kappa \ll \gamma_{\perp}$, the relaxation oscillation frequency is given by $\omega_{\rm ro}$, as defined in Eq.~\rf{-omega_ro-}~\ct{1071726}. Such resonances cause well-known sidebands in the intensity fluctuation spectra~\ct{1071726,6603264}, see also Fig.~\ref{Fig6_rev}(a,b) below. Analogous sideband peaks due to relaxation oscillations are not resolved in optical spectra  of the non-superradiant laser with $2\kappa/\gamma_{\perp}=0.2 \ll 1$ in Fig.~\ref{Fig5_rev}(b). We attribute this to the fact that such a laser has smaller population fluctuations than a superradiant laser, as we have seen above in the analysis of the mean photon numbers, comparing Figs.~\ref{Fig3_rev}(b) with \ref{Fig4_rev}(b). 

For comparison, the dashed versions of Curves~1, 2 and 3 show the corresponding spectra if population fluctuations are neglected. While Curves~1 with and without population fluctuations are practically identical (apart from a small structure in the center that we will discuss below), the dashed and solid Curves~2 and~3 in Fig.~\ref{Fig5_rev}(a)
are qualitatively different: no  sideband peaks are observed when population fluctuations are neglected. By contrast, solid and dashed curves are quite close to each other in Fig.~\ref{Fig5_rev}(b) for the conventional (non-superradiant) laser, where population fluctuations are smaller than in the superradiant laser.

The two-peak structure in $n(\omega)$ for Curve~1 in Fig.~\ref{Fig5_rev}(a) has a different origin than the sideband peaks in curves 2-5 in this Figure. The two broad peaks in Curve 1 are due to  collective Rabi splitting (CRS),  which occurs when a large number of emitters exhibit Stark shifts in the lasing field. The parameters used for Curve~1 satisfy the conditions for CRS,  in particular $P<P_c$, see Sec.~\ref{Below_th} after Eq.~\rf{splt_b_th}. For Curves~2 to 5, we have $P>P_c$, so CRS  is absent. We described CRS in more detail in Ref.~\ct{Andre:19}, but without taking population fluctuations into account.  

\begin{figure}[thb]\bc
\includegraphics[width=8cm]{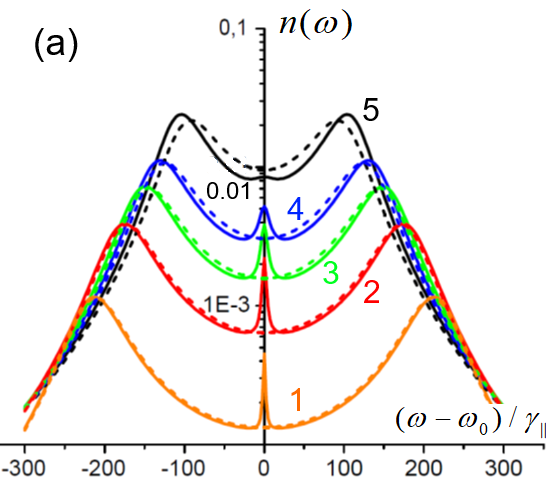}\\
\includegraphics[width=8cm]{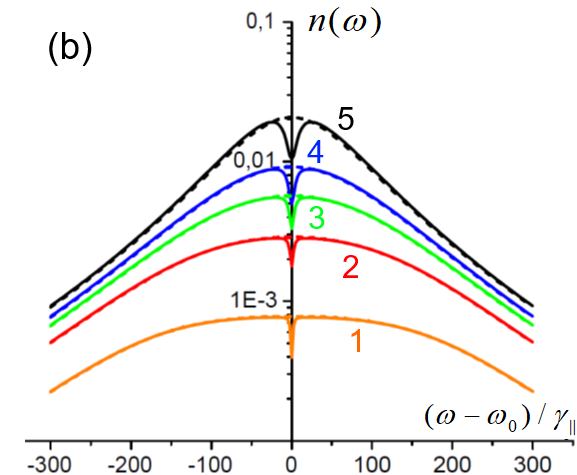}
\caption{Spectra when weakly pumping (a) a superradiant laser with $\gamma_{\perp}^{\rm min}=5\cdot 10^{10}~s^{-1}$ and $2\kappa/\gamma_{\perp} = 2$  and (b) a non-superradiant laser with $\gamma_{\perp}=5\cdot 10^{11}~s^{-1}$ and $2\kappa/\gamma_{\perp} = 0.2$. The  pump rates are $P=0.16$ (curves 1);  0.48 (curves 2); 0.8 (curves 3);  1.12 (curves 4) and  2 (curves 5). Solid curves represent spectra with population fluctuations. For comparison, the corresponding dashed curves are found upon neglecting population  fluctuations. Small structures (peaks and dips) in the centers of the spectra are due to nonlinear polarization induced by population fluctuations. }
\label{Fig5w}\ec
\end{figure}

Figs.~\ref{Fig5w}(a,b) show optical spectra for low pump rates $P\leq 2$.  The two broad peaks in the curves in Fig.~\ref{Fig5w}~(a) are due to CRS. At these lower pump rates than previously considered in Fig.~\ref{Fig5_rev}, population fluctuations lead to small features in the center of the optical spectra, namely a peak in Fig.~\ref{Fig5w}(a) and a dip in Fig.~\ref{Fig5w}(b). 
%
%
%
%
These small features arise due to nonlinearities. Mathematically, 
the denominator of Eq.~\rf{sp_x00} for $n_S(\omega)$ is a cubic polynomial of $\omega^2$ and the spectrum $n(\omega)$ in Eq.~\rf{full_las_sp} is a complicated function of $\omega^2$. 

Physically, the photons emitted at the frequencies of the two CRS peaks participate in a nonlinear scattering process: they induce  population fluctuations, which couple to the polarization, and thereby  have a  back 
action on photon emission. In other words, photons may be 
re-absorbed and then emitted again by other emitters, which constitutes a nonlinear photon scattering process. 
Fig.~\ref{Fig5w}~(a) shows that photons from both CRS peaks are re-absorbed and re-emitted most effectively near the center of the spectrum, 
leading to the small central peak 
in the optical spectra in Fig.~\ref{Fig5w}(a). 

Fig.~\ref{Fig5w}(b) shows the modification of the spectrum $n(\omega)$  due to the nonlinear photon scattering in a conventional (i.e. non-superradiant) laser, when conditions for  CRS are not satisfied and CRS peaks are absent. Here, most  emitters are near the center of the spectrum, and photons emitted in this spectral region  are absorbed and re-emitted by all other emitters away from the center. By nonlinear scattering the energy of the field is thus taken from the center of the spectrum, where we see a dip, and
re-emitted far from the center. Indeed we see that away from the dip, the solid curves in Fig.~\ref{Fig5w}(b) lie slightly above the dashed curves, which do not take into account the nonlinear photon scattering.  

For increasing pump rates, the main lasing peak grows, and the tiny nonlinear structures in the center of the lasing spectrum disappear.
%
\subsection{Spectra of intensity fluctuations}\label{Sec:nRF}
%
The radio-frequency spectrum of laser intensity fluctuations is important for applications of lasers in  optical communications and can be measured by direct photodetection of the lasing field as fluctuations of the photocurrent~\ct{1071726}. Here we will
derive an approximate expression for the intensity fluctuation spectra of our nanolasers, restricting ourselves to the high-pump limit, where almost all energy resides in the A-combinations of the quadratures. 

Slow fluctuations of A-combinations, or equivalently phase fluctuations according to Fig.~\ref{Fig2_3}, can be neglected in calculations of photon number fluctuations. In this high-excitation limit, according to Eqs.~\rf{int_app_1} and~\rf{sym_ass},
\beq
    \hat{a} \approx (\sqrt{n}+\hat{a}'_S)e^{i\pi/4}. \lb{a_for_popfl}
\eeq
In Eq.~\rf{a_for_popfl} the operator  $\hat{a}'_S$ describes broadband amplitude fluctuations, and we  take $\hat{a}'_S = \hat{a}_S$ as in Eq.~\rf{app_prime}. 
Then from Eq.~\rf{a_for_popfl} we obtain the approximation $\hat{a} \approx (\sqrt{n}+\hat{a}_S)e^{i\pi/4}$, 
which leads to the  photon number (or intensity) fluctuation spectrum
\beq
n_{\rm RF}(\omega) = 4n n_S(\omega), \lb{Pn_fl_sp}
\eeq
where $n_S(\omega)$ is given by Eq.~\rf{sp_x00}. In
%
%
\begin{figure}[thb]\bc
\includegraphics[width=8cm]{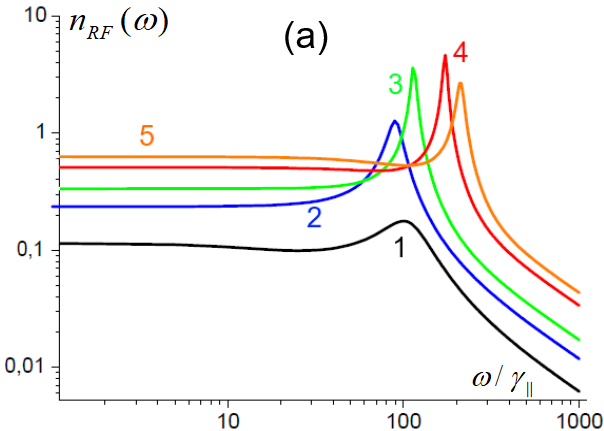}\\\includegraphics[width=8cm]{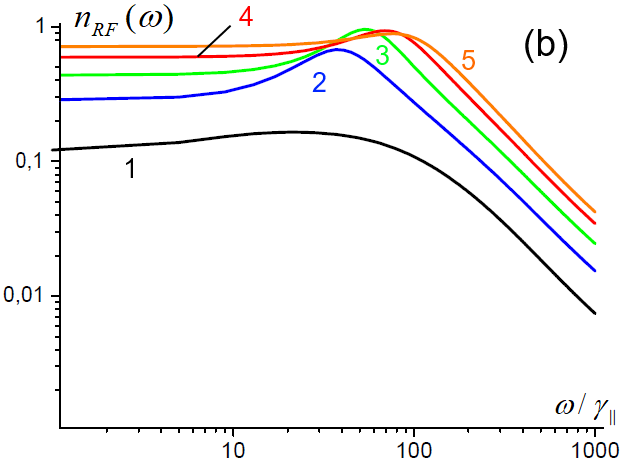}
\caption{Photon number fluctuations in the high-pump limit for (a) a superradiant laser and (b) a non-superradiant laser. The parameters for the lasers and for the Curves~1 to 5 are the same as in Fig.~\ref{Fig5_rev}(a,b). Photon number fluctuations and relaxation oscillation peaks for the non-superradiant laser are much smaller than for the superradiant laser. 
}
\label{Fig6_rev}\ec
\end{figure}
%
%
Fig.~\ref{Fig6_rev},  photon number fluctuation spectra are shown  for the same parameter values  as in Fig.~\ref{Fig5_rev}. In Figs.~\ref{Fig6_rev}(a,b) one observes the  well-known relaxation oscillation peaks~\ct{McKinstrie:20, 1071729}. When $\gamma_{\perp}$ increases
corresponding to the transition from a superradiant laser [Fig.~\ref{Fig6_rev}~(a)] to a conventional laser [Fig.~\ref{Fig6_rev}~(b)], the population fluctuations are reduced and the maxima of the relaxation oscillation peaks in $n_{\rm RF}(\omega)$ decrease. Similarly, the  sideband peaks in the optical spectra of the superradiant laser shown in   Fig.~\ref{Fig5_rev}(a) disappear 
upon increasing $\gamma_{\perp}$, finally arriving to the case of the conventional laser with optical spectra as shown in Fig.~\ref{Fig5_rev}(b). This similarity confirms that sideband peaks in optical spectra of a superradiant laser are caused by strong population fluctuations leading to strong relaxation oscillations. 
\section{Discussion}\label{Dis}
In this section we discuss our approach to linearize the Maxwell-Bloch equations, followed by a discussion of the main results.

In the high-pump limit, Eqs.~\rf{1} can be linearized around their mean values, when the field and polarization are sums of their coherent parts, which is a solution of the semiclassical Equations~\rf{nn21}, plus small fluctuations~\ct{RevModPhys.68.127,  PhysRevA.47.1431, 1071726}. Such a linearization is equivalent to the small-signal analysis that is well-known in electrical engineering~\ct{Math_el_en} and which has also been applied to laser rate equations before~\ct{McKinstrie:20,8506446}. Rate equations can not be applied for superradiant lasers, where polarization is a dynamical variable.

The standard small-signal analysis of Eqs.~\rf{1} cannot be applied in the  low-pump limit, where the mean values of the lasing field and polarization vanish. 
In order to linearize the Maxwell-Bloch equations in the low-pump limit, we made a first approximation within our  new method in Sec.~\ref{Below_th} by  neglecting  population fluctuations with respect to the large mean-value population of the large number of emitters. 
This approximation gives satisfactory results when calculating  mean values. An important example 
is the mean photon number, which is well described  
both below, near and above the semiclassical laser threshold. Also, the laser linewidth is accurately accounted for, at least below threshold. However, it is one of the main points of this paper that 
accounting for population fluctuations is necessary in order to correctly account for the linewidth of the laser above threshold, as well as the detailed structure of the laser spectrum below as well as above threshold.

We therefore made an improved analysis, where both the population fluctuations and the dynamics of the material polarization are taken into account. 
First we considered the  high-pump limit in Sec.~\ref{LLE_Lf}, but again using an approach that differs from and extends the conventional small-signal analysis.  The first difference is that we separate the mean values and fluctuations only in the {\it non}linear terms in the laser equations~\rf{1}. A further difference is that we use the symmetric (S) and the anti-symmetric (A) combinations~\rf{sym_ass} of the quadratures.   In Fig.~\ref{Fig2_3} we showed that S-combinations are related to the amplitude and  A-combinations to the phase of the lasing field. We arrived at two linear sets of equations: a set of two equations~\rf{sym_antisym_eq_1} and  \rf{sym_antisym_eq_3} for A-combinations and a set of three equations~\rf{sym_antisym_eq_1}, \rf{sym_antisym_eq_2} and  \rf{sym_antisym_eq_4} for S-combinations of the quadratures coupled to population fluctuations. Our finding  that population fluctuations are explicitly coupled to the S- but not to the A-combinations, is in  agreement with the well-known fact~\ct{Coldren} that variations in the carrier density change the gain, which subsequently changes the amplitude. If the alpha-parameter (linewidth enhancement factor) is zero, as in our case, the gain changes do not affect the phase. 

Next we put forward the hypothesis that the linearized equations~\rf{sym_antisym_eq} as derived in the high-pump limit can be used in the intermediate- and in the low-pump limits as well, where they give satisfactory approximate predictions. First of all, if we neglect population fluctuations, then Eqs.~\rf{sym_antisym_eq} become identical to the equations~\rf{1n} that we derived for the  low-pump limit in Sec.~\ref{Below_th}. 

Having discussed our new approach in detail, in the remaining part of this section we will  discuss our new results. 

We demonstrated that the well-known factor $1/2$ difference between the laser linewidth Eq.~\rf{lw_bel_th} for the low-pump limit and Eq.~\rf{nn19} for the high-pump limit arises due to population fluctuations. If we would neglect population fluctuations, then  we would  obtain the linewidth~\rf{lw_bel_th} for arbitrary pump rates.
Furthermore, 
we calculated the laser linewidth at arbitrary
pump rates, taking population fluctuations into account, and demonstrated  a smooth transition between 
the limits for low and high pump rates, as shown in Figs.~\ref{Fig3_rev}(a) and~\ref{Fig4_rev}(a).

We calculated the full optical spectrum, given by Eqs.~\rf{fullspectrum} or \rf{full_las_sp}. 
In the high-pump limit, the optical spectrum features a narrow peaked spectrum $n_A(\omega)$, with a width determined by the phase fluctuations, on top of a broad background spectrum $n_S(\omega)$, related to the amplitude fluctuations. Similar sharp ``coherent'' peaks and ``incoherent'' wings are seen in experiments, for example in Fig.~1 of Ref.~\ct{PhysRevLett.72.3815} or, for superradiant lasers, in Fig.~4(b) of Ref.~\ct{PhysRevX.6.011025}. The usual analysis completely neglects this broad background emission in the optical spectra~\ct{1071986}. 

We found sidebands in the {optical spectrum $n(\omega)$, shown in  Fig.~\ref{Fig5_rev}(a) for a superradiant laser, characterized by a high beta factor and a low-quality cavity ($2\kappa/\gamma_{\perp} \geq 1$). In this case, population fluctuations are especially important and the laser displays strong relaxation oscillations, which show up as sidebands in the optical spectra.

Another mechanism that gives rise to satellite peaks in optical spectra  is well-known for semiconductor lasers. It arises because the refractive index depends on the level of excitation, as described by the so-called $\alpha$-parameter in the rate equations~\ct{Henry1986}. However, this is not the mechanism leading to the sideband peaks in our model, where the frequency of the lasing field is exactly on resonance with the lasing transition and the $\alpha$-parameter is zero. 

For weak pumping, when the main lasing peak has not yet appeared, we predict small features in the center of the lasing spectra as displayed in Fig.~\ref{Fig5w}.  These small spectral peaks and dips only appear when taking population fluctuations into account. With our present theory we only make  qualitative predictions about these small features, because  we neglect the second-order correlations~\rf{soc_1}, although at low excitation these are of the same  order of magnitude as the linear terms~\rf{app_prime} in Eq.~\rf{sym_antisym_eq_4}.  
From calculations that do account for these correlations~\rf{soc_1}, not shown in this paper, we find that  the small central peaks and dips are preserved, but their heights and widths are changed.  Below we give an interpretation why the central peak [dip] appears for the lasers parameters of Fig~\ref{Fig5w}(a) [Fig~\ref{Fig5w}(b)]. 

Population fluctuations (PF) at low excitation have a relatively high spectral power density in the range $\gamma_{\parallel} \ll \gamma_{\perp}, 2\kappa$ near zero frequency, where the PF are mostly due to the pump and the population decay noise. Far from the center of the spectrum on the other hand, the PF are caused by the interaction with the field and polarization, and the spectral density of PF is relatively small. Since $\gamma_{\parallel}$ is much smaller than the width of the spectra in Fig.~\ref{Fig5w}, the PF can produce either a peak or a dip of width $\sim \gamma_{\parallel}$ in the center of the optical spectrum.} 
 
When collective Rabi splitting (CRS) is absent, as in Fig.~\ref{Fig5w}(b), then the maxima in the linear and nonlinear parts of the polarization coincide and interfere destructively in the frequency region of width $\gamma_{\parallel}$ around the center of the optical spectrum. The linear part of the polarization (i.e. the part that is independent of population fluctuations) induces the birth of a photon. This leads to a PF, which decreases the population of the upper level. Such a PF, in turn, leads to a polarization fluctuation, which 
suppress the birth of the next  photon. This is a reason for a dip in the center of the optical spectra in Fig.~\ref{Fig5w}(b).  

The situation is different when the maxima of the linear part of the polarization are not in the center of the optical spectrum, as it happens when collective Rabi splitting occurs.
Then the PF in the frequency region near the center of the spectrum are less affected by the lasing field, compared to the case without CRS. The field now resides mostly in  CRS peaks far from the center,  where the spectral power density of the PF is relatively small. The pump, therefore, increases the  PF near the center of the spectrum, leading to a central peak, as shown in Fig.~\ref{Fig5w}(a).

The new sideband peaks and fine structures of spectra of superradiant laser have not been seen in the experiments of Refs.~\ct{PhysRevX.6.011025, PhysRevA.96.013847,PhysRevA.81.033847,Bohnet,Jahnke}. Based on our theory, we 
predict the kind of lasers and the parameter region, given after Eq.~\rf{eq_N_fin}, in which to observe such features.  
The  active medium should 
work in a three-level (effectively two-level) scheme. The conditions for a laser to be superradiant must be satisfied, i.e.  $\tilde{\beta}_c \gg 1$ and $2\kappa \geq \gamma_{\perp}$. For example, one can reduce $\gamma_{\perp}$, so that $2\kappa \geq \gamma_{\perp}$, by lowering the temperature of a nanolaser with q-dots as the active medium~\ct{PhysRevB.46.15574}.
Or {\it vice versa}, by increasing the temperature one can go from the superradiant to the non-superradiant  regime in the same laser. The pump rate must be in the range from $\gamma_{\parallel}$ to $20\gamma_{\parallel}$,   
corresponding to the intermediate region [see Fig.~\ref{Fig2_rev}(b)], where population fluctuations have their maximal effect, and the sideband peaks in optical spectrum [see Fig.~\ref{Fig5_rev}(a)] can be observed. 
Lasers with possibly larger relaxation oscillation peaks in the intensity fluctuation spectrum, as in Fig.~\ref{Fig6_rev}(a), are good candidates for observing  sideband peaks in the optical spectrum as predicted here.

In our theory we have neglected inhomogeneous broadening of the active medium. This implies that for our current theory to be directly appliccable, the actual inhomogeneous broadening must be much smaller than the distance between sideband peaks in the field spectra in Fig.~\ref{Fig5_rev}(a). It was shown in Ref.~\ct{PhysRevA.100.053821} that  collective effects synchronize  emitters, even if the cavity linewidth is smaller than the inhomogeneous broadening. So we expect that some  features in the spectra of SR lasers as discussed here will also show up in the presence of inhomogeneous broadening. 
More precise criteria for the observation of sideband peaks in the optical spectra of superradiant lasers will be derived from detailed investigations of the optical spectrum~\rf{full_las_sp} in the future. 
\section{Conclusion}\label{Conc}
We presented a quantum theory for the spectra and fluctuations of a single-mode homogeneously broadened two-level  laser. 
We developed a new approach, which does not employ the common approximation of eliminating the polarization adiabatically, meaning that the theory also applies to
superradiant  lasers with a low-quality cavity and a high beta factor. We linearised the equations, solved them and obtained analytical results.

We identify the LED region, where the laser linewidth described by Eq.~\rf{lw_bel_th} is wide as in Refs.~\ct{PhysRevLett.72.3815, Yariv}, the lasing region with a narrow laser linewidth Eq.~\rf{nn19} in agreement with Ref.~\ct{van_Exter}, and the intermediate region in between them. 

Different from the rate-equation approaches of Refs.~\ct{Henry1986,1072058,1071986,1071726}, we describe a smooth transition between the linewidth Eq.~\rf{lw_bel_th} in the LED region and the linewidth Eq.~\rf{nn19} in the lasing region, 
as  in Figures~\ref{Fig3_rev}(a) and~\ref{Fig4_rev}(a). Furthermore, we calculated optical spectra, including phase and amplitude fluctuations, which respectively lead to a narrow peak and a  wide background of the spectra at high excitation, as in Fig.~\ref{Fig5_rev}. 

For superradiant lasers in the intermediate region, we predict two sideband peaks in  the optical spectra $n(\omega)$, as a consequence of strong relaxation oscillations induced by strong population fluctuations. The sideband structure in $n(\omega)$ dissapears for a non-superradiant laser with smaller relaxation oscillations and weaker population fluctuations. 

In the LED region, we predict a structure (peak or dip) in the center of the lasing spectrum that is caused by the interference of the linear and nonlinear parts of the polarisation. The interference is constructive, leading to a small peak in the center of the spectrum, if the linear part of polarisation displays collective Rabi splitting~\ct{Andre:19}. Otherwise the interference is destructive and then we predict a corresponding small dip in the center of the spectrum.

We expect that our approach and results will be useful also for further theoretical and experimental studies of spectra, fluctuations and correlations in lasers for a wide range of parameters, at any pump values and when the active-medium polarisation cannot be adiabatically eliminated, as in  superradiant lasers.

In this paper we used our method only to calculate mean values and lasing spectra. In the future, our approach 
could also be used to calculate higher-order correlations, for example the second-order correlation function $g_2$ of the lasing field.

\begin{acknowledgments}
This work has been funded by Villum Fonden (VKR Center of Excellence NATEC-II, grant 8692), and by the Danish National Research Foundation through NanoPhoton - Center for Nanophotonics (grant number DNRF147). M.W. acknowledges support from the  Independent Research Fund Denmark – Natural  Sciences  (Project  No.  0135-00403B).
Igor Protsenko wishes to acknowledge the support of  an Otto Moensted Visiting Professorship grant (19-12-1159).
\end{acknowledgments}

\appendix
\section{Diffusion coefficients}\label{App_A}
%
The Heisenberg-Langevin equation for an  operator $\hat{Q}_{\alpha}$ is 
\beq
{d\hat{Q}_{\alpha}}/{dt} = \hat{M}_{\alpha} +  \hat{F}_{\alpha}, \lb{Ap1}
\eeq
where $\hat{F}_{\alpha}$ is the Langevin force, with properties%
\[ \langle\hat{F}_{\alpha}\rangle = 0, \hspace{1cm} \langle\hat{F}_{\alpha}(t)\hat{F}_{\beta}(t')\rangle =2D_{\alpha\beta}\delta(t-t').
\]
Here the diffusion coefficient $2{{D}_{\alpha \beta }}$ is determined from the ``generalized Einstein formula''
\beq
2D_{\alpha \beta} =  d\langle\hat{Q}_{\alpha}\hat{Q}_{\beta}\rangle/dt - \langle\hat{M}_{\alpha}\hat{Q}_{\beta}\rangle - \langle\hat{Q}_{\alpha}\hat{M}_{\beta}\rangle. \lb{Ap2}
\eeq
If we denote the upper (lower) state of the $i$-th emitter by $\left|e\right>_i$  ($\left|g\right>_i$), then $\hat{\sigma}_i = \left|g\right>_i\left<e\right|_i$, $\hat{\sigma}_i^+ = \left|e\right>_i\left< g\right|_i$ and $\hat{n}_i^{(\alpha)} = \left|\alpha\right>_i\left<\alpha\right|_i$, for $\alpha = \{e,g\}$. Using the orthogonality of the states $\left<\alpha\right|_i\left|\beta\right>_j = \delta_{\alpha\beta}\delta_{ij}$, we obtain $\hat{\sigma}_i^+\hat{\sigma}_j = \hat{n}_i^{(e)}\delta_{ij}$  and therefore $\hat{v}^+\hat{v} = f\hat{N}_e$, where $\hat{v}$ and $f$  are the same as in Eqs.~\rf{1}.

From Eq.~\rf{Ap2} we find
\beq
    2D_{v^+v} = f[\gamma_{\perp}N_e+\gamma_{\parallel}(PN_g-N_e)]. \lb{Ap3}
\eeq
Taking $\hat{a}(\omega)$ given by Eq.~\rf{nn13}, the diffusion coefficient~\rf{Ap3} and $2D_{a^+a} =0$ we find the commutator
\[
\langle[\hat{a},\hat{a}^+]\rangle = 1-\frac{4\kappa n}{\gamma_{\perp}(N_{\rm th}+N_0)}\left(2n+\frac{1}{1+2\kappa/\gamma_{\perp}}\right).
\]
The mismatch with $\langle[\hat{a},\hat{a}^+]\rangle = 1$ comes from the second term on the right in Eq.~\rf{Ap3}, which must be removed in the approximation $\hat{N}_e = N_e$ (i.e. when neglecting population fluctuations), so that we take instead $2D_{v^+v} = f\gamma_{\perp}N_e$  as in Eq.~\rf{nn320} of the main text. This example shows that  usage of exact diffusion coefficients in combination with approximate equations is an excess of accuracy and may lead to the breaking of commutation relations. 

If $\hat{Q}_{\gamma_i} = \hat{Q}_{\alpha_i} + \hat{Q}_{\beta_i}$ then
\beq
    2D_{\gamma_1\gamma_2} = 2D_{\alpha_1\alpha_2} + 2D_{\beta_1\alpha_2} + 2D_{\beta_2\alpha_1} + 2D_{\beta_1\beta_2}, \lb{Ap4}
\eeq
as it follows from Eq.~\rf{Ap2}. From Eq.~\rf{Ap2} we find
\beq
2D_{vv^+} = f[\gamma_{\perp}N_g-\gamma_{\parallel}(PN_g-N_e)], \hspace{0.25cm}2D_{aa^+} = 2\kappa, \lb{Ap5}
\eeq
while the remaining diffusion coefficients for $\hat{a}$, $\hat{a}^+$,  $\hat{v}$, $\hat{v}^+$ all vanish. From Eqs.~\rf{Ap3} -- \rf{Ap5} we obtain the diffusion coefficients~\rf{nn14} of the main text.

Using the operator relations $\hat{N}_e\hat{N}_e = \hat{N}_e$ and  $\hat{N}_e\hat{N}_g = 0$, we obtain $2D_{N_eN_e}$ in Eq.~\rf{dif1_x} of the main text.
%
\section{Stationary solutions}\label{App_0}
%
In this Appendix we derive expressions for the stationary population inversion and photon number both below and above threshold. 

First we consider the situation  below threshold. We solve Eq.~\rf{nn36a} and find the stationary population inversion 
\begin{subequations}\lb{N_bel_th}\beq
N(P)  =  \frac{N_{\rm th}}{2(P+1)}\left[2(P-1)N_0/N_{\rm th} -M_b-\sqrt{Q_b}\right], \lb{N_bel_th_1}
\eeq
where
$ M_b=(P-1)N_0/N_{\rm th}-P-1-\tilde{\beta}_c$ and
$ Q_b = M_b^2+8P\tilde{\beta}_cN_0/N_{\rm th}$. After that we determine the stationary photon number $n(P)$ from the energy conservation law Eq.~\rf{nn2} and find 
\beq
n(P)  =  \frac{1}{4\tilde{\beta}}\left(M_b+\sqrt{Q_b}\right).\lb{N_bel_th_2}
\eeq
\end{subequations}

Next we do the analogous analysis in the
high-excitation limit. In this case, almost all photons are in the A-combination of quadratures, so in Eq.~\rf{nn18} we set $n_A=n$  and we neglect the constant term $2\kappa/\gamma_{\perp}$ with respect to the term $\sim 1/(N_{\rm th}-N)$ that gets large when $N\rightarrow N_{\rm th}$ at high pump. 
By inserting this photon number $n$ into Eq.~\rf{nn2} we again arrive at a quadratic equation for $N$,
\[
	 \tilde{\beta}_c(N_0+N_{\rm th})/2 = [P(N_0-N)-N_0-N](1-N/N_{\rm th}). 
\]
Solving this equation and using the energy conservation law Eq.~\rf{nn2}, we find solutions $N$ and $n$ that are again given by Eqs.~\rf{N_bel_th_1}, but now with the coefficients
\beqr
M_a&=&(P-1)N_0/N_{\rm th}-P-1, \nonumber\\
Q_a &= &M_a^2+2(P+1)\tilde{\beta}_c(N_0/N_{\rm th}+1),\nonumber
\eeqr
instead of  $M_b$ and $Q_b$, respectively.
\section{Derivation of equations \rf{sym_antisym_eq}}\label{App_C0}
We insert $\hat{a}_{x,p} = \sqrt{n}+\hat{a}_{x,p}'$ and $\hat{v}_{x,p} = V+\hat{v}_{x,p}'$, defined in  Eqs.~\rf{int_app}, into the nonlinear terms  in Eqs.~\rf{quadratures_eq} and neglect nonlinear contributions in these terms, so that
\beqr
\hat{a}_{x,p}\delta\hat{N}_e &\approx& \sqrt{n}\delta\hat{N}_e,\lb{C_01}\\  {{{\hat{a}}}_{p}}{{{\hat{v}}}_{x}}+{{{\hat{v}}}_{p}}{{{\hat{a}}}_{x}}-\left\langle {{{\hat{a}}}_{p}}{{{\hat{v}}}_{x}}+{{{\hat{v}}}_{p}}{{{\hat{a}}}_{x}} \right\rangle &\approx& \nonumber \sqrt{n}(\hat{v}_x' + \hat{v}_p') + V(\hat{a}_x' + \hat{a}_p').
\eeqr
Inserting approximation \rf{C_01} into Eqs.~\rf{quadratures_eq} we arrive at the linear equations 
\begin{subequations}\lb{C3_2}\beqr
   \dot{\hat{a}}_{x,p}&=&-\kappa {{{\hat{a}}}_{x,p}}+{{\Omega }_{0}}{{{\hat{v}}}_{x,p}}+{{{\hat{F}}}_{{{a}_{x,p}}}}, \lb{C3_2_1}\\ 
 {{{\dot{\hat{v}}}}_{x,p}}&=&-({{\gamma }_{\bot }}/2){{{\hat{v}}}_{x,p}}+{{\Omega }_{0}}f\left( {{{\hat{a}}}_{x,p}}N+2\sqrt{n}\delta {{{\hat{N}}}_{e}} \right)\nonumber\\ & & +{{{\hat{F}}}_{{{v}_{x,p}}}}, \lb{C3_2_2}\\ 
 \delta\dot{\hat{N}}_e&=&-\Omega_0\left[\sqrt{n}(\hat{v}_x' + \hat{v}_p') + V(\hat{a}_x' + \hat{a}_p')\right]\nonumber\\& &-{{\gamma }_{\parallel }}(P+1)\delta {{{\hat{N}}}_{e}}+{{{\hat{F}}}_{{{N}_{e}}}}. \lb{C3_2_3}
 \eeqr\end{subequations}
We rewrite Eqs.~\rf{C3_2}  in terms of $\hat{a}_{A,S}$, $\hat{v}_{A,S}$ and $\hat{a}_{A,S}'$, $\hat{v}_{A,S}'$  introduced in Eq.~\rf{sym_ass}. In Eq.~\rf{C3_2_3} we use expression \rf{7_wd} for $V$ and replace $\hat{v}_x' + \hat{v}_p' = 2\hat{v}_S'$, $\hat{a}_x' + \hat{a}_p' = 2\hat{a}_S'$. After this we obtain from Eqs.~\rf{C3_2} 
\begin{subequations}\lb{C4_2}
\beqr
   \dot{\hat{a}}_{A,S}&=&-\kappa {{{\hat{a}}}_{A,S}}+{{\Omega }_{0}}{{{\hat{v}}}_{A,S}}+\hat{F}_{a_{A,S}}, \lb{C4_2_1}\\ 
 \dot{\hat{v}}_A&=&-(\gamma _{\bot}/2)\hat{v}_A+\Omega _0f\hat{a}_AN  +\hat{F}_{v_A}, \lb{C4_2_2}\\
  \dot{\hat{v}}_S&=&-(\gamma _{\bot}/2)\hat{v}_S+\lb{C4_2_3}\\ & & \hspace{1cm}\Omega _0f(\hat{a}_SN+2\sqrt{n}\delta \hat{N}_e)  +\hat{F}_{v_S}, \nonumber\\ 
  \delta\dot{\hat{N}}_e&=&-2\sqrt{n}\left(\Omega_0\hat{v}_S' + \kappa\hat{a}_S'\right)\lb{C4_2_4}\\& & \hspace{2cm}-\gamma _{\parallel}(P+1)\delta \hat{N}_e+\hat{F}_{N_e}.\nonumber 
 \eeqr\end{subequations}
Equation~\rf{C4_2_4} depends on $\hat{a}_{S}'$ and $\hat{v}_{S}'$, so we must find them. According to Eq.~\rf{sym_ass}, the Fourier-components of $\hat{a}_{S}'$ and $\hat{v}_{S}'$ are  zero in the frequency region $|\omega| < \Delta\omega/2$ and otherwise are identical to the Fourier-components of $\hat{a}_{S}$ and $\hat{v}_{S}$ for $|\omega| > \Delta\omega/2$.  Therefore, in order to find $\hat{a}_{S}'$ and $\hat{v}_{S}'$ we solve Eqs.~\rf{C4_2_1} and \rf{C4_2_3}  by Fourier-transform and set the  Fourier components $\hat{a}_{S}'(\omega)$ and $\hat{v}_{S}'(\omega)$ to zero for $|\omega| < \Delta\omega/2$. After that the inverse Fourier transform gives $\hat{a}_{S}'$ and $\hat{v}_{S}'$. The ``cut-off'' frequency $\Delta\omega$ is the same as in the approximation~\rf{int_app} and as shown in Fig.~\ref{Fig5lfc_ab}. 

Unlike $\hat{a}_{S}'$ and $\hat{v}_{S}'$, the exact operators $\hat{a}_{S}$ and $\hat{v}_{S}$  have non-zero Fourier-components also for $|\omega|<\Delta\omega$.  We anticipate  that  $\hat{a}_{S}(\omega)$ and $\hat{v}_{S}(\omega)$ do not have sharp changes for $|\omega|<\Delta\omega$ [unlike $\hat{a}_{A}(\omega)$ and $\hat{v}_{A}(\omega)$], as shown in Fig.~\ref{Fig5lfc_ab}, so we approximately extend $\hat{a}_{S}'$ and $\hat{v}_{S}'$ to the frequency region $|\omega|<\Delta\omega$. We do this by calculating the Fourier-components of $\hat{a}_{S}'$ and $\hat{v}_{S}'$ using equations, identical to Eqs.~\rf{C4_2_1} and~\rf{C4_2_3}, for all frequencies, including $|\omega|<\Delta\omega$.  This means that we approximate $\hat{a}_{S}'\approx \hat{a}_{S}$ and $\hat{v}_{S}'\approx \hat{v}_{S}$. With this approximation we make the replacement~\rf{app_prime} in Eq.~\rf{C4_2_4} and arrive from Eqs.~\rf{C4_2} to Eqs.~\rf{sym_antisym_eq}.

\bibliography{myrefs}

\begin{thebibliography}{71}%
\makeatletter
\providecommand \@ifxundefined [1]{%
 \@ifx{#1\undefined}
}%
\providecommand \@ifnum [1]{%
 \ifnum #1\expandafter \@firstoftwo
 \else \expandafter \@secondoftwo
 \fi
}%
\providecommand \@ifx [1]{%
 \ifx #1\expandafter \@firstoftwo
 \else \expandafter \@secondoftwo
 \fi
}%
\providecommand \natexlab [1]{#1}%
\providecommand \enquote  [1]{``#1''}%
\providecommand \bibnamefont  [1]{#1}%
\providecommand \bibfnamefont [1]{#1}%
\providecommand \citenamefont [1]{#1}%
\providecommand \href@noop [0]{\@secondoftwo}%
\providecommand \href [0]{\begingroup \@sanitize@url \@href}%
\providecommand \@href[1]{\@@startlink{#1}\@@href}%
\providecommand \@@href[1]{\endgroup#1\@@endlink}%
\providecommand \@sanitize@url [0]{\catcode `\\12\catcode `\$12\catcode
  `\&12\catcode `\#12\catcode `\^12\catcode `\_12\catcode `\%12\relax}%
\providecommand \@@startlink[1]{}%
\providecommand \@@endlink[0]{}%
\providecommand \url  [0]{\begingroup\@sanitize@url \@url }%
\providecommand \@url [1]{\endgroup\@href {#1}{\urlprefix }}%
\providecommand \urlprefix  [0]{URL }%
\providecommand \Eprint [0]{\href }%
\providecommand \doibase [0]{https://doi.org/}%
\providecommand \selectlanguage [0]{\@gobble}%
\providecommand \bibinfo  [0]{\@secondoftwo}%
\providecommand \bibfield  [0]{\@secondoftwo}%
\providecommand \translation [1]{[#1]}%
\providecommand \BibitemOpen [0]{}%
\providecommand \bibitemStop [0]{}%
\providecommand \bibitemNoStop [0]{.\EOS\space}%
\providecommand \EOS [0]{\spacefactor3000\relax}%
\providecommand \BibitemShut  [1]{\csname bibitem#1\endcsname}%
\let\auto@bib@innerbib\@empty
\bibitem [{\citenamefont {{Noda}}\ \emph {et~al.}(2017)\citenamefont {{Noda}},
  \citenamefont {{Kitamura}}, \citenamefont {{Okino}}, \citenamefont
  {{Yasuda}},\ and\ \citenamefont {{Tanaka}}}]{7907235}%
  \BibitemOpen
  \bibfield  {author} {\bibinfo {author} {\bibfnamefont {S.}~\bibnamefont
  {{Noda}}}, \bibinfo {author} {\bibfnamefont {K.}~\bibnamefont {{Kitamura}}},
  \bibinfo {author} {\bibfnamefont {T.}~\bibnamefont {{Okino}}}, \bibinfo
  {author} {\bibfnamefont {D.}~\bibnamefont {{Yasuda}}}, and\ \bibinfo {author}
  {\bibfnamefont {Y.}~\bibnamefont {{Tanaka}}},\ }\bibfield  {title} {\bibinfo
  {title} {Photonic-crystal surface-emitting lasers: Review and introduction of
  modulated-photonic crystals},\ }\href
  {https://doi.org/10.1109/JSTQE.2017.2696883} {\bibfield  {journal} {\bibinfo
  {journal} {IEEE Journal of Selected Topics in Quantum Electronics}\ }\textbf
  {\bibinfo {volume} {23}},\ \bibinfo {pages} {1} (\bibinfo {year}
  {2017})}\BibitemShut {NoStop}%
\bibitem [{\citenamefont {Noda}(2006)}]{Noda260}%
  \BibitemOpen
  \bibfield  {author} {\bibinfo {author} {\bibfnamefont {S.}~\bibnamefont
  {Noda}},\ }\bibfield  {title} {\bibinfo {title} {Seeking the ultimate
  nanolaser},\ }\href {https://doi.org/10.1126/science.1131322} {\bibfield
  {journal} {\bibinfo  {journal} {Science}\ }\textbf {\bibinfo {volume}
  {314}},\ \bibinfo {pages} {260} (\bibinfo {year} {2006})}\BibitemShut
  {NoStop}%
\bibitem [{\citenamefont {Prieto}\ \emph {et~al.}(2015)\citenamefont {Prieto},
  \citenamefont {Llorens}, \citenamefont {Mu{\~{n}}oz-Cam\'{u}{\~{n}}ez},
  \citenamefont {Taboada}, \citenamefont {Canet-Ferrer}, \citenamefont
  {Ripalda}, \citenamefont {Robles}, \citenamefont {Mu{\~{n}}oz-Matutano},
  \citenamefont {Mart{\'{i}}nez-Pastor},\ and\ \citenamefont
  {Postigo}}]{Prieto:15}%
  \BibitemOpen
  \bibfield  {author} {\bibinfo {author} {\bibfnamefont {I.}~\bibnamefont
  {Prieto}}, \bibinfo {author} {\bibfnamefont {J.~M.}\ \bibnamefont {Llorens}},
  \bibinfo {author} {\bibfnamefont {L.~E.}\ \bibnamefont
  {Mu{\~{n}}oz-Cam\'{u}{\~{n}}ez}}, \bibinfo {author} {\bibfnamefont {A.~G.}\
  \bibnamefont {Taboada}}, \bibinfo {author} {\bibfnamefont {J.}~\bibnamefont
  {Canet-Ferrer}}, \bibinfo {author} {\bibfnamefont {J.~M.}\ \bibnamefont
  {Ripalda}}, \bibinfo {author} {\bibfnamefont {C.}~\bibnamefont {Robles}},
  \bibinfo {author} {\bibfnamefont {G.}~\bibnamefont {Mu{\~{n}}oz-Matutano}},
  \bibinfo {author} {\bibfnamefont {J.~P.}\ \bibnamefont
  {Mart{\'{i}}nez-Pastor}}, and\ \bibinfo {author} {\bibfnamefont {P.~A.}\
  \bibnamefont {Postigo}},\ }\bibfield  {title} {\bibinfo {title} {Near
  thresholdless laser operation at room temperature},\ }\href
  {https://doi.org/10.1364/OPTICA.2.000066} {\bibfield  {journal} {\bibinfo
  {journal} {Optica}\ }\textbf {\bibinfo {volume} {2}},\ \bibinfo {pages} {66}
  (\bibinfo {year} {2015})}\BibitemShut {NoStop}%
\bibitem [{\citenamefont {Takiguchi}\ \emph {et~al.}(2016)\citenamefont
  {Takiguchi}, \citenamefont {Taniyama}, \citenamefont {Sumikura},
  \citenamefont {Birowosuto}, \citenamefont {Kuramochi}, \citenamefont
  {Shinya}, \citenamefont {Sato}, \citenamefont {Takeda}, \citenamefont
  {Matsuo},\ and\ \citenamefont {Notomi}}]{Takiguchi:16}%
  \BibitemOpen
  \bibfield  {author} {\bibinfo {author} {\bibfnamefont {M.}~\bibnamefont
  {Takiguchi}}, \bibinfo {author} {\bibfnamefont {H.}~\bibnamefont {Taniyama}},
  \bibinfo {author} {\bibfnamefont {H.}~\bibnamefont {Sumikura}}, \bibinfo
  {author} {\bibfnamefont {M.~D.}\ \bibnamefont {Birowosuto}}, \bibinfo
  {author} {\bibfnamefont {E.}~\bibnamefont {Kuramochi}}, \bibinfo {author}
  {\bibfnamefont {A.}~\bibnamefont {Shinya}}, \bibinfo {author} {\bibfnamefont
  {T.}~\bibnamefont {Sato}}, \bibinfo {author} {\bibfnamefont {K.}~\bibnamefont
  {Takeda}}, \bibinfo {author} {\bibfnamefont {S.}~\bibnamefont {Matsuo}}, and\
  \bibinfo {author} {\bibfnamefont {M.}~\bibnamefont {Notomi}},\ }\bibfield
  {title} {\bibinfo {title} {Systematic study of thresholdless oscillation in
  high-$\beta$ buried multiple-quantum-well photonic crystal nanocavity
  lasers},\ }\href {https://doi.org/10.1364/OE.24.003441} {\bibfield  {journal}
  {\bibinfo  {journal} {Opt. Express}\ }\textbf {\bibinfo {volume} {24}},\
  \bibinfo {pages} {3441} (\bibinfo {year} {2016})}\BibitemShut {NoStop}%
\bibitem [{\citenamefont {Ota}\ \emph {et~al.}(2017)\citenamefont {Ota},
  \citenamefont {Kakuda}, \citenamefont {Watanabe}, \citenamefont {Iwamoto},\
  and\ \citenamefont {Arakawa}}]{Ota:17}%
  \BibitemOpen
  \bibfield  {author} {\bibinfo {author} {\bibfnamefont {Y.}~\bibnamefont
  {Ota}}, \bibinfo {author} {\bibfnamefont {M.}~\bibnamefont {Kakuda}},
  \bibinfo {author} {\bibfnamefont {K.}~\bibnamefont {Watanabe}}, \bibinfo
  {author} {\bibfnamefont {S.}~\bibnamefont {Iwamoto}}, and\ \bibinfo {author}
  {\bibfnamefont {Y.}~\bibnamefont {Arakawa}},\ }\bibfield  {title} {\bibinfo
  {title} {Thresholdless quantum dot nanolaser},\ }\href
  {https://doi.org/10.1364/OE.25.019981} {\bibfield  {journal} {\bibinfo
  {journal} {Opt. Express}\ }\textbf {\bibinfo {volume} {25}},\ \bibinfo
  {pages} {19981} (\bibinfo {year} {2017})}\BibitemShut {NoStop}%
\bibitem [{\citenamefont {Nozaki}\ \emph {et~al.}(2007)\citenamefont {Nozaki},
  \citenamefont {Kita},\ and\ \citenamefont {Baba}}]{Nozaki:07}%
  \BibitemOpen
  \bibfield  {author} {\bibinfo {author} {\bibfnamefont {K.}~\bibnamefont
  {Nozaki}}, \bibinfo {author} {\bibfnamefont {S.}~\bibnamefont {Kita}}, and\
  \bibinfo {author} {\bibfnamefont {T.}~\bibnamefont {Baba}},\ }\bibfield
  {title} {\bibinfo {title} {Room temperature continuous wave operation and
  controlled spontaneous emission in ultrasmall photonic crystal nanolaser},\
  }\href {https://doi.org/10.1364/OE.15.007506} {\bibfield  {journal} {\bibinfo
   {journal} {Opt. Express}\ }\textbf {\bibinfo {volume} {15}},\ \bibinfo
  {pages} {7506} (\bibinfo {year} {2007})}\BibitemShut {NoStop}%
\bibitem [{\citenamefont {Yu}\ \emph {et~al.}(2017)\citenamefont {Yu},
  \citenamefont {Xue}, \citenamefont {Semenova}, \citenamefont {Yvind},\ and\
  \citenamefont {Mork}}]{Yu2017}%
  \BibitemOpen
  \bibfield  {author} {\bibinfo {author} {\bibfnamefont {Y.}~\bibnamefont
  {Yu}}, \bibinfo {author} {\bibfnamefont {W.}~\bibnamefont {Xue}}, \bibinfo
  {author} {\bibfnamefont {E.}~\bibnamefont {Semenova}}, \bibinfo {author}
  {\bibfnamefont {K.}~\bibnamefont {Yvind}}, and\ \bibinfo {author}
  {\bibfnamefont {J.}~\bibnamefont {Mork}},\ }\bibfield  {title} {\bibinfo
  {title} {Demonstration of a self-pulsing photonic crystal fano laser},\
  }\href {https://doi.org/10.1038/nphoton.2016.248} {\bibfield  {journal}
  {\bibinfo  {journal} {Nature Photonics}\ }\textbf {\bibinfo {volume} {11}},\
  \bibinfo {pages} {81} (\bibinfo {year} {2017})}\BibitemShut {NoStop}%
\bibitem [{\citenamefont {Li}\ \emph {et~al.}(2019)\citenamefont {Li},
  \citenamefont {Wang}, \citenamefont {Li},\ and\ \citenamefont
  {Tong}}]{Li2019}%
  \BibitemOpen
  \bibfield  {author} {\bibinfo {author} {\bibfnamefont {Y.}~\bibnamefont
  {Li}}, \bibinfo {author} {\bibfnamefont {L.}~\bibnamefont {Wang}}, \bibinfo
  {author} {\bibfnamefont {L.}~\bibnamefont {Li}}, and\ \bibinfo {author}
  {\bibfnamefont {L.}~\bibnamefont {Tong}},\ }\bibfield  {title} {\bibinfo
  {title} {Optical microfiber-based ultrafast fiber lasers},\ }\href
  {https://doi.org/10.1007/s00340-019-7303-z} {\bibfield  {journal} {\bibinfo
  {journal} {Appl. Phys. B}\ }\textbf {\bibinfo {volume} {125}},\ \bibinfo
  {pages} {192} (\bibinfo {year} {2019})}\BibitemShut {NoStop}%
\bibitem [{\citenamefont {Lermer}\ \emph {et~al.}(2013)\citenamefont {Lermer},
  \citenamefont {Gregersen}, \citenamefont {Lorke}, \citenamefont {Schild},
  \citenamefont {Gold}, \citenamefont {M{\o}rk}, \citenamefont {Schneider},
  \citenamefont {Forchel}, \citenamefont {Reitzenstein}, \citenamefont
  {H{\"o}fling},\ and\ \citenamefont {Kamp}}]{doi:10.1063/1.4791563}%
  \BibitemOpen
  \bibfield  {author} {\bibinfo {author} {\bibfnamefont {M.}~\bibnamefont
  {Lermer}}, \bibinfo {author} {\bibfnamefont {N.}~\bibnamefont {Gregersen}},
  \bibinfo {author} {\bibfnamefont {M.}~\bibnamefont {Lorke}}, \bibinfo
  {author} {\bibfnamefont {E.}~\bibnamefont {Schild}}, \bibinfo {author}
  {\bibfnamefont {P.}~\bibnamefont {Gold}}, \bibinfo {author} {\bibfnamefont
  {J.}~\bibnamefont {M{\o}rk}}, \bibinfo {author} {\bibfnamefont
  {C.}~\bibnamefont {Schneider}}, \bibinfo {author} {\bibfnamefont
  {A.}~\bibnamefont {Forchel}}, \bibinfo {author} {\bibfnamefont
  {S.}~\bibnamefont {Reitzenstein}}, \bibinfo {author} {\bibfnamefont
  {S.}~\bibnamefont {H{\"o}fling}}, and\ \bibinfo {author} {\bibfnamefont
  {M.}~\bibnamefont {Kamp}},\ }\bibfield  {title} {\bibinfo {title} {High beta
  lasing in micropillar cavities with adiabatic layer design},\ }\href
  {https://doi.org/10.1063/1.4791563} {\bibfield  {journal} {\bibinfo
  {journal} {Appl. Phys. Lett.}\ }\textbf {\bibinfo {volume} {102}},\ \bibinfo
  {pages} {052114} (\bibinfo {year} {2013})}\BibitemShut {NoStop}%
\bibitem [{\citenamefont {Kreinberg}\ \emph {et~al.}(2017)\citenamefont
  {Kreinberg}, \citenamefont {Chow}, \citenamefont {Wolters}, \citenamefont
  {Schneider}, \citenamefont {Gies}, \citenamefont {Jahnke}, \citenamefont
  {H{\"o}fling}, \citenamefont {Kamp},\ and\ \citenamefont
  {Reitzenstein}}]{Kreinberg2017}%
  \BibitemOpen
  \bibfield  {author} {\bibinfo {author} {\bibfnamefont {S.}~\bibnamefont
  {Kreinberg}}, \bibinfo {author} {\bibfnamefont {W.~W.}\ \bibnamefont {Chow}},
  \bibinfo {author} {\bibfnamefont {J.}~\bibnamefont {Wolters}}, \bibinfo
  {author} {\bibfnamefont {C.}~\bibnamefont {Schneider}}, \bibinfo {author}
  {\bibfnamefont {C.}~\bibnamefont {Gies}}, \bibinfo {author} {\bibfnamefont
  {F.}~\bibnamefont {Jahnke}}, \bibinfo {author} {\bibfnamefont
  {S.}~\bibnamefont {H{\"o}fling}}, \bibinfo {author} {\bibfnamefont
  {M.}~\bibnamefont {Kamp}}, and\ \bibinfo {author} {\bibfnamefont
  {S.}~\bibnamefont {Reitzenstein}},\ }\bibfield  {title} {\bibinfo {title}
  {Emission from quantum-dot high-$\beta$ microcavities: transition from
  spontaneous emission to lasing and the effects of superradiant emitter
  coupling},\ }\href {https://doi.org/10.1038/lsa.2017.30} {\bibfield
  {journal} {\bibinfo  {journal} {Light Sci. Appl.}\ }\textbf {\bibinfo
  {volume} {6}},\ \bibinfo {pages} {e17030} (\bibinfo {year}
  {2017})}\BibitemShut {NoStop}%
\bibitem [{\citenamefont {Suh}\ \emph {et~al.}(2012)\citenamefont {Suh},
  \citenamefont {Kim}, \citenamefont {Zhou}, \citenamefont {Huntington},
  \citenamefont {Co}, \citenamefont {Wasielewski},\ and\ \citenamefont
  {Odom}}]{Suh2012}%
  \BibitemOpen
  \bibfield  {author} {\bibinfo {author} {\bibfnamefont {J.~Y.}\ \bibnamefont
  {Suh}}, \bibinfo {author} {\bibfnamefont {C.~H.}\ \bibnamefont {Kim}},
  \bibinfo {author} {\bibfnamefont {W.}~\bibnamefont {Zhou}}, \bibinfo {author}
  {\bibfnamefont {M.~D.}\ \bibnamefont {Huntington}}, \bibinfo {author}
  {\bibfnamefont {D.~T.}\ \bibnamefont {Co}}, \bibinfo {author} {\bibfnamefont
  {M.~R.}\ \bibnamefont {Wasielewski}}, and\ \bibinfo {author} {\bibfnamefont
  {T.~W.}\ \bibnamefont {Odom}},\ }\bibfield  {title} {\bibinfo {title}
  {Plasmonic bowtie nanolaser arrays},\ }\href
  {https://doi.org/10.1021/nl303086r} {\bibfield  {journal} {\bibinfo
  {journal} {Nano Lett.}\ }\textbf {\bibinfo {volume} {12}},\ \bibinfo {pages}
  {5769} (\bibinfo {year} {2012})}\BibitemShut {NoStop}%
\bibitem [{\citenamefont {Khajavikhan}\ \emph {et~al.}(2012)\citenamefont
  {Khajavikhan}, \citenamefont {Simic}, \citenamefont {Katz}, \citenamefont
  {Lee}, \citenamefont {Slutsky}, \citenamefont {Mizrahi}, \citenamefont
  {Lomakin},\ and\ \citenamefont {Fainman}}]{Khajavikhan2012}%
  \BibitemOpen
  \bibfield  {author} {\bibinfo {author} {\bibfnamefont {M.}~\bibnamefont
  {Khajavikhan}}, \bibinfo {author} {\bibfnamefont {A.}~\bibnamefont {Simic}},
  \bibinfo {author} {\bibfnamefont {M.}~\bibnamefont {Katz}}, \bibinfo {author}
  {\bibfnamefont {J.~H.}\ \bibnamefont {Lee}}, \bibinfo {author} {\bibfnamefont
  {B.}~\bibnamefont {Slutsky}}, \bibinfo {author} {\bibfnamefont
  {A.}~\bibnamefont {Mizrahi}}, \bibinfo {author} {\bibfnamefont
  {V.}~\bibnamefont {Lomakin}}, and\ \bibinfo {author} {\bibfnamefont
  {Y.}~\bibnamefont {Fainman}},\ }\bibfield  {title} {\bibinfo {title}
  {Thresholdless nanoscale coaxial lasers},\ }\href
  {https://doi.org/10.1038/nature10840} {\bibfield  {journal} {\bibinfo
  {journal} {Nature}\ }\textbf {\bibinfo {volume} {482}},\ \bibinfo {pages}
  {204 EP } (\bibinfo {year} {2012})}\BibitemShut {NoStop}%
\bibitem [{\citenamefont {Kurosaka}\ \emph {et~al.}(2010)\citenamefont
  {Kurosaka}, \citenamefont {Iwahashi}, \citenamefont {Liang}, \citenamefont
  {Sakai}, \citenamefont {Miyai}, \citenamefont {Kunishi}, \citenamefont
  {Ohnishi},\ and\ \citenamefont {Noda}}]{Kurosaka2010}%
  \BibitemOpen
  \bibfield  {author} {\bibinfo {author} {\bibfnamefont {Y.}~\bibnamefont
  {Kurosaka}}, \bibinfo {author} {\bibfnamefont {S.}~\bibnamefont {Iwahashi}},
  \bibinfo {author} {\bibfnamefont {Y.}~\bibnamefont {Liang}}, \bibinfo
  {author} {\bibfnamefont {K.}~\bibnamefont {Sakai}}, \bibinfo {author}
  {\bibfnamefont {E.}~\bibnamefont {Miyai}}, \bibinfo {author} {\bibfnamefont
  {W.}~\bibnamefont {Kunishi}}, \bibinfo {author} {\bibfnamefont
  {D.}~\bibnamefont {Ohnishi}}, and\ \bibinfo {author} {\bibfnamefont
  {S.}~\bibnamefont {Noda}},\ }\bibfield  {title} {\bibinfo {title} {On-chip
  beam-steering photonic-crystal lasers},\ }\href
  {https://doi.org/10.1038/nphoton.2010.118} {\bibfield  {journal} {\bibinfo
  {journal} {Nature Photon.}\ }\textbf {\bibinfo {volume} {4}},\ \bibinfo
  {pages} {447} (\bibinfo {year} {2010})}\BibitemShut {NoStop}%
\bibitem [{\citenamefont {{Zhou}}\ \emph {et~al.}(2019)\citenamefont {{Zhou}},
  \citenamefont {{Liu}}, \citenamefont {{Ge}}, \citenamefont {{Zhao}},
  \citenamefont {{Yang}}, \citenamefont {{Reuterskiöld-Hedlund}},\ and\
  \citenamefont {{Hammar}}}]{8658143}%
  \BibitemOpen
  \bibfield  {author} {\bibinfo {author} {\bibfnamefont {W.}~\bibnamefont
  {{Zhou}}}, \bibinfo {author} {\bibfnamefont {S.}~\bibnamefont {{Liu}}},
  \bibinfo {author} {\bibfnamefont {X.}~\bibnamefont {{Ge}}}, \bibinfo {author}
  {\bibfnamefont {D.}~\bibnamefont {{Zhao}}}, \bibinfo {author} {\bibfnamefont
  {H.}~\bibnamefont {{Yang}}}, \bibinfo {author} {\bibfnamefont
  {C.}~\bibnamefont {{Reuterskiöld-Hedlund}}}, and\ \bibinfo {author}
  {\bibfnamefont {M.}~\bibnamefont {{Hammar}}},\ }\bibfield  {title} {\bibinfo
  {title} {On-chip photonic crystal surface-emitting membrane lasers},\ }\href
  {https://doi.org/10.1109/JSTQE.2019.2902904} {\bibfield  {journal} {\bibinfo
  {journal} {IEEE JSTQE}\ }\textbf {\bibinfo {volume} {25}},\ \bibinfo {pages}
  {1} (\bibinfo {year} {2019})}\BibitemShut {NoStop}%
\bibitem [{\citenamefont {Crosnier}\ \emph {et~al.}(2017)\citenamefont
  {Crosnier}, \citenamefont {Sanchez}, \citenamefont {Bouchoule}, \citenamefont
  {Monnier}, \citenamefont {Beaudoin}, \citenamefont {Sagnes}, \citenamefont
  {Raj},\ and\ \citenamefont {Raineri}}]{Crosnier2017}%
  \BibitemOpen
  \bibfield  {author} {\bibinfo {author} {\bibfnamefont {G.}~\bibnamefont
  {Crosnier}}, \bibinfo {author} {\bibfnamefont {D.}~\bibnamefont {Sanchez}},
  \bibinfo {author} {\bibfnamefont {S.}~\bibnamefont {Bouchoule}}, \bibinfo
  {author} {\bibfnamefont {P.}~\bibnamefont {Monnier}}, \bibinfo {author}
  {\bibfnamefont {G.}~\bibnamefont {Beaudoin}}, \bibinfo {author}
  {\bibfnamefont {I.}~\bibnamefont {Sagnes}}, \bibinfo {author} {\bibfnamefont
  {R.}~\bibnamefont {Raj}}, and\ \bibinfo {author} {\bibfnamefont
  {F.}~\bibnamefont {Raineri}},\ }\bibfield  {title} {\bibinfo {title} {Hybrid
  indium phosphide-on-silicon nanolaser diode},\ }\href
  {https://doi.org/10.1038/nphoton.2017.56} {\bibfield  {journal} {\bibinfo
  {journal} {Nature Photon.}\ }\textbf {\bibinfo {volume} {11}},\ \bibinfo
  {pages} {297} (\bibinfo {year} {2017})}\BibitemShut {NoStop}%
\bibitem [{\citenamefont {Purcell}(1946)}]{PhysRev.69.674}%
  \BibitemOpen
  \bibfield  {author} {\bibinfo {author} {\bibfnamefont {E.~M.}\ \bibnamefont
  {Purcell}},\ }\bibfield  {title} {\bibinfo {title} {Spontaneus emission
  probabilities at radio frequencies},\ }\href
  {https://doi.org/10.1103/PhysRev.69.674} {\bibfield  {journal} {\bibinfo
  {journal} {Phys. Rev.}\ }\textbf {\bibinfo {volume} {69}},\ \bibinfo {pages}
  {681} (\bibinfo {year} {1946})}\BibitemShut {NoStop}%
\bibitem [{\citenamefont {Khanin}(2005)}]{Khanin}%
  \BibitemOpen
  \bibfield  {author} {\bibinfo {author} {\bibfnamefont {Y.~I.}\ \bibnamefont
  {Khanin}},\ }\href@noop {} {\emph {\bibinfo {title} {Fundamentals of laser
  dynamics}}}\ (\bibinfo  {publisher} {Cambridge International Science Pub},\
  \bibinfo {year} {2005})\BibitemShut {NoStop}%
\bibitem [{\citenamefont {Belyanin}\ \emph {et~al.}(1998)\citenamefont
  {Belyanin}, \citenamefont {Kocharovsky},\ and\ \citenamefont
  {Kocharovsky}}]{Belyanin_1998}%
  \BibitemOpen
  \bibfield  {author} {\bibinfo {author} {\bibfnamefont {A.~A.}\ \bibnamefont
  {Belyanin}}, \bibinfo {author} {\bibfnamefont {V.~V.}\ \bibnamefont
  {Kocharovsky}}, and\ \bibinfo {author} {\bibfnamefont {V.~V.}\ \bibnamefont
  {Kocharovsky}},\ }\bibfield  {title} {\bibinfo {title} {Superradiant
  generation of femtosecond pulses in quantum-well heterostructures},\ }\href
  {https://doi.org/10.1088/1355-5111/10/2/002} {\bibfield  {journal} {\bibinfo
  {journal} {Quant. Semiclass. Opt.: JEOS Part B}\ }\textbf {\bibinfo {volume}
  {10}},\ \bibinfo {pages} {L13} (\bibinfo {year} {1998})}\BibitemShut
  {NoStop}%
\bibitem [{\citenamefont {Temnov}(2005)}]{PhysRevA.71.053818}%
  \BibitemOpen
  \bibfield  {author} {\bibinfo {author} {\bibfnamefont {V.~V.}\ \bibnamefont
  {Temnov}},\ }\bibfield  {title} {\bibinfo {title} {Superradiance and
  subradiance in the overdamped many-atom micromaser},\ }\href
  {https://doi.org/10.1103/PhysRevA.71.053818} {\bibfield  {journal} {\bibinfo
  {journal} {Phys. Rev. A}\ }\textbf {\bibinfo {volume} {71}},\ \bibinfo
  {pages} {053818} (\bibinfo {year} {2005})}\BibitemShut {NoStop}%
\bibitem [{\citenamefont {{M. A. Norcia }}\ and\ \citenamefont
  {Thompson}(2016)}]{PhysRevX.6.011025}%
  \BibitemOpen
  \bibfield  {author} {\bibinfo {author} {\bibnamefont {{M. A. Norcia }}}and\
  \bibinfo {author} {\bibfnamefont {J.~K.}\ \bibnamefont {Thompson}},\
  }\bibfield  {title} {\bibinfo {title} {Cold-strontium laser in the
  superradiant crossover regime},\ }\href
  {https://doi.org/10.1103/PhysRevX.6.011025} {\bibfield  {journal} {\bibinfo
  {journal} {Phys. Rev. X}\ }\textbf {\bibinfo {volume} {6}},\ \bibinfo {pages}
  {011025} (\bibinfo {year} {2016})}\BibitemShut {NoStop}%
\bibitem [{\citenamefont {Sch\"affer}\ \emph {et~al.}(2017)\citenamefont
  {Sch\"affer}, \citenamefont {Christensen}, \citenamefont {Henriksen},\ and\
  \citenamefont {Thomsen}}]{PhysRevA.96.013847}%
  \BibitemOpen
  \bibfield  {author} {\bibinfo {author} {\bibfnamefont {S.~A.}\ \bibnamefont
  {Sch\"affer}}, \bibinfo {author} {\bibfnamefont {B.~T.~R.}\ \bibnamefont
  {Christensen}}, \bibinfo {author} {\bibfnamefont {M.~R.}\ \bibnamefont
  {Henriksen}}, and\ \bibinfo {author} {\bibfnamefont {J.~W.}\ \bibnamefont
  {Thomsen}},\ }\bibfield  {title} {\bibinfo {title} {Dynamics of
  bad-cavity-enhanced interaction with cold {S}r atoms for laser
  stabilization},\ }\href {https://doi.org/10.1103/PhysRevA.96.013847}
  {\bibfield  {journal} {\bibinfo  {journal} {Phys. Rev. A}\ }\textbf {\bibinfo
  {volume} {96}},\ \bibinfo {pages} {013847} (\bibinfo {year}
  {2017})}\BibitemShut {NoStop}%
\bibitem [{\citenamefont {{D. Meiser }}\ and\ \citenamefont
  {Holland}(2010)}]{PhysRevA.81.033847}%
  \BibitemOpen
  \bibfield  {author} {\bibinfo {author} {\bibnamefont {{D. Meiser }}}and\
  \bibinfo {author} {\bibfnamefont {M.~J.}\ \bibnamefont {Holland}},\
  }\bibfield  {title} {\bibinfo {title} {Steady-state superradiance with
  alkaline-earth-metal atoms},\ }\href
  {https://doi.org/10.1103/PhysRevA.81.033847} {\bibfield  {journal} {\bibinfo
  {journal} {Phys. Rev. A}\ }\textbf {\bibinfo {volume} {81}},\ \bibinfo
  {pages} {033847} (\bibinfo {year} {2010})}\BibitemShut {NoStop}%
\bibitem [{\citenamefont {Debnath}\ \emph {et~al.}(2018)\citenamefont
  {Debnath}, \citenamefont {Zhang},\ and\ \citenamefont
  {M\o{}lmer}}]{PhysRevA.98.063837}%
  \BibitemOpen
  \bibfield  {author} {\bibinfo {author} {\bibfnamefont {K.}~\bibnamefont
  {Debnath}}, \bibinfo {author} {\bibfnamefont {Y.}~\bibnamefont {Zhang}}, and\
  \bibinfo {author} {\bibfnamefont {K.}~\bibnamefont {M\o{}lmer}},\ }\bibfield
  {title} {\bibinfo {title} {Lasing in the superradiant crossover regime},\
  }\href {https://doi.org/10.1103/PhysRevA.98.063837} {\bibfield  {journal}
  {\bibinfo  {journal} {Phys. Rev. A}\ }\textbf {\bibinfo {volume} {98}},\
  \bibinfo {pages} {063837} (\bibinfo {year} {2018})}\BibitemShut {NoStop}%
\bibitem [{\citenamefont {Bohnet}\ \emph {et~al.}(2012)\citenamefont {Bohnet},
  \citenamefont {Chen}, \citenamefont {Weiner}, \citenamefont {Meiser},
  \citenamefont {Holland},\ and\ \citenamefont {Thompson}}]{Bohnet}%
  \BibitemOpen
  \bibfield  {author} {\bibinfo {author} {\bibfnamefont {J.~G.}\ \bibnamefont
  {Bohnet}}, \bibinfo {author} {\bibfnamefont {Z.}~\bibnamefont {Chen}},
  \bibinfo {author} {\bibfnamefont {J.~M.}\ \bibnamefont {Weiner}}, \bibinfo
  {author} {\bibfnamefont {D.}~\bibnamefont {Meiser}}, \bibinfo {author}
  {\bibfnamefont {M.~J.}\ \bibnamefont {Holland}}, and\ \bibinfo {author}
  {\bibfnamefont {J.~K.}\ \bibnamefont {Thompson}},\ }\bibfield  {title}
  {\bibinfo {title} {A steady-state superradiant laser with less than one
  intracavity photon},\ }\href {https://doi.org/10.1038/nature10920} {\bibfield
   {journal} {\bibinfo  {journal} {Nature}\ }\textbf {\bibinfo {volume}
  {484}},\ \bibinfo {pages} {78} (\bibinfo {year} {2012})}\BibitemShut
  {NoStop}%
\bibitem [{\citenamefont {Jahnke}\ \emph {et~al.}(2016)\citenamefont {Jahnke},
  \citenamefont {Gies}, \citenamefont {A{\ss}mann}, \citenamefont {Bayer},
  \citenamefont {Leymann}, \citenamefont {Foerster}, \citenamefont {Wiersig},
  \citenamefont {Schneider}, \citenamefont {Kamp},\ and\ \citenamefont
  {H{\"o}fling}}]{Jahnke}%
  \BibitemOpen
  \bibfield  {author} {\bibinfo {author} {\bibfnamefont {F.}~\bibnamefont
  {Jahnke}}, \bibinfo {author} {\bibfnamefont {C.}~\bibnamefont {Gies}},
  \bibinfo {author} {\bibfnamefont {M.}~\bibnamefont {A{\ss}mann}}, \bibinfo
  {author} {\bibfnamefont {M.}~\bibnamefont {Bayer}}, \bibinfo {author}
  {\bibfnamefont {H.~A.~M.}\ \bibnamefont {Leymann}}, \bibinfo {author}
  {\bibfnamefont {A.}~\bibnamefont {Foerster}}, \bibinfo {author}
  {\bibfnamefont {J.}~\bibnamefont {Wiersig}}, \bibinfo {author} {\bibfnamefont
  {C.}~\bibnamefont {Schneider}}, \bibinfo {author} {\bibfnamefont
  {M.}~\bibnamefont {Kamp}}, and\ \bibinfo {author} {\bibfnamefont
  {S.}~\bibnamefont {H{\"o}fling}},\ }\bibfield  {title} {\bibinfo {title}
  {Giant photon bunching, superradiant pulse emission and excitation trapping
  in quantum-dot nanolasers},\ }\href {https://doi.org/10.1038/ncomms11540}
  {\bibfield  {journal} {\bibinfo  {journal} {Nature Commun.}\ }\textbf
  {\bibinfo {volume} {7}},\ \bibinfo {pages} {11540} (\bibinfo {year}
  {2016})}\BibitemShut {NoStop}%
\bibitem [{\citenamefont {Bhatti}\ \emph {et~al.}(2015)\citenamefont {Bhatti},
  \citenamefont {von Zanthier},\ and\ \citenamefont {Agarwal}}]{corr}%
  \BibitemOpen
  \bibfield  {author} {\bibinfo {author} {\bibfnamefont {D.}~\bibnamefont
  {Bhatti}}, \bibinfo {author} {\bibfnamefont {J.}~\bibnamefont {von
  Zanthier}}, and\ \bibinfo {author} {\bibfnamefont {G.~S.}\ \bibnamefont
  {Agarwal}},\ }\bibfield  {title} {\bibinfo {title} {Superbunchipoll.},\
  }\href {https://doi.org/10.1038/srep17335} {\ \textbf {\bibinfo {volume}
  {5}},\ \bibinfo {pages} {17335} (\bibinfo {year} {2015})}\BibitemShut
  {NoStop}%
\bibitem [{\citenamefont {Zhou}\ \emph {et~al.}(2017)\citenamefont {Zhou},
  \citenamefont {Li}, \citenamefont {Bai}, \citenamefont {Chen}, \citenamefont
  {Liu}, \citenamefont {Xu},\ and\ \citenamefont {Zheng}}]{PhysRevA.95.053809}%
  \BibitemOpen
  \bibfield  {author} {\bibinfo {author} {\bibfnamefont {Y.}~\bibnamefont
  {Zhou}}, \bibinfo {author} {\bibfnamefont {F.-l.}\ \bibnamefont {Li}},
  \bibinfo {author} {\bibfnamefont {B.}~\bibnamefont {Bai}}, \bibinfo {author}
  {\bibfnamefont {H.}~\bibnamefont {Chen}}, \bibinfo {author} {\bibfnamefont
  {J.}~\bibnamefont {Liu}}, \bibinfo {author} {\bibfnamefont {Z.}~\bibnamefont
  {Xu}}, and\ \bibinfo {author} {\bibfnamefont {H.}~\bibnamefont {Zheng}},\
  }\bibfield  {title} {\bibinfo {title} {Superbunching pseudothermal light},\
  }\href {https://doi.org/10.1103/PhysRevA.95.053809} {\bibfield  {journal}
  {\bibinfo  {journal} {Phys. Rev. A}\ }\textbf {\bibinfo {volume} {95}},\
  \bibinfo {pages} {053809} (\bibinfo {year} {2017})}\BibitemShut {NoStop}%
\bibitem [{\citenamefont {Schawlow}(1958)}]{PhysRev.112.1940}%
  \BibitemOpen
  \bibfield  {author} {\bibinfo {author} {\bibfnamefont {C.~H.}\ \bibnamefont
  {Schawlow}, \bibfnamefont {A.~L.~Townes}},\ }\bibfield  {title} {\bibinfo
  {title} {Infrared and optical masers},\ }\href
  {https://doi.org/10.1103/PhysRev.112.1940} {\bibfield  {journal} {\bibinfo
  {journal} {Phys. Rev.}\ }\textbf {\bibinfo {volume} {112}},\ \bibinfo {pages}
  {1940} (\bibinfo {year} {1958})}\BibitemShut {NoStop}%
\bibitem [{\citenamefont {Haken}(1964)}]{PhysRevLett.13.329}%
  \BibitemOpen
  \bibfield  {author} {\bibinfo {author} {\bibfnamefont {H.}~\bibnamefont
  {Haken}},\ }\bibfield  {title} {\bibinfo {title} {Theory of coherence of
  laser light},\ }\href {https://doi.org/10.1103/PhysRevLett.13.329} {\bibfield
   {journal} {\bibinfo  {journal} {Phys. Rev. Lett.}\ }\textbf {\bibinfo
  {volume} {13}},\ \bibinfo {pages} {329} (\bibinfo {year} {1964})}\BibitemShut
  {NoStop}%
\bibitem [{\citenamefont {Scully}(1967)}]{PhysRev.159.208}%
  \BibitemOpen
  \bibfield  {author} {\bibinfo {author} {\bibfnamefont {W.~E.}\ \bibnamefont
  {Scully}, \bibfnamefont {M.~O.~Lamb}},\ }\bibfield  {title} {\bibinfo {title}
  {Quantum theory of an optical maser. {I}. {G}eneral theory},\ }\href
  {https://doi.org/10.1103/PhysRev.159.208} {\bibfield  {journal} {\bibinfo
  {journal} {Phys. Rev.}\ }\textbf {\bibinfo {volume} {159}},\ \bibinfo {pages}
  {208} (\bibinfo {year} {1967})}\BibitemShut {NoStop}%
\bibitem [{\citenamefont {Haken}(1984)}]{Haken_book1984}%
  \BibitemOpen
  \bibfield  {author} {\bibinfo {author} {\bibfnamefont {H.}~\bibnamefont
  {Haken}},\ }\href@noop {} {\emph {\bibinfo {title} {Laser theory}}}\
  (\bibinfo  {publisher} {Springer-Verlag Berlin Heidelberg},\ \bibinfo {year}
  {1984})\BibitemShut {NoStop}%
\bibitem [{\citenamefont {Kelley}\ \emph {et~al.}(1966)\citenamefont {Kelley},
  \citenamefont {Lax},\ and\ \citenamefont {Tannenwald}}]{Lax_book1966}%
  \BibitemOpen
  \bibfield  {author} {\bibinfo {author} {\bibfnamefont {P.~L.}\ \bibnamefont
  {Kelley}}, \bibinfo {author} {\bibfnamefont {B.}~\bibnamefont {Lax}}, and\
  \bibinfo {author} {\bibfnamefont {P.~E.}\ \bibnamefont {Tannenwald}},\
  }\href@noop {} {\emph {\bibinfo {title} {{P}hysics of {Q}uantum
  {E}lectronics}}}\ (\bibinfo  {publisher} {McGraw-Hill, Inc.},\ \bibinfo
  {year} {1966})\BibitemShut {NoStop}%
\bibitem [{\citenamefont {Henry}(1986)}]{Henry1986}%
  \BibitemOpen
  \bibfield  {author} {\bibinfo {author} {\bibfnamefont {C.~H.}\ \bibnamefont
  {Henry}},\ }\bibfield  {title} {\bibinfo {title} {Phase noise in
  semiconductor lasers},\ }\href@noop {} {\bibfield  {journal} {\bibinfo
  {journal} {J. Lightware Techn.}\ }\textbf {\bibinfo {volume} {LT-4}},\
  \bibinfo {pages} {298} (\bibinfo {year} {1986})}\BibitemShut {NoStop}%
\bibitem [{\citenamefont {{Henry}}(1983)}]{1072058}%
  \BibitemOpen
  \bibfield  {author} {\bibinfo {author} {\bibfnamefont {C.}~\bibnamefont
  {{Henry}}},\ }\bibfield  {title} {\bibinfo {title} {Theory of the phase noise
  and power spectrum of a single mode injection laser},\ }\href
  {https://doi.org/10.1109/JQE.1983.1072058} {\bibfield  {journal} {\bibinfo
  {journal} {IEEE J. Quant. Electron.}\ }\textbf {\bibinfo {volume} {19}},\
  \bibinfo {pages} {1391} (\bibinfo {year} {1983})}\BibitemShut {NoStop}%
\bibitem [{\citenamefont {Vahala}(1983)}]{1071986}%
  \BibitemOpen
  \bibfield  {author} {\bibinfo {author} {\bibfnamefont {A.}~\bibnamefont
  {Vahala}, \bibfnamefont {K.~Yariv}},\ }\bibfield  {title} {\bibinfo {title}
  {Semiclassical theory of noise in semiconductor lasers - {P}art {I}},\
  }\href@noop {} {\bibfield  {journal} {\bibinfo  {journal} {IEEE J. Quant.
  Electron.}\ }\textbf {\bibinfo {volume} {19}},\ \bibinfo {pages} {1096}
  (\bibinfo {year} {1983})}\BibitemShut {NoStop}%
\bibitem [{\citenamefont {{Yamamoto}}(1983)}]{1071726}%
  \BibitemOpen
  \bibfield  {author} {\bibinfo {author} {\bibfnamefont {Y.}~\bibnamefont
  {{Yamamoto}}},\ }\bibfield  {title} {\bibinfo {title} {{AM} and {FM} quantum
  noise in semiconductor lasers - {P}art {I}: {T}heoretical analysis},\ }\href
  {https://doi.org/10.1109/JQE.1983.1071726} {\bibfield  {journal} {\bibinfo
  {journal} {IEEE J. Quant. Electron.}\ }\textbf {\bibinfo {volume} {19}},\
  \bibinfo {pages} {34} (\bibinfo {year} {1983})}\BibitemShut {NoStop}%
\bibitem [{\citenamefont {McKinstrie}(2020)}]{McKinstrie:20}%
  \BibitemOpen
  \bibfield  {author} {\bibinfo {author} {\bibfnamefont {C.~J.}\ \bibnamefont
  {McKinstrie}},\ }\bibfield  {title} {\bibinfo {title} {Stochastic and
  probabilistic equations for three- and four-level lasers: tutorial},\ }\href
  {https://doi.org/10.1364/JOSAB.379976} {\bibfield  {journal} {\bibinfo
  {journal} {J. Opt. Soc. Am. B}\ }\textbf {\bibinfo {volume} {37}},\ \bibinfo
  {pages} {1333} (\bibinfo {year} {2020})}\BibitemShut {NoStop}%
\bibitem [{\citenamefont {{ J. M{\o}rk }}\ and\ \citenamefont
  {Lippi}(2018)}]{doi:10.1063/1.5022958}%
  \BibitemOpen
  \bibfield  {author} {\bibinfo {author} {\bibnamefont {{ J. M{\o}rk }}}and\
  \bibinfo {author} {\bibfnamefont {G.~L.}\ \bibnamefont {Lippi}},\ }\bibfield
  {title} {\bibinfo {title} {Rate equation description of quantum noise in
  nanolasers with few emitters},\ }\href@noop {} {\bibfield  {journal}
  {\bibinfo  {journal} {Appl. Phys. Lett.}\ }\textbf {\bibinfo {volume}
  {112}},\ \bibinfo {pages} {141103} (\bibinfo {year} {2018})}\BibitemShut
  {NoStop}%
\bibitem [{\citenamefont {Auff{\`{e}}ves}\ \emph {et~al.}(2011)\citenamefont
  {Auff{\`{e}}ves}, \citenamefont {Gerace}, \citenamefont {Portolan},
  \citenamefont {Drezet},\ and\ \citenamefont {Santos}}]{Auff_ves_2011}%
  \BibitemOpen
  \bibfield  {author} {\bibinfo {author} {\bibfnamefont {A.}~\bibnamefont
  {Auff{\`{e}}ves}}, \bibinfo {author} {\bibfnamefont {D.}~\bibnamefont
  {Gerace}}, \bibinfo {author} {\bibfnamefont {S.}~\bibnamefont {Portolan}},
  \bibinfo {author} {\bibfnamefont {A.}~\bibnamefont {Drezet}}, and\ \bibinfo
  {author} {\bibfnamefont {M.~F.}\ \bibnamefont {Santos}},\ }\bibfield  {title}
  {\bibinfo {title} {Few emitters in a cavity: from cooperative emission to
  individualization},\ }\href {https://doi.org/10.1088/1367-2630/13/9/093020}
  {\bibfield  {journal} {\bibinfo  {journal} {New J. Phys.}\ }\textbf {\bibinfo
  {volume} {13}},\ \bibinfo {pages} {093020} (\bibinfo {year}
  {2011})}\BibitemShut {NoStop}%
\bibitem [{\citenamefont {Mascarenhas}\ \emph {et~al.}(2013)\citenamefont
  {Mascarenhas}, \citenamefont {Gerace}, \citenamefont {Santos},\ and\
  \citenamefont {Auff{\`e}ves}}]{PhysRevA.88.063825}%
  \BibitemOpen
  \bibfield  {author} {\bibinfo {author} {\bibfnamefont {E.}~\bibnamefont
  {Mascarenhas}}, \bibinfo {author} {\bibfnamefont {D.}~\bibnamefont {Gerace}},
  \bibinfo {author} {\bibfnamefont {M.~F.}\ \bibnamefont {Santos}}, and\
  \bibinfo {author} {\bibfnamefont {A.}~\bibnamefont {Auff{\`e}ves}},\
  }\bibfield  {title} {\bibinfo {title} {Cooperativity of a few quantum
  emitters in a single-mode cavity},\ }\href
  {https://doi.org/10.1103/PhysRevA.88.063825} {\bibfield  {journal} {\bibinfo
  {journal} {Phys. Rev. A}\ }\textbf {\bibinfo {volume} {88}},\ \bibinfo
  {pages} {063825} (\bibinfo {year} {2013})}\BibitemShut {NoStop}%
\bibitem [{\citenamefont {{Moelbjerg}}\ \emph {et~al.}(2013)\citenamefont
  {{Moelbjerg}}, \citenamefont {{Kaer}}, \citenamefont {{Lorke}}, \citenamefont
  {{Tromborg}},\ and\ \citenamefont {{Mørk}}}]{6603264}%
  \BibitemOpen
  \bibfield  {author} {\bibinfo {author} {\bibfnamefont {A.}~\bibnamefont
  {{Moelbjerg}}}, \bibinfo {author} {\bibfnamefont {P.}~\bibnamefont {{Kaer}}},
  \bibinfo {author} {\bibfnamefont {M.}~\bibnamefont {{Lorke}}}, \bibinfo
  {author} {\bibfnamefont {B.}~\bibnamefont {{Tromborg}}}, and\ \bibinfo
  {author} {\bibfnamefont {J.}~\bibnamefont {{Mørk}}},\ }\bibfield  {title}
  {\bibinfo {title} {Dynamical properties of nanolasers based on few discrete
  emitters},\ }\href {https://doi.org/10.1109/JQE.2013.2282464} {\bibfield
  {journal} {\bibinfo  {journal} {IEEE J. Quant. Electron.}\ }\textbf {\bibinfo
  {volume} {49}},\ \bibinfo {pages} {945} (\bibinfo {year} {2013})}\BibitemShut
  {NoStop}%
\bibitem [{\citenamefont {Gies}\ \emph {et~al.}(2007)\citenamefont {Gies},
  \citenamefont {Wiersig}, \citenamefont {Lorke},\ and\ \citenamefont
  {Jahnke}}]{PhysRevA.75.013803}%
  \BibitemOpen
  \bibfield  {author} {\bibinfo {author} {\bibfnamefont {C.}~\bibnamefont
  {Gies}}, \bibinfo {author} {\bibfnamefont {J.}~\bibnamefont {Wiersig}},
  \bibinfo {author} {\bibfnamefont {M.}~\bibnamefont {Lorke}}, and\ \bibinfo
  {author} {\bibfnamefont {F.}~\bibnamefont {Jahnke}},\ }\bibfield  {title}
  {\bibinfo {title} {Semiconductor model for quantum-dot-based microcavity
  lasers},\ }\href {https://doi.org/10.1103/PhysRevA.75.013803} {\bibfield
  {journal} {\bibinfo  {journal} {Phys. Rev. A}\ }\textbf {\bibinfo {volume}
  {75}},\ \bibinfo {pages} {013803} (\bibinfo {year} {2007})}\BibitemShut
  {NoStop}%
\bibitem [{\citenamefont {Meiser}\ \emph {et~al.}(2009)\citenamefont {Meiser},
  \citenamefont {Ye}, \citenamefont {Carlson},\ and\ \citenamefont
  {Holland}}]{PhysRevLett.102.163601}%
  \BibitemOpen
  \bibfield  {author} {\bibinfo {author} {\bibfnamefont {D.}~\bibnamefont
  {Meiser}}, \bibinfo {author} {\bibfnamefont {J.}~\bibnamefont {Ye}}, \bibinfo
  {author} {\bibfnamefont {D.~R.}\ \bibnamefont {Carlson}}, and\ \bibinfo
  {author} {\bibfnamefont {M.~J.}\ \bibnamefont {Holland}},\ }\bibfield
  {title} {\bibinfo {title} {Prospects for a millihertz-linewidth laser},\
  }\href {https://doi.org/10.1103/PhysRevLett.102.163601} {\bibfield  {journal}
  {\bibinfo  {journal} {Phys. Rev. Lett.}\ }\textbf {\bibinfo {volume} {102}},\
  \bibinfo {pages} {163601} (\bibinfo {year} {2009})}\BibitemShut {NoStop}%
\bibitem [{\citenamefont {Kirton}\ and\ \citenamefont
  {Keeling}(2018)}]{Kirton_2018}%
  \BibitemOpen
  \bibfield  {author} {\bibinfo {author} {\bibfnamefont {P.}~\bibnamefont
  {Kirton}}and\ \bibinfo {author} {\bibfnamefont {J.}~\bibnamefont {Keeling}},\
  }\bibfield  {title} {\bibinfo {title} {Superradiant and lasing states in
  driven-dissipative {D}icke models},\ }\href
  {https://doi.org/10.1088/1367-2630/aaa11d} {\bibfield  {journal} {\bibinfo
  {journal} {New J. Phys.}\ }\textbf {\bibinfo {volume} {20}},\ \bibinfo
  {pages} {015009} (\bibinfo {year} {2018})}\BibitemShut {NoStop}%
\bibitem [{\citenamefont {J\"ager}\ \emph {et~al.}(2020)\citenamefont
  {J\"ager}, \citenamefont {Holland},\ and\ \citenamefont
  {Morigi}}]{PhysRevA.101.023616}%
  \BibitemOpen
  \bibfield  {author} {\bibinfo {author} {\bibfnamefont {S.~B.}\ \bibnamefont
  {J\"ager}}, \bibinfo {author} {\bibfnamefont {M.~J.}\ \bibnamefont
  {Holland}}, and\ \bibinfo {author} {\bibfnamefont {G.}~\bibnamefont
  {Morigi}},\ }\bibfield  {title} {\bibinfo {title} {Superradiant
  optomechanical phases of cold atomic gases in optical resonators},\ }\href
  {https://doi.org/10.1103/PhysRevA.101.023616} {\bibfield  {journal} {\bibinfo
   {journal} {Phys. Rev. A}\ }\textbf {\bibinfo {volume} {101}},\ \bibinfo
  {pages} {023616} (\bibinfo {year} {2020})}\BibitemShut {NoStop}%
\bibitem [{\citenamefont {Zhang}\ \emph {et~al.}(2018)\citenamefont {Zhang},
  \citenamefont {Zhang},\ and\ \citenamefont {M{\o}lmer}}]{Zhang_2018}%
  \BibitemOpen
  \bibfield  {author} {\bibinfo {author} {\bibfnamefont {Y.}~\bibnamefont
  {Zhang}}, \bibinfo {author} {\bibfnamefont {Y.-X.}\ \bibnamefont {Zhang}},
  and\ \bibinfo {author} {\bibfnamefont {K.}~\bibnamefont {M{\o}lmer}},\
  }\bibfield  {title} {\bibinfo {title} {Monte-{C}arlo simulations of
  superradiant lasing},\ }\href {https://doi.org/10.1088/1367-2630/aaec36}
  {\bibfield  {journal} {\bibinfo  {journal} {New J. Phys.}\ }\textbf {\bibinfo
  {volume} {20}},\ \bibinfo {pages} {112001} (\bibinfo {year}
  {2018})}\BibitemShut {NoStop}%
\bibitem [{\citenamefont {{P. P. Vasil'ev }}\ and\ \citenamefont
  {Penty}(2020)}]{Vasil_ev_2020}%
  \BibitemOpen
  \bibfield  {author} {\bibinfo {author} {\bibnamefont {{P. P. Vasil'ev }}}and\
  \bibinfo {author} {\bibfnamefont {R.~V.}\ \bibnamefont {Penty}},\ }\bibfield
  {title} {\bibinfo {title} {Wigner function and photon number distribution of
  a superradiant state in semiconductor heterostructures},\ }\href
  {https://doi.org/10.1088/1367-2630/aba3e8} {\bibfield  {journal} {\bibinfo
  {journal} {New J. Phys.}\ }\textbf {\bibinfo {volume} {22}},\ \bibinfo
  {pages} {083046} (\bibinfo {year} {2020})}\BibitemShut {NoStop}%
\bibitem [{\citenamefont {Ostermann}\ \emph {et~al.}(2019)\citenamefont
  {Ostermann}, \citenamefont {Meignant}, \citenamefont {Genes},\ and\
  \citenamefont {Ritsch}}]{Ostermann_2019}%
  \BibitemOpen
  \bibfield  {author} {\bibinfo {author} {\bibfnamefont {L.}~\bibnamefont
  {Ostermann}}, \bibinfo {author} {\bibfnamefont {C.}~\bibnamefont {Meignant}},
  \bibinfo {author} {\bibfnamefont {C.}~\bibnamefont {Genes}}, and\ \bibinfo
  {author} {\bibfnamefont {H.}~\bibnamefont {Ritsch}},\ }\bibfield  {title}
  {\bibinfo {title} {Super- and subradiance of clock atoms in multimode optical
  waveguides},\ }\href {https://doi.org/10.1088/1367-2630/ab05fb} {\bibfield
  {journal} {\bibinfo  {journal} {New J. Phys.}\ }\textbf {\bibinfo {volume}
  {21}},\ \bibinfo {pages} {025004} (\bibinfo {year} {2019})}\BibitemShut
  {NoStop}%
\bibitem [{\citenamefont {Suarez}\ \emph {et~al.}(2019)\citenamefont {Suarez},
  \citenamefont {Auw{\" a}rter}, \citenamefont {Arruda}, \citenamefont
  {Bachelard}, \citenamefont {Courteille}, \citenamefont {Zimmermann},\ and\
  \citenamefont {Slama}}]{Suarez_2019}%
  \BibitemOpen
  \bibfield  {author} {\bibinfo {author} {\bibfnamefont {E.}~\bibnamefont
  {Suarez}}, \bibinfo {author} {\bibfnamefont {D.}~\bibnamefont {Auw{\"
  a}rter}}, \bibinfo {author} {\bibfnamefont {T.~J.}\ \bibnamefont {Arruda}},
  \bibinfo {author} {\bibfnamefont {R.}~\bibnamefont {Bachelard}}, \bibinfo
  {author} {\bibfnamefont {P.~W.}\ \bibnamefont {Courteille}}, \bibinfo
  {author} {\bibfnamefont {C.}~\bibnamefont {Zimmermann}}, and\ \bibinfo
  {author} {\bibfnamefont {S.}~\bibnamefont {Slama}},\ }\bibfield  {title}
  {\bibinfo {title} {Photon-antibunching in the fluorescence of statistical
  ensembles of emitters at an optical nanofiber-tip},\ }\href
  {https://doi.org/10.1088/1367-2630/ab0a99} {\bibfield  {journal} {\bibinfo
  {journal} {New J. Phys.}\ }\textbf {\bibinfo {volume} {21}},\ \bibinfo
  {pages} {035009} (\bibinfo {year} {2019})}\BibitemShut {NoStop}%
\bibitem [{\citenamefont {Scully}(1997)}]{Scully}%
  \BibitemOpen
  \bibfield  {author} {\bibinfo {author} {\bibfnamefont {M.~S.}\ \bibnamefont
  {Scully}, \bibfnamefont {M.~O.~Zubairy}},\ }\href@noop {} {\emph {\bibinfo
  {title} {Quantum {O}ptics}}}\ (\bibinfo  {publisher} {Cambridge University
  Press},\ \bibinfo {year} {1997})\BibitemShut {NoStop}%
\bibitem [{\citenamefont {Duan}\ \emph {et~al.}(2000)\citenamefont {Duan},
  \citenamefont {Giedke}, \citenamefont {Cirac},\ and\ \citenamefont
  {Zoller}}]{PhysRevLett.84.2722}%
  \BibitemOpen
  \bibfield  {author} {\bibinfo {author} {\bibfnamefont {L.-M.}\ \bibnamefont
  {Duan}}, \bibinfo {author} {\bibfnamefont {G.}~\bibnamefont {Giedke}},
  \bibinfo {author} {\bibfnamefont {J.~I.}\ \bibnamefont {Cirac}}, and\
  \bibinfo {author} {\bibfnamefont {P.}~\bibnamefont {Zoller}},\ }\bibfield
  {title} {\bibinfo {title} {Inseparability criterion for continuous variable
  systems},\ }\href {https://doi.org/10.1103/PhysRevLett.84.2722} {\bibfield
  {journal} {\bibinfo  {journal} {Phys. Rev. Lett.}\ }\textbf {\bibinfo
  {volume} {84}},\ \bibinfo {pages} {2722} (\bibinfo {year}
  {2000})}\BibitemShut {NoStop}%
\bibitem [{\citenamefont {Feyisa}(2020)}]{Feyisa2020}%
  \BibitemOpen
  \bibfield  {author} {\bibinfo {author} {\bibfnamefont {C.~G.}\ \bibnamefont
  {Feyisa}},\ }\bibfield  {title} {\bibinfo {title} {Enhanced {CV} entanglement
  quantification in a {CEL} with parametric amplifier and coupled to squeezed
  vacuum},\ }\href {https://doi.org/10.1007/s13538-020-00759-6} {\bibfield
  {journal} {\bibinfo  {journal} {Braz. J. Phys.}\ }\textbf {\bibinfo {volume}
  {50}},\ \bibinfo {pages} {379} (\bibinfo {year} {2020})}\BibitemShut
  {NoStop}%
\bibitem [{\citenamefont {Pegg}\ and\ \citenamefont
  {Barnett}(1989)}]{PhysRevA.39.1665}%
  \BibitemOpen
  \bibfield  {author} {\bibinfo {author} {\bibfnamefont {D.~T.}\ \bibnamefont
  {Pegg}}and\ \bibinfo {author} {\bibfnamefont {S.~M.}\ \bibnamefont
  {Barnett}},\ }\bibfield  {title} {\bibinfo {title} {Phase properties of the
  quantized single-mode electromagnetic field},\ }\href
  {https://doi.org/10.1103/PhysRevA.39.1665} {\bibfield  {journal} {\bibinfo
  {journal} {Phys. Rev. A}\ }\textbf {\bibinfo {volume} {39}},\ \bibinfo
  {pages} {1665} (\bibinfo {year} {1989})}\BibitemShut {NoStop}%
\bibitem [{\citenamefont {Davidovich}(1996)}]{RevModPhys.68.127}%
  \BibitemOpen
  \bibfield  {author} {\bibinfo {author} {\bibfnamefont {L.}~\bibnamefont
  {Davidovich}},\ }\bibfield  {title} {\bibinfo {title} {Sub-{P}oissonian
  processes in quantum optics},\ }\href
  {https://doi.org/10.1103/RevModPhys.68.127} {\bibfield  {journal} {\bibinfo
  {journal} {Rev. Mod. Phys.}\ }\textbf {\bibinfo {volume} {68}},\ \bibinfo
  {pages} {127} (\bibinfo {year} {1996})}\BibitemShut {NoStop}%
\bibitem [{\citenamefont {Kolobov}\ \emph {et~al.}(1993)\citenamefont
  {Kolobov}, \citenamefont {Davidovich}, \citenamefont {Giacobino},\ and\
  \citenamefont {Fabre}}]{PhysRevA.47.1431}%
  \BibitemOpen
  \bibfield  {author} {\bibinfo {author} {\bibfnamefont {M.~I.}\ \bibnamefont
  {Kolobov}}, \bibinfo {author} {\bibfnamefont {L.}~\bibnamefont {Davidovich}},
  \bibinfo {author} {\bibfnamefont {E.}~\bibnamefont {Giacobino}}, and\
  \bibinfo {author} {\bibfnamefont {C.}~\bibnamefont {Fabre}},\ }\bibfield
  {title} {\bibinfo {title} {Role of pumping statistics and dynamics of atomic
  polarization in quantum fluctuations of laser sources},\ }\href
  {https://doi.org/10.1103/PhysRevA.47.1431} {\bibfield  {journal} {\bibinfo
  {journal} {Phys. Rev. A}\ }\textbf {\bibinfo {volume} {47}},\ \bibinfo
  {pages} {1431} (\bibinfo {year} {1993})}\BibitemShut {NoStop}%
\bibitem [{\citenamefont {Butylkin}\ \emph {et~al.}(1989)\citenamefont
  {Butylkin}, \citenamefont {Khronopulo}, \citenamefont {Kaplan},\ and\
  \citenamefont {Yakubovich}}]{Butylkin}%
  \BibitemOpen
  \bibfield  {author} {\bibinfo {author} {\bibfnamefont {V.}~\bibnamefont
  {Butylkin}}, \bibinfo {author} {\bibfnamefont {Y.}~\bibnamefont
  {Khronopulo}}, \bibinfo {author} {\bibfnamefont {A.~E.}\ \bibnamefont
  {Kaplan}}, and\ \bibinfo {author} {\bibfnamefont {E.~I.}\ \bibnamefont
  {Yakubovich}},\ }\href@noop {} {\emph {\bibinfo {title} {Resonant Nonlinear
  Interactions of Light with Matter}}}\ (\bibinfo  {publisher} {Springer},\
  \bibinfo {address} {Berlin, Heidelberg},\ \bibinfo {year} {1989})\BibitemShut
  {NoStop}%
\bibitem [{\citenamefont {Yariv}(1967)}]{Yariv}%
  \BibitemOpen
  \bibfield  {author} {\bibinfo {author} {\bibfnamefont {A.}~\bibnamefont
  {Yariv}},\ }\href@noop {} {\emph {\bibinfo {title} {Quantum {E}lectronics}}}\
  (\bibinfo  {publisher} {John Wiley and Sons, Inc.},\ \bibinfo {year}
  {1967})\BibitemShut {NoStop}%
\bibitem [{\citenamefont {Protsenko}\ \emph {et~al.}(1999)\citenamefont
  {Protsenko}, \citenamefont {Domokos}, \citenamefont {Lef\`evre-Seguin},
  \citenamefont {Hare}, \citenamefont {Raimond},\ and\ \citenamefont
  {Davidovich}}]{PhysRevA.59.1667}%
  \BibitemOpen
  \bibfield  {author} {\bibinfo {author} {\bibfnamefont {I.}~\bibnamefont
  {Protsenko}}, \bibinfo {author} {\bibfnamefont {P.}~\bibnamefont {Domokos}},
  \bibinfo {author} {\bibfnamefont {V.}~\bibnamefont {Lef\`evre-Seguin}},
  \bibinfo {author} {\bibfnamefont {J.}~\bibnamefont {Hare}}, \bibinfo {author}
  {\bibfnamefont {J.~M.}\ \bibnamefont {Raimond}}, and\ \bibinfo {author}
  {\bibfnamefont {L.}~\bibnamefont {Davidovich}},\ }\bibfield  {title}
  {\bibinfo {title} {Quantum theory of a thresholdless laser},\ }\href
  {https://doi.org/10.1103/PhysRevA.59.1667} {\bibfield  {journal} {\bibinfo
  {journal} {Phys. Rev. A}\ }\textbf {\bibinfo {volume} {59}},\ \bibinfo
  {pages} {1667} (\bibinfo {year} {1999})}\BibitemShut {NoStop}%
\bibitem [{\citenamefont {Andr\'{e}}\ \emph {et~al.}(2019)\citenamefont
  {Andr\'{e}}, \citenamefont {Protsenko}, \citenamefont {Uskov}, \citenamefont
  {M{\o}rk},\ and\ \citenamefont {Wubs}}]{Andre:19}%
  \BibitemOpen
  \bibfield  {author} {\bibinfo {author} {\bibfnamefont {E.~C.}\ \bibnamefont
  {Andr\'{e}}}, \bibinfo {author} {\bibfnamefont {I.~E.}\ \bibnamefont
  {Protsenko}}, \bibinfo {author} {\bibfnamefont {A.~V.}\ \bibnamefont
  {Uskov}}, \bibinfo {author} {\bibfnamefont {J.}~\bibnamefont {M{\o}rk}}, and\
  \bibinfo {author} {\bibfnamefont {M.}~\bibnamefont {Wubs}},\ }\bibfield
  {title} {\bibinfo {title} {On collective {R}abi splitting in nanolasers and
  nano-{LED}s},\ }\href {https://doi.org/10.1364/OL.44.001415} {\bibfield
  {journal} {\bibinfo  {journal} {Opt. Lett.}\ }\textbf {\bibinfo {volume}
  {44}},\ \bibinfo {pages} {1415} (\bibinfo {year} {2019})}\BibitemShut
  {NoStop}%
\bibitem [{\citenamefont {Siegman}(1986)}]{Siegman}%
  \BibitemOpen
  \bibfield  {author} {\bibinfo {author} {\bibfnamefont {A.~E.}\ \bibnamefont
  {Siegman}},\ }\href@noop {} {\emph {\bibinfo {title} {Lasers}}}\ (\bibinfo
  {publisher} {University Science Books; Sausalito, CA},\ \bibinfo {year}
  {1986})\BibitemShut {NoStop}%
\bibitem [{\citenamefont {{ P. R. Rice }}\ and\ \citenamefont
  {Carmichael}(1994)}]{PhysRevA.50.4318}%
  \BibitemOpen
  \bibfield  {author} {\bibinfo {author} {\bibnamefont {{ P. R. Rice }}}and\
  \bibinfo {author} {\bibfnamefont {H.~J.}\ \bibnamefont {Carmichael}},\
  }\bibfield  {title} {\bibinfo {title} {Photon statistics of a cavity-{QED}
  laser: A comment on the laser--phase-transition analogy},\ }\href
  {https://doi.org/10.1103/PhysRevA.50.4318} {\bibfield  {journal} {\bibinfo
  {journal} {Phys. Rev. A}\ }\textbf {\bibinfo {volume} {50}},\ \bibinfo
  {pages} {4318} (\bibinfo {year} {1994})}\BibitemShut {NoStop}%
\bibitem [{\citenamefont {Sargent}\ \emph {et~al.}(1974)\citenamefont
  {Sargent}, \citenamefont {Scully},\ and\ \citenamefont
  {Lamb}}]{trove.nla.gov.au/work/21304573}%
  \BibitemOpen
  \bibfield  {author} {\bibinfo {author} {\bibfnamefont {M.}~\bibnamefont
  {Sargent}}, \bibinfo {author} {\bibfnamefont {M.~O.}\ \bibnamefont {Scully}},
  and\ \bibinfo {author} {\bibfnamefont {W.~E.}\ \bibnamefont {Lamb}},\
  }\href@noop {} {{\selectlanguage {English}\emph {\bibinfo {title} {Laser
  Physics}}}}\ (\bibinfo  {publisher} {London : Addison-Wesley},\ \bibinfo
  {year} {1974})\BibitemShut {NoStop}%
\bibitem [{\citenamefont {Kuppens}\ \emph {et~al.}(1994)\citenamefont
  {Kuppens}, \citenamefont {van Exter},\ and\ \citenamefont
  {Woerdman}}]{PhysRevLett.72.3815}%
  \BibitemOpen
  \bibfield  {author} {\bibinfo {author} {\bibfnamefont {S.~J.~M.}\
  \bibnamefont {Kuppens}}, \bibinfo {author} {\bibfnamefont {M.~P.}\
  \bibnamefont {van Exter}}, and\ \bibinfo {author} {\bibfnamefont {J.~P.}\
  \bibnamefont {Woerdman}},\ }\bibfield  {title} {\bibinfo {title}
  {Quantum-limited linewidth of a bad-cavity laser},\ }\href
  {https://doi.org/10.1103/PhysRevLett.72.3815} {\bibfield  {journal} {\bibinfo
   {journal} {Phys. Rev. Lett.}\ }\textbf {\bibinfo {volume} {72}},\ \bibinfo
  {pages} {3815} (\bibinfo {year} {1994})}\BibitemShut {NoStop}%
\bibitem [{\citenamefont {{M. Pollnau }}\ and\ \citenamefont
  {Eichhorn}(2020)}]{POLLNAU2020100255}%
  \BibitemOpen
  \bibfield  {author} {\bibinfo {author} {\bibnamefont {{M. Pollnau }}}and\
  \bibinfo {author} {\bibfnamefont {M.}~\bibnamefont {Eichhorn}},\ }\bibfield
  {title} {\bibinfo {title} {Spectral coherence, part {I}: Passive-resonator
  linewidth, fundamental laser linewidth, and {S}chawlow-{T}ownes
  approximation},\ }\href
  {https://doi.org/https://doi.org/10.1016/j.pquantelec.2020.100255} {\bibfield
   {journal} {\bibinfo  {journal} {Prog. Quant. Electron.}\ }\textbf {\bibinfo
  {volume} {72}},\ \bibinfo {pages} {100255} (\bibinfo {year}
  {2020})}\BibitemShut {NoStop}%
\bibitem [{\citenamefont {van Exter}\ \emph {et~al.}(1995)\citenamefont {van
  Exter}, \citenamefont {Kuppens},\ and\ \citenamefont {Woerdman}}]{van_Exter}%
  \BibitemOpen
  \bibfield  {author} {\bibinfo {author} {\bibfnamefont {M.~P.}\ \bibnamefont
  {van Exter}}, \bibinfo {author} {\bibfnamefont {S.~J.~M.}\ \bibnamefont
  {Kuppens}}, and\ \bibinfo {author} {\bibfnamefont {J.~P.}\ \bibnamefont
  {Woerdman}},\ }\bibfield  {title} {\bibinfo {title} {Theory for the linewidth
  of a bad-cavity laser},\ }\href@noop {} {\bibfield  {journal} {\bibinfo
  {journal} {Phys. Rev. A}\ }\textbf {\bibinfo {volume} {51}},\ \bibinfo
  {pages} {809} (\bibinfo {year} {1995})}\BibitemShut {NoStop}%
\bibitem [{\citenamefont {{U. Bockelmann }}\ and\ \citenamefont
  {Egeler}(1992)}]{PhysRevB.46.15574}%
  \BibitemOpen
  \bibfield  {author} {\bibinfo {author} {\bibnamefont {{U. Bockelmann }}}and\
  \bibinfo {author} {\bibfnamefont {T.}~\bibnamefont {Egeler}},\ }\bibfield
  {title} {\bibinfo {title} {Electron relaxation in quantum dots by means of
  {A}uger processes},\ }\href {https://doi.org/10.1103/PhysRevB.46.15574}
  {\bibfield  {journal} {\bibinfo  {journal} {Phys. Rev. B}\ }\textbf {\bibinfo
  {volume} {46}},\ \bibinfo {pages} {15574} (\bibinfo {year}
  {1992})}\BibitemShut {NoStop}%
\bibitem [{\citenamefont {Yamamoto}\ \emph {et~al.}(1983)\citenamefont
  {Yamamoto}, \citenamefont {Saito},\ and\ \citenamefont {Mukai}}]{1071729}%
  \BibitemOpen
  \bibfield  {author} {\bibinfo {author} {\bibfnamefont {Y.}~\bibnamefont
  {Yamamoto}}, \bibinfo {author} {\bibfnamefont {S.}~\bibnamefont {Saito}},
  and\ \bibinfo {author} {\bibfnamefont {T.}~\bibnamefont {Mukai}},\ }\bibfield
   {title} {\bibinfo {title} {{AM} and {FM} quantum noise in semiconductor
  lasers - {P}art {II}: {C}omparison of theoretical and experimental results
  for {A}l{G}a{A}s lasers},\ }\href {https://doi.org/10.1109/JQE.1983.1071729}
  {\bibfield  {journal} {\bibinfo  {journal} {IEEE Journal of Quantum
  Electronics}\ }\textbf {\bibinfo {volume} {19}},\ \bibinfo {pages} {47}
  (\bibinfo {year} {1983})}\BibitemShut {NoStop}%
\bibitem [{\citenamefont {Attenborough}(2003)}]{Math_el_en}%
  \BibitemOpen
  \bibfield  {author} {\bibinfo {author} {\bibfnamefont {M.}~\bibnamefont
  {Attenborough}},\ }\href@noop {} {\emph {\bibinfo {title} {Mathematics for
  Electrical Engineering...}}}\ (\bibinfo  {publisher} {Newnes},\ \bibinfo
  {address} {Amsterdam},\ \bibinfo {year} {2003})\BibitemShut {NoStop}%
\bibitem [{\citenamefont {{Zali}}\ \emph {et~al.}(2018)\citenamefont {{Zali}},
  \citenamefont {{Moravvej-Farshi}}, \citenamefont {{Yu}},\ and\ \citenamefont
  {{M{\o}rk}}}]{8506446}%
  \BibitemOpen
  \bibfield  {author} {\bibinfo {author} {\bibfnamefont {A.~R.}\ \bibnamefont
  {{Zali}}}, \bibinfo {author} {\bibfnamefont {M.~K.}\ \bibnamefont
  {{Moravvej-Farshi}}}, \bibinfo {author} {\bibfnamefont {Y.}~\bibnamefont
  {{Yu}}}, and\ \bibinfo {author} {\bibfnamefont {J.}~\bibnamefont
  {{M{\o}rk}}},\ }\bibfield  {title} {\bibinfo {title} {Small and large signal
  analysis of photonic crystal {F}ano laser},\ }\href
  {https://doi.org/10.1109/JLT.2018.2877816} {\bibfield  {journal} {\bibinfo
  {journal} {J. Lightwave Techn.}\ }\textbf {\bibinfo {volume} {36}},\ \bibinfo
  {pages} {5611} (\bibinfo {year} {2018})}\BibitemShut {NoStop}%
\bibitem [{\citenamefont {Coldren}\ \emph {et~al.}(2012)\citenamefont
  {Coldren}, \citenamefont {Corzine},\ and\ \citenamefont
  {Masanovic}}]{Coldren}%
  \BibitemOpen
  \bibfield  {author} {\bibinfo {author} {\bibfnamefont {L.~A.}\ \bibnamefont
  {Coldren}}, \bibinfo {author} {\bibfnamefont {S.~W.}\ \bibnamefont
  {Corzine}}, and\ \bibinfo {author} {\bibfnamefont {M.~L.}\ \bibnamefont
  {Masanovic}},\ }\href@noop {} {\emph {\bibinfo {title} {Diode lasers and
  photonic integrated circuits}}}\ (\bibinfo  {publisher} {Wiley, 2nd ed.},\
  \bibinfo {year} {2012})\BibitemShut {NoStop}%
\bibitem [{\citenamefont {Debnath}\ \emph {et~al.}(2019)\citenamefont
  {Debnath}, \citenamefont {Zhang},\ and\ \citenamefont
  {M\o{}lmer}}]{PhysRevA.100.053821}%
  \BibitemOpen
  \bibfield  {author} {\bibinfo {author} {\bibfnamefont {K.}~\bibnamefont
  {Debnath}}, \bibinfo {author} {\bibfnamefont {Y.}~\bibnamefont {Zhang}}, and\
  \bibinfo {author} {\bibfnamefont {K.}~\bibnamefont {M\o{}lmer}},\ }\bibfield
  {title} {\bibinfo {title} {Collective dynamics of inhomogeneously broadened
  emitters coupled to an optical cavity with narrow linewidth},\ }\href
  {https://doi.org/10.1103/PhysRevA.100.053821} {\bibfield  {journal} {\bibinfo
   {journal} {Phys. Rev. A}\ }\textbf {\bibinfo {volume} {100}},\ \bibinfo
  {pages} {053821} (\bibinfo {year} {2019})}\BibitemShut {NoStop}%
\end{thebibliography}%

\end{document}